\pdfoutput=1
\documentclass[11pt]{article}
\usepackage{mathtools,amssymb,amsfonts,graphicx}

\DeclareFontFamily{U}{bbold}{}
\DeclareFontShape{U}{bbold}{m}{n}
 {  <-5.5> s*[1.04] bbold5
    <5.5-6.5> s*[1.04] bbold6
    <6.5-7.5> s*[1.04] bbold7
    <7.5-8.5> s*[1.04] bbold8
    <8.5-9.5> s*[1.04] bbold9
    <9.5-11.5> s*[1.04] bbold10
    <11.5-16> s*[1.04] bbold12
    <16-> s*[1.04] bbold17
 }{}

\usepackage[bbgreekl]{mathbbol}
\DeclareSymbolFontAlphabet{\mathbbm}{bbold}
\DeclareSymbolFontAlphabet{\mathbb}{AMSb}%

\usepackage{hyperref}
\usepackage[nosort]{cite}
\usepackage{verbatim}
\usepackage{multicol,color,xcolor,longtable}

\DeclareFontFamily{U}{matha}{\hyphenchar\font45}
\DeclareFontShape{U}{matha}{m}{n}{
      <5> <6> <7> <8> <9> <10> gen * matha
      <10.95> matha10 <12> <14.4> <17.28> <20.74> <24.88> matha12
      }{}
\DeclareSymbolFont{matha}{U}{matha}{m}{n}

\DeclareMathSymbol{\oleft}{2}{matha}{"68}
\DeclareMathSymbol{\oright}{2}{matha}{"69}

\usepackage[small,bf,hang]{caption}
\usepackage{slashed}
\usepackage{latexsym,epsfig}

\definecolor{darkred}{rgb}{0.65,0.15,0}
\definecolor{darkgreen}{rgb}{.05,.5,.05}
\hypersetup{pdfborder={0 0 0},colorlinks=true,urlcolor=darkred,citecolor=blue,linkcolor=darkred,linktocpage=true}

\usepackage{mathrsfs}
\usepackage{dsfont}
\usepackage[Symbolsmallscale]{upgreek}

\makeatletter
\g@addto@macro\bfseries{\boldmath}
\makeatother

\newcommand{\vir}{\mathfrak{vir}}
\newcommand{\hevir}{\hat{\mathfrak{e}}_{8}\oleft\mathfrak{vir}}

\newcommand{\virm}{\mathfrak{vir}^{-}}
\newcommand{\hevirm}{\hat{\mathfrak{e}}_{8}\oleft\mathfrak{vir}^-}
\newcommand{\Virm}{\mathrm{Vir}^{-}}

\newcommand{\nn}{\nonumber}

\newcommand{\ints}{\mathds{Z}}
\newcommand{\reals}{\mathds{R}}

\def\bea{\begin{eqnarray}}\def\eea{\end{eqnarray}}
\newcommand{\be}{\begin{equation}}
\newcommand{\ee}{\end{equation}}
\newcommand{\CR}{\nonumber \\*}

\newcommand{\cL}{{\mathcal L}}
\newcommand{\cD}{{\mathcal D}}

\newcommand{\cF}{{\mathcal{F}}}
\newcommand{\cG}{{\mathcal{G}}}
\newcommand{\cA}{{\mathcal{A}}}
\newcommand{\cB}{{\mathcal{B}}}
\newcommand{\cM}{{\mathcal{M}}}
\newcommand{\cV}{{\mathcal{V}}}
\newcommand{\cJ}{{\mathcal{J}}}
\newcommand{\cS}{\mathcal{S}}
\newcommand{\cP}{{\mathcal{P}}}
\newcommand{\cQ}{{\mathcal{Q}}}
\newcommand{\cC}{{\mathcal{C}}}

\newcommand{\bbA}{\mathbbm{A}}
\newcommand{\bbF}{\mathbbm{F}}

\newcommand{\ord}[1]{{\scriptscriptstyle (#1)}}

\newcommand{\dK}{{\mathsf{K}}}
\newcommand{\dd}{{\mathrm{d}}}

\newcommand{\mf}[1]{{\mathfrak{#1}}}

\makeatletter

\@addtoreset{equation}{section}
\makeatother

\newcommand{\Tr}{{\mathrm{Tr}}}

\newcommand{\ket}[1]{ \lvert #1 \rangle}
\newcommand{\bra}[1]{ \langle #1 \rvert}
\newcommand{\braket}[2]{ \langle #1 | #2 \rangle}

\newcommand{\Ket}[1]{ \big\lvert #1 \big\rangle}
\newcommand{\Bra}[1]{ \big\langle #1 \big\rvert}
\newcommand{\Braket}[2]{ \big\langle #1 \big| #2 \big\rangle}

\newcommand{\sugrarho}{\varrho}
\newcommand{\sugrasigma}{\sigma}

\def\hE8{\widehat{E}_{8}}
\def\he8{\hat{\mathfrak{e}}_{8}}

\makeatletter
\def\sbm{\@ifstar\sbm@starred\sbm@unstarred}
\newcommand{\sbm@unstarred}[1]{%
\raisebox{1.1pt}{\scalebox{1.}[.88]{\ensuremath{[}}}%
\mathbbm{#1}\raisebox{1.1pt}{\scalebox{1.}[.88]{\ensuremath{]}}}%
}
\newcommand{\sbm@starred}[1]{%
\raisebox{1.1pt}{\scalebox{1.}[.88]{\ensuremath{[}}}%
{#1}\raisebox{1.1pt}{\scalebox{1.}[.88]{\ensuremath{]}}}%
}
\makeatother

\newcommand{\hbm}[1]{\widehat{\mathbbm{#1}}}

\makeatletter
\DeclareRobustCommand\fJ{\@ifnextchar({\@shiftfJ}{\mathfrak{J}}}
\def\@shiftfJ(#1){\mathfrak{J}^{(#1)}}
\DeclareRobustCommand\fP{\@ifnextchar({\@shiftfP}{\mathcal{P}}}
\def\@shiftfP(#1){\mathcal{P}^{(#1)}}
\makeatother

\newcommand{\bbLambda}{\mathbbm{\Lambda}}

\newcommand{\bbGamma}{\mathbbm{\Gamma}}

\newcommand{\cU}{\mathcal{U}}

\usepackage[shadow,textwidth=2.7cm]{todonotes}
\usepackage{ifthen}
\reversemarginpar
\newcounter{todocounter}

\colorlet{gicolor}{red!30!white}

\newcommand{\giinline}[2][]{
  \ifthenelse { \equal {#1} {} }
    { \def\temp {#2} }  
    { \def\temp {#1} }   
  \refstepcounter{todocounter}\todo[color=gicolor,inline,caption={\textbf{\thetodocounter. GI} \temp}]{\textbf{\thetodocounter. GI:} #2}{}}

\colorlet{fccolor}{blue!40!white}

\newcommand{\fcinline}[2][]{
  \ifthenelse { \equal {#1} {} }
    { \def\temp {#2} }  
    { \def\temp {#1} }   
  \refstepcounter{todocounter}\todo[color=fccolor,inline,caption={\textbf{\thetodocounter. FC} \temp}]{\textbf{\thetodocounter. FC:} #2}{}}

\colorlet{akcolor}{green!40!white}

\newcommand{\akinline}[2][]{
  \ifthenelse { \equal {#1} {} }
    { \def\temp {#2} }  
    { \def\temp {#1} }   
  \refstepcounter{todocounter}\todo[color=akcolor,inline,caption={\textbf{\thetodocounter. AK} \temp}]{\textbf{\thetodocounter. AK:} #2}{}}

\colorlet{hscolor}{orange!20!white}

\newcommand{\hsinline}[2][]{
  \ifthenelse { \equal {#1} {} }
    { \def\temp {#2} }  
    { \def\temp {#1} }   
  \refstepcounter{todocounter}\todo[color=hscolor,inline,caption={\textbf{\thetodocounter. HS} \temp}]{\textbf{\thetodocounter. HS:} #2}{}}

\colorlet{gbcolor}{yellow!40!white}

\newcommand{\gbinline}[2][]{
  \ifthenelse { \equal {#1} {} }
    { \def\temp {#2} }  
    { \def\temp {#1} }   
  \refstepcounter{todocounter}\todo[color=gbcolor,inline,caption={\textbf{\thetodocounter. GB} \temp}]{\textbf{\thetodocounter. GB:} #2}{}}

\usepackage{graphicx}

\makeatletter
\def\widebreve{\mathpalette\wide@breve}
\def\wide@breve#1#2{\sbox\z@{$#1#2$}%
     \mathop{\vbox{\m@th\ialign{##\crcr
\kern0.08em\brevefill#1{0.7\wd\z@}\crcr\noalign{\nointerlineskip}%
                    $\hss#1#2\hss$\crcr}}}\limits}
\def\brevefill#1#2{$\m@th\sbox\tw@{$#1($}%
  \hss\resizebox{#2}{\wd\tw@}{\rotatebox[origin=c]{90}{\upshape(}}\kern.7pt\hss$}
\makeatother

\newcommand{\sugraGamma}{{\widebreve{\Gamma}}}

\newcommand{\sugraupgamma}{{\hspace{.02em}\breve{\hspace{-.02em}\upgamma\hspace{.02em}}\hspace{-.02em}}}

\newcommand{\sugrachi}{\breve{\chi}}

\newcommand{\prepi}{{}^{\pi}\!}

\begin{document}

{\flushright {CPHT-RR061.092023}\\[5mm]}

\vspace{15mm}

\begin{center}
    {\LARGE \sc Maximal $D=2$ supergravities\\[4mm]{}from higher dimensions}
    \\[13mm]

{\large
Guillaume Bossard${}^{1}$, Franz Ciceri${}^2$,\\[1ex] Gianluca Inverso${}^{3}$ and Axel Kleinschmidt${}^{2,4}$}

\vspace{8mm}
${}^1${\it Centre de Physique Th\'eorique, CNRS, Institut Polytechnique de Paris, \\ FR-91128 Palaiseau cedex, France}
\vskip 1.2 ex
${}^2${\it Max-Planck-Institut f\"{u}r Gravitationsphysik (Albert-Einstein-Institut)\\
Am M\"{u}hlenberg 1, DE-14476 Potsdam, Germany}
\vskip 1.2 ex
${}^3${\it INFN, Sezione di Padova \\
    Via Marzolo 8, 35131 Padova, Italy}
\vskip 1.2 ex
${}^4${\it International Solvay Institutes\\
ULB-Campus Plaine CP231, BE-1050 Brussels, Belgium}

\end{center}

\vspace{10mm}

\begin{center} 
\hrule

\vspace{6mm}

\begin{tabular}{p{14cm}}
{\small%
We develop in detail the general framework of consistent  Kaluza--Klein truncations from $D{=}11$ and type II supergravities to gauged maximal supergravities in two dimensions.
In particular, we unveil the complete bosonic dynamics of all gauged maximal supergravities that admit a geometric uplift.
Our construction relies on generalised Scherk--Schwarz reductions of E$_9$ exceptional field theory.
The application to the reduction of $D{=}11$ supergravity on $S^8\times S^1$ to SO(9) gauged supergravity is presented in a companion paper.
}
\end{tabular}
\vspace{5mm}
\hrule
\end{center}

\thispagestyle{empty}

\newpage
\setcounter{page}{1}


\setcounter{tocdepth}{2}
 \tableofcontents

\vspace{5mm}
\noindent\rule{\textwidth}{1.5pt}


\section{Introduction and summary}
\label{sec:intro}

Gauged supergravity theories with maximal supersymmetry represent rich sources of theories that are especially relevant in the context of black hole physics and holography.
One is particularly interested in those gauged supergravities in $D$ dimensions that can arise from ten- or eleven-dimensional supergravity by consistent truncations.
Here, `consistent'  means that any solution of the $D$-dimensional theory can be viewed as (or \emph{uplifted to}) a solution of a higher-dimensional theory.

Gauged supergravities generally possess an intricate structure of  classical solutions with different amounts of residual (super-)symmetry, and a special role is played by vacuum solutions.
Constructing the so-called scalar potential governing the landscape of vacua 
has been achieved for all maximal supergravity theories in $D\geq 3$ space-time dimensions~\cite{Nicolai:2000sc,Nicolai:2001sv,deWit:2004nw,Samtleben:2005bp,deWit:2007kvg,Bergshoeff:2007ef}.
While some results for $D=2$ are available  in~\cite{Samtleben:2007an}, the general construction of gauged supergravities in $D=2$ dimensions has not been achieved.
In particular, the general form of the scalar potential in $D=2$ maximal supergravity, which is the focus of the present paper, was not known until recently.
Its form was announced in~\cite{Bossard:2022wvi} and we here explain how to derive it along with the full bosonic dynamics of all $D=2$ gauged maximal supergravities that admit consistent uplifts to ten or eleven dimensions.

Ungauged maximal supergravity in $D=2$ dimensions has an infinite-dimensional global symmetry that includes the affine Kac--Moody group E$_9$ and that is realised on-shell~\cite{Geroch:1972yt,Geroch:1970nt,Belinsky:1971nt,Julia:1980gr,Breitenlohner:1986um,Nicolai:1987kz,Julia:1996nu,Paulot:2006zp}.
The realisation of the symmetry relies on the introduction of an infinite tower of dual scalar fields that propagate altogether the 128 bosonic physical degrees of freedom. In this sense, the E$_9$ symmetry is similar to the global E$_7$ symmetry in $D=4$ dimensions. In a \textit{gauged} supergravity theory, a part of the global symmetry is promoted to a local symmetry, for instance using the embedding tensor formalism~\cite{Nicolai:2000sc,Nicolai:2001sv,deWit:2002vt}.

The main obstacle to the construction of $D=2$ gauged maximal supergravities, compared to $D>2$, is that E$_9$ is an infinite-dimensional affine Kac--Moody symmetry and the fermions must transform under its maximal compact subgroup $K(\mathrm{E}_9)$ that is not of Kac--Moody type and whose representation theory is largely unknown.\footnote{See~\cite{Nicolai:2004nv} for a construction of the action of $K(\mathrm{E}_9)$ on the fermions of supergravity and~\cite{Kleinschmidt:2021agj} for more general representations. For the usual analysis of gauged supergravities one requires moreover invariant tensors mediating between the bosons and fermions and their construction and properties are a challenge.}
The general form of the scalar potential is usually determined using supersymmetry and therefore was not accessible with the methods of~\cite{Samtleben:2007an}. 
One can also work in specific duality frames in which only a finite-dimensional symmetry group is manifest and a standard supersymmetry analysis can in principle be carried out case by case. The only  example worked out in this way is  SO(9) gauged supergravity~\cite{Ortiz:2012ib}.
This theory notably admits a half-supersymmetric domain wall solution~\cite{Ortiz:2012ib}, which can be interpreted as a solution of IIA supergravity that is conformal to AdS$_2 \times S^8$. The latter describes the near-horizon geometry of D0-branes \cite{Horowitz:1991cd}, and the corresponding solution in eleven dimensions is the SO(9)-invariant pp-wave of~\cite{Nicolai:2000zt}. However, whether or not the full SO(9) theory itself arises as a consistent truncation of eleven-dimensional supergravity on $S^8\times S^1$ was not established at the time of its construction.
We prove this is the case in the companion paper~\cite{SO9} by applying the methods developed in the present paper. The existence of such a consistent truncation is particularly relevant for the holographic study of the conjectured dual quantum mechanics, which is the supermembrane matrix model~\cite{Hoppe,deWit:1988wri,Banks:1996vh}.

\medskip

Arguments based on generalised structure groups for supersymmetric backgrounds \cite{Lee:2014mla,Cassani:2019vcl} strongly imply that any consistent truncation leading to a gauged maximal supergravity should be expressible as a so-called generalised Scherk--Schwarz reduction (gSS) of exceptional field theory (ExFT)~\cite{Grana:2008yw,Aldazabal:2011nj,Geissbuhler:2011mx,Grana:2012rr,Berman:2012uy,Musaev:2013rq,Aldazabal:2013mya,Berman:2013cli,Aldazabal:2013via,Lee:2014mla,Hohm:2014qga,Hohm:2017wtr}.
Exceptional field theory is a framework that rewrites maximal supergravity in a form that makes manifest the lower-dimensional hidden rigid Cremmer--Julia symmetry E$_{11-D}$~\cite{Cremmer:1978ds,Cremmer:1997ct}.\footnote{With E$_{11-D}$ we denote the split real form of the Lie (or Kac--Moody) group of Cartan type E$_{11-D}$. Its Lie algebra will be written as $\mathfrak{e}_{11-D}$.}
In this formalism, diffeomorphisms and $p$-form gauge symmetries along an internal space are captured in a unified fashion in terms of so-called generalised diffeomorphisms~\cite{Coimbra:2011nw,Berman:2012vc}.
Both eleven-dimensional and type IIB supergravity can be described in one common framework.
Exceptional field theory is also powerful for testing perturbative stability of solutions arising from consistent Kaluza--Klein truncations~\cite{Malek:2019eaz,Guarino:2020flh} and thus represents a useful tool for studying certain aspects of the swampland conjectures \cite{Ooguri:2016pdq,Palti:2019pca}. 
The construction of E$_{11-D}$ exceptional field theory has been completed for all $D\geq 2$~\cite{Berman:2010is,Hohm:2013pua,Hohm:2013vpa,Hohm:2013uia,Hohm:2014fxa,Abzalov:2015ega,Musaev:2015ces,Berman:2015rcc,Bossard:2018utw,Bossard:2021jix}, see also~\cite{Hohm:2019bba,Berman:2020tqn} for reviews.\footnote{A construction of $E_{11}$ exceptional field theory, up to some algebraic conjectures, was presented in~\cite{Bossard:2017wxl,Bossard:2019ksx,Bossard:2021ebg}, building partly on earlier work in~\cite{West:2001as,West:2003fc,Tumanov:2016abm}.}

Generalised Scherk--Schwarz reductions are based on an object called the \textit{twist matrix} that encodes the dependence on the internal coordinates of all fields and gauge parameters.\footnote{We always impose that this dependence respects the exceptional field theory section constraint, such that the reduction can be interpreted geometrically in terms of ten- or eleven-dimensional supergravity.} 
The twist matrix is constructed in such a way that the internal coordinate dependence effectively factors out of the equations of motion. 
The latter then reduce to those of a $D$-dimensional gauged maximal supergravity.
The gauging is encoded in an embedding tensor which is identified with a certain generalised notion of torsion associated to the twist matrix.

In this paper, we apply the method of generalised Scherk--Schwarz reduction to the recently constructed E$_9$ exceptional field theory~\cite{Bossard:2018utw,Bossard:2021jix} and formulate the full bosonic dynamics of the resulting gauged supergravity, which we now summarise.
The gauged supergravity pseudo-Lagrangian is the sum of a topological term and a potential term:
\begin{align}
\label{eq:Lintro}
    \mathcal{L}^{\text{pseudo}}_{\text{gsugra}}= \mathcal{L}^{\text{top}}_{\text{gsugra}}- V_{\text{gsugra}}^{\phantom{top}}\,.
\end{align}
This is a pseudo-Lagrangian in the sense that its Euler--Lagrange equations do not determine the full dynamics but one has to additionally impose a (gauged) twisted self-duality equation as is standard for manifest duality invariance in even space-time dimensions. Moreover, we shall adopt the conformal gauge for the $D=2$ metric for convenience in our analysis, so that we must impose separately the associated  Virasoro constraint.

The result for the scalar potential was already announced in~\cite{Bossard:2022wvi} and takes the compact form
\begin{align}
\label{eq:VsugraIntro}
V_{\text{gsugra}}=\frac{1}{2\varrho^3}\,\bra{\theta}M^{-1}\ket{\theta}+\frac{1}{2\varrho}\eta_{-2\,\alpha\beta}\,\bra{\theta}T^{\alpha}M^{-1} T^{\beta\dagger}\ket{\theta}\,,
\end{align}
where $\bra{\theta}$ represents the constant embedding tensor, taking values in the so-called basic representation of E$_9$. 
The generators of the global symmetry are denoted by $T^\alpha$ and $\eta_{-2\,\alpha\beta}$ yields a particular contraction of the generators. The scalar $\varrho$ is the dilaton arising in the reduction to $D=2$ and $M$ is a Hermitian element in an extension of E$_9$ (including the negative mode number Virasoro generators) that encodes the scalar fields of the theory. 
The `matrix' $M$ involves infinitely many components, but the scalar products in the potential above depend only on finitely many fields for any possible embedding tensor.
The bra-ket notation and field content will be reviewed in detail in Section~\ref{sec:revsec}.

The topological term is given as a top-form by
\begin{align}\label{eq:Ltop gsugra intro}
\cL^{\rm top}_{\rm gsugra}\ \dd x^0\wedge\dd x^1 
=
2\varrho\,\mathbf{D}\breve\chi_1 + \varrho\,\langle\theta|\mathbf{O}(M)|F\rangle\,,
\end{align}
where $\mathbf{D}\breve\chi_1$ is the gauged  $K(\mathfrak{e}_9)$ covariant differential of an auxiliary one-form  $\breve\chi_1$, that is on-shell dual to the central charge component of the scalar field current. It belongs to an indecomposable representation of $K(\mathfrak{e}_9)$, and the covariant differential  $\mathbf{D}\breve\chi_1$ involves a Wess--Zumino type term for the scalar fields. The term $\langle\theta|\mathbf{O}(M)|F\rangle$ represents a certain scalar field dependent contraction of the embedding tensor $\bra{\theta}$ and the gauge field strength $\ket{F}$.
Both schematic terms in~\eqref{eq:Ltop gsugra intro} are displayed explicitly in~\eqref{eq:sugra Ltop}.

Unlike for gauged supergravity theories in higher-dimensions there are no kinetic terms for the scalar or vector fields. These terms are subsumed in a twisted self-duality equation that is compatible with the Euler--Lagrange equations derived from~\eqref{eq:Lintro}.
This twisted self-duality equation is given by
\begin{align}
\label{eq:twsdintro}
    \star P = P^{(1)}\,,
\end{align}
where $P$ is a current one-form built out of the scalar fields $M$ of the theory and $P^{(1)}$ is the same current with mode number shifted by 1 and the centre component given by the one-form $\breve\chi_1$ that enters in the topological term. Their precise definitions are given in \eqref{eq: gauged sugra P and Q} and \eqref{eq:gsugra shifted P}. As in the ungauged theory, this duality constraint ensures that among the scalar fields only $128$ degrees of freedom are propagating as necessary for maximal supergravity.

\medskip

The results above will be derived from a complete generalised Scherk--Schwarz reduction of E$_9$ ExFT and we perform the calculations in the so-called extended formulation of E$_9$ ExFT. The qualifying adjective refers to the fact the theory is not only invariant under the affine Kac--Moody E$_9$ symmetry but also under half a Virasoro algebra~\cite{Bossard:2021jix}. The equivalent so-called minimal formulation of~\cite{Bossard:2021jix} requires fewer fields but has slightly more complicated transformation laws. 

The reduction ansatz to $D=2$ must also take into account certain constrained scalar fields. We present a careful analysis of the (absence of) potential ambiguities arising from this extra complication that does not exist for exceptional field theories with $D>2$.
With the complete ansatz, and after checking the structure of the resulting gauge transformations, one has to perform the reduction of all the individual pieces of E$_9$ ExFT.
These are also given by a potential term, a topological term and a twisted self-duality equation.
In this  paper we restrict to gaugings that do not involve the trombone~\cite{LeDiffon:2008sh,LeDiffon:2011wt}, which in $D=2$ is realised by the $L_0$ scaling symmetry. 
We present the reduction of the various pieces separately, leading to~\eqref{eq:Lintro} and~\eqref{eq:twsdintro}.
We also derive the reduction of the non-Lagrangian duality equation for the vector field strength.
For any non-trombone gauging and an appropriate duality frame one can always write a physical Lagrangian involving only finitely many fields. This will be exemplified in the companion paper \cite{SO9}.
There we also show that the requirement that such a gauging admit an uplift to ten or eleven dimensions reduces the number of independent components of the embedding tensor to a finite amount.
This set of admissible gauge couplings is nonetheless very large and as of yet mostly unexplored.
In fact, even for $D>2$ gauged maximal supergravities, a classification of all models admitting an uplift is a difficult task that has not been achieved yet.

The structure of the present article is as follows. We first give a general review of our notation and the expected structure of gauged supergravity in $D=2$ dimensions in Section~\ref{sec:rev}. Then, in Section~\ref{sec:kinematics}, we present the generalised diffeomorphisms of E$_9$ ExFT and the gSS ansatz for them. The requirement that this give rise to a consistent gauge algebra for gauged supergravity leads to the gSS condition and already allows us to study uplift conditions for Lagrangian gaugings. In Section~\ref{sec:top}, we then reduce the topological term of E$_9$ ExFT and in Section~\ref{sec:pot} its potential term. 
The role of constrained two-forms in the duality equations for the vector field strength is discussed in Section~\ref{sec:nonLag} before we make some concluding remarks in Section~\ref{sec:conclusions}. Several appendices contain additional details on some more technical derivations.


\section{\texorpdfstring{Symmetries and gaugings of $D=2$ maximal supergravity}{Symmetries and gaugings of D=2 maximal supergravity}}
\label{sec:rev}

This section reviews the known and expected structures in the bosonic sector of two-dimensional maximal supergravity, including its gauged extensions.

\subsection{\texorpdfstring{Ungauged $D=2$ supergravity and its algebraic structures}{Ungauged D=2 supergravity and its algebraic structures}}
\label{sec:revsec}

We begin with a very brief summary of the symmetries of (the bosonic sector of) unguaged $D=2$ maximal supergravity and of the phrasing of its dynamics in terms of a covariant twisted self-duality constraint. A more detailed discussion in our conventions can be found in Section~2 of\cite{Bossard:2021jix}.

\subsubsection{\texorpdfstring{Dilaton-gravity coupled to a coset model in $D=2$}{Dilaton-gravity coupled to a coset model in D=2}}

The bosonic sector of $D=2$ (ungauged) maximal supergravity can be described in terms of a physical Lagrangian with manifest off-shell E$_8$ global symmetries.
We specify {\it physical} Lagrangian to distinguish a proper Lagrangian from a pseudo-Lagrangian for which the Euler--Lagrange equations must be supplemented by duality equations.
These symmetries arise from a non-linear sigma model on $\mathrm{E}_8/(\mathrm{Spin}(16)/\ints_2)$, coupled to a dilaton $\varrho>0$ and to $D=2$ gravity:
\begin{equation}
\label{eq:sugun}
\cL_{\text{sugra}} = \sqrt{-g}\,\varrho\,\left(
R- g^{\mu\nu}\eta^{AB} \mathring{P}_{\mu\,A} \mathring{P}_{\nu\,B}
\right)\,.
\end{equation}
Here, $\mathring{P}_{\mu\,A}$ are the (Hermitian) currents of the non-linear sigma model, with $A=1,\ldots,248$ denoting the adjoint representation of $\mathfrak{e}_8$.
These currents are constructed from a coset representative $\mathring{V}\in\mathrm{E}_8$ via
\begin{equation}
\mathring{P}_{\mu\,A} T^A_0  = \frac12\partial_\mu \mathring{V}\,\mathring{V}^{-1}+\mathrm{h.c.}
\end{equation}
where the $T^A_0 $ form a basis of the $\mathfrak{e}_8$ generators and the subset of anti-Hermitian generators spans the $\mathfrak{so}(16)$ subalgebra that is being quotiented out in the sigma model.
The subscript $_0$ will be motivated momentarily when we introduce the relevant $\mathfrak{e}_9$ structures.
The invariant metric over $\mathfrak{e}_8$ is denoted $\eta^{AB}=\frac{1}{60}f^{AC}{}_D f^{BD}{}_C$ and its inverse $\eta_{AB}$, where $f^{AB}{}_C$ are the structure constants.

The equations of motion read
\begin{equation}\label{eq:ungauged physical eom}
\nabla_\mu\partial^\mu\varrho = 0\,,\qquad
\nabla_{\!\mu}\big( \varrho\, \mathring{V}^{-1}\mathring{P}_{A}^\mu\,T^A_0  \mathring{V} \big) = 0 \,, \qquad 
R =  g^{\mu\nu}\eta^{AB} \mathring{P}_{\mu\,A} \mathring{P}_{\nu\,B}
\; .
\end{equation}
They have to be supplemented by the Einstein equations obtained by varying with respect to the uni-modular metric $\tilde{g}_{\mu\nu}$ that appears in the decomposition $g_{\mu\nu} = e^{2\sigma} \tilde{g}_{\mu\nu}$.
The variation with respect to $\sigma$ yields the first equation above.
It is common to choose the conformal gauge $\tilde{g}_{\mu\nu} = \eta_{\mu\nu}$,
in which case the equations formerly associated to the variation of $\tilde g_{\mu\nu}$ reduce to the Virasoro constraint
\begin{equation}\label{eq:vir cstr ungauged}
\delta \tilde{g}^{\mu\nu}\big(2\partial_\mu \varrho \,\partial_\nu\sigma - \partial_\mu \partial_\nu \varrho - \varrho\,\textbf{} \eta^{AB}\mathring{P}_{\mu\,A}\mathring{P}_{\nu\,B}\big)= 0\,,
\end{equation}
with $\delta\tilde{g}^{\mu\nu}$ an arbitrary symmetric traceless matrix.
In this paper we will work exclusively in conformal gauge.

The equations of motion enjoy a much richer set of global symmetries than the original Lagrangian~\eqref{eq:sugun} that is invariant under $\mathfrak{e}_8$. This richer set is generated by the affine Lie algebra $\mathfrak{e}_9$. Notice that this is true regardless of the choice of conformal gauge. 
There are several ways to capture such on-shell symmetries \cite{Geroch:1972yt,Geroch:1970nt,Belinsky:1971nt,Julia:1980gr,Breitenlohner:1986um,Nicolai:1987kz,Julia:1996nu,Paulot:2006zp}.
In Section~\ref{sec:extsym} below, we summarise the presentation of \cite{Bossard:2021jix} based on a twisted self-duality equation and refer to it for the relation to earlier formulations and to \cite{Breitenlohner:1986um} in particular.

\subsubsection{Algebras and their representations}
\label{sec:alg}

We begin by describing the affine symmetry algebra and its Virasoro extension.  We refer the reader also to~\cite{Bossard:2018utw,Bossard:2021jix} for more details on our notation.
First, one extends the $T^A_0$ generators to the loop algebra $\hat{\mathfrak{e}}_8$ over $\mathfrak{e}_8$, with generators $T^A_m$, $m\in\ints$, central charge $\dK$ and commutation relations:
\begin{align}
\big[ T^A_m\,,\,T^B_n  \big] &= f^{AB}{}_C T^C_{m+n} + m \,\eta^{AB} \delta_{m+n,0} \dK\,,
\intertext{where $f^{AB}{}_{C}$ are the structure constants of $\mathfrak{e}_8$.
Associated to  $\hat{\mathfrak{e}}_8$ there is a Virasoro algebra with generators $L_m$ and commutation relations}
\big[ L_m\,,\,L_n  \big] &= (m-n) L_{m+n} + \frac{c_{\vir}}{12} m(m^2-1) \delta_{m+n,0} \dK\,,
\\
\label{e:virT}
\big[ L_m\,,\,T^A_n  \big] &= -n T^A_{m+n}\,.
\end{align}
More precisely, the Virasoro generators are defined in a highest/lowest weight representation of $\hat{\mathfrak{e}}_8$ through the Sugawara construction~\cite{Sugawara:1967rw,Goddard:1986bp}.
Several relevant objects in gauged supergravity as well as in ExFT transform in the so-called basic representation that will be defined momentarily and in this representation of $\mathfrak{e}_9$ we have that $c_\vir=8$ and $\dK$ is the identity. 
We will leave $c_\vir$ symbolic as the formulation carries over to other affine algebras in the basic representation.
The Lie algebra $\mathfrak{e}_9$ is the affine extension of $\mathfrak{e}_8$, obtained by combining $\hat{\mathfrak{e}}_8$ with a derivation that we identify with $L_0$. 
We will combine the generators presented above into a basis $T^\alpha$ of $\hevir$ (the symbol $\oleft$ reflects that $\vir$ normalises $\hat{\mathfrak{e}}_8$, see~\eqref{e:virT}), with structure constants $f^{\alpha\beta}{}_\gamma$:
\begin{equation}
\hevir = \langle\{ T^\alpha \}\rangle = \langle\{ T^A_m\,,\ L_m\,,\ \dK\}\rangle\,,\qquad
[T^\alpha\,,\,T^\beta] = f^{\alpha\beta}{}_\gamma T^\gamma\,.
\end{equation}

The algebras $\hat{\mathfrak e}_8$ and $\mathfrak{e}_9$ are associated to the (centrally extended) loop group $\widehat{\rm E}_8$ and to the affine Kac--Moody group E$_9$, respectively.
These groups can be further extended by the exponential of the algebra $\virm$ of non-positive Virasoro generators $L_m$, $m\le0$. We denote by $\Virm$ the group obtained by this exponential map.\footnote{The exponential map defining $\Virm$ is to be interpreted in terms of formal power series in an auxiliary spectral parameter as described in Section~2 of~\cite{Bossard:2021jix}. For details on the rigorous definition of E$_9$ as a group, see Appendix~F~of~\cite{Bossard:2021jix} and references therein.}

There is no invariant pairing on all of $\hevir$, but non-degenerate invariant pairings can be defined on the subalgebra $\hat{\mathfrak{e}}_8\oplus \langle L_k\rangle$ for any $k\in \ints$, where the centrally extended loop algebra is supplemented by a single Virasoro generator $L_k$. These pairings will play a recurring role in our construction and are defined by
\begin{align}
\label{eq:etak}
    \eta_{k\,\alpha \beta} T^\alpha \otimes T^\beta = \sum_{m\in\ints} \eta_{AB} T_m^A \otimes T_{k-m}^B - L_k \otimes \dK - \dK \otimes L_k\,.
\end{align}
While the sum on the left-hand side is over all generators, the ones not belonging to $\hat{\mathfrak{e}}_8\oplus \langle L_k\rangle$ do not appear on the right-hand side.
These objects are invariant under $\hat{\mathfrak{e}}_8$ and transform under $\vir$ as follows
\begin{equation}
\label{eq:eta trf}
\eta_{k\,\alpha \beta} \Big([L_m,T^\alpha] {\otimes} T^\beta {+} T^\alpha {\otimes} [L_m,T^\beta]  \Big)
=
(m-k) \eta_{k{+}m\,\alpha \beta} T^\alpha {\otimes} T^\beta -\frac{c_\vir}{6}(m^3{-}m)\delta_{m+k,0}\dK{\otimes}\dK\,.
\end{equation}
Notice that $\eta_{k\,\alpha\beta}$ is invariant under $L_k$ and carries weight $k$ under $L_0$. We also write  $\eta_{\alpha\beta}$ for the unshifted $\mf{e}_9$-invariant form $\eta_{0\, \alpha\beta}$.

We moreover introduce the shift operators $\cS_m$, $m\in\ints$, which shift the mode number of loop and Virasoro generators and annihilate the central charge:
\begin{equation}\label{eq:shift op def}
\cS_m(T^A_n) = T^A_{m+n}\,,\quad \cS_m(L_n) = L_{m+n}\,,\quad \cS_{m}(\dK) = 0\,.
\end{equation}
This implies the following identity
\begin{align}
\eta_{n\,\alpha\beta} T^\alpha\otimes \cS_{m}(T^\beta)=\eta_{n+m\,\alpha\beta} T^\alpha\otimes T^\beta+L_{m+n}\otimes \mathsf{K}\,,\;\;\;\;\;\;\;\;\;\;\;\;\text{for}\;n,m\in\mathds{Z}\,.\label{eq:etaSid}
\end{align}
It follows from this formula that the shift operators do not commute with $\hevir$ because of the central element.

Besides E$_9$ we also require its maximal unitary subgroup that we denote by $K(\mathrm{E}_9)$ and that is generated by the subalgebra $K(\mathfrak{e}_9)$ of anti-Hermitian generators inside $\mathfrak{e}_9$. We define Hermitian conjugation on $\hevir$ by
\begin{equation}
(T^A_m)^\dagger = \eta_{AB} T^B_{-m}\,,\quad
(L_m)^\dagger = L_{-m}\,,\quad
\dK^\dagger=\dK\,.
\end{equation} 

The basic representation, denoted $\overline{R(\Lambda_0)_s}$, of $\hevir$ is the irreducible lowest weight representation built on top of an $\mathfrak{e}_8$- and $\mf{sl}_2$-invariant lowest-weight vector~\cite{Kac,Goddard:1986bp}. The subscript $s$ on the representation is called its (conformal) weight and corresponds to the eigenvalue of the so-called standard derivation $L_0{+}s \dK$ of the affine algebra. In this paper, we will use Fock space notation for the elements of the basic representation such that the lowest-weight vector $\bra{0}$ satisfies
\begin{align}
\bra{0} T_m^A = 0\quad\text{(for $m\leq 0$)} \,,\quad  \bra{0} L_m = 0 \quad\text{(for $m\leq 1$)} \,,\quad  \bra{0} \dK = \bra{0}\; , 
\end{align}
and other states of the representation are generated by the repeated action of the raising operators $T_m^A$ for $m>0$.
We will introduce  a basis $\bra{e^M}$ specifically for bra vectors with $s=-1$, such that $\bra{W}\in\overline{R(\Lambda_0)_{-1}}$ is expanded as $\bra{W} = W_M\bra{e^M}$. 

There is also a dual representation $R(\Lambda_0)_s$ with `ket' vectors $\ket{V}$. 
We write the action of a generator $X= X_\alpha T^\alpha\in\hevirm$ on a vector $\ket{V}\in R(\Lambda_0)_s$ as
\begin{align}
\label{eq:trmV}
 \delta_X \ket{V} =    X_\alpha\, \bbdelta^\alpha\ket{V} = -\Big(X_\alpha T^\alpha \ket{V} +  s  X_{0}\ket{V}\Big)\,,
\end{align}
where $X_0$ is the coefficient of $X$ along $L_0$.
More generally, we write the components of $X=X_\alpha T^\alpha$ along $L_n$ as $X_n$.

\subsubsection{\texorpdfstring{Extended symmetries in $D=2$}{Extended symmetries in D=2}}
\label{sec:extsym}

In order to realise the rigid E$_9$ symmetry as a local action on fields, one must introduce an infinite set of scalar fields dual to the original E$_8$ currents $\mathring{P}_{\mu\,A}$.
Together with the dilaton $\varrho$ and the conformal scale factor $\sigma$, these fields parametrise the coset space E$_9/K(\mathrm{E}_9)$. 

A further extension of field space turns out to be convenient in order to phrase the duality relations for $\mathring{P}_{\mu\,A}$ in terms of an E$_9$ invariant and $K(\mathrm{E}_9)$ covariant twisted self-duality equation.
In particular, in order to reproduce the dilaton factor in the second equation of \eqref{eq:ungauged physical eom}, we introduce another infinite set of fields $\varphi_m$, $m\ge1$, dual to $\varrho$ and associated to the negative Virasoro generators.\footnote{One can alternatively introduce a single auxiliary vector field rather than a series of dual dilatons. This would lead to the \textit{minimal} formulation of $D=2$ gravity following the same construction as Section~5 of~\cite{Bossard:2021jix}.}
In total, the scalar fields parametrise the infinite-dimensional coset space
\begin{equation}\label{eq:HeVirm coset}
\frac{\widehat{\mathrm E}_8\rtimes\Virm}{K(\mathrm{E}_9)}\,.
\end{equation}
We parametrise \eqref{eq:HeVirm coset} in terms of a coset representative $V$, transforming under infinitesimal $\hevirm$ elements as follows
\begin{equation}\label{eq:ungauged coset trf}
\bbdelta^\alpha V(x) = \breve{k}^\alpha(x) V(x) + V(x) T^\alpha\,, \quad \breve{k}^\alpha(x) = - ( \breve{k}^\alpha(x) )^\dagger \in K(\mathfrak{e}_9)
\,,\quad T^\alpha\in\hevirm
\end{equation}
where $\breve{k}^\alpha(x) $ is the compensating transformation needed to preserve a choice of $K(\mathrm{E}_9)$ gauge for~$V$.
The coset representative can be decomposed as
\begin{equation}
V = \sugraGamma V_{\scalebox{0.6}{loop}} \,,\quad \sugraGamma\in\Virm\,,\ V_{\scalebox{0.6}{loop}}\in\widehat{\rm E}_8\,,
\end{equation}
where $V_{\scalebox{0.6}{loop}}$ parametrises $\widehat{\rm E}_8/K(\mathrm{E}_9)$ and the $\Virm$ element is expanded as
\begin{equation}
\sugraGamma = \sugrarho^{-L_0} e^{-\varphi_1 L_{-1}} e^{-\varphi_2 L_{-2}} e^{-\varphi_3 L_{-3}}\cdots \,.
\end{equation}
The loop element $V_{\scalebox{0.6}{loop}}$ also contains the conformal factor $\sigma$ through  $e^{-\sigma\dK}$:
\begin{align}
    V= e^{-\sugrasigma\dK} \mathring{V} e^{Y_{1\,A}T^A_{-1}} e^{Y_{2\,A}T^A_{-2}}\cdots \,, \label{GaugeNegative}
\end{align}
where $\mathring{V}$ represents the E$_8$ coset and the $Y_n^A$ for $n>0$ are the dual scalar fields.

For several objects defined above in $D=2$ supergravity we add a breve accent $\breve{\cdot}$ to distinguish them from the equivalent objects defined in exceptional field theory in \cite{Bossard:2021jix}, while keeping manifest the similarity between the two theories. All the group-theoretic algebraic identities demonstrated in \cite{Bossard:2021jix} also apply to the equivalent gauged supergravity objects. 

We introduce the $\hevir$ valued Hermitian current and its component expansion
\begin{equation}\label{eq:ungauged P}
P_\mu = \frac12 \partial_\mu  V V^{-1} + \mathrm{h.c.}
= \sum_{m\in \ints}  P_{\mu\,A}^{\hspace{2mm}m}\, T^A_m + \sum_{m\in \ints} P_{\mu\,m} L_m +P_{\mu\,\dK} \dK\,,
\end{equation}
that transforms non-linearly under $\hevirm$ through the compensating transformation $\breve{k}^\alpha$.
We can now present the twisted self-duality equation
\begin{equation}\label{eq:twsd ungauged}
\star P = \cS_1\big(P\big) + \breve\chi_1\dK\,.
\end{equation}
In this equation, we moved to form notation by setting $P=P_\mu\dd x^\mu$. The Hodge dual $\star$ is defined with respect to the Minkowski metric $\eta_{\mu\nu}$. 
We also introduced a new auxiliary one-form $\breve\chi_1=\breve\chi_{1\,\mu} \dd x^\mu$ to compensate for the lack of a central element in $\mathcal{S}_1(P)$ according to~\eqref{eq:shift op def}. 
Together with the appropriate transformation of $\breve\chi_1$ this restores covariance of the right-hand side of~\eqref{eq:twsd ungauged} under $K(\mathrm{E}_9)$ transformations.\footnote{In fact, to capture the symmetries and dynamics of $D=2$ ungauged maximal supergravity we can choose to work in the adjoint representation of $\hevir$, so that $\dK$ is represented trivially and $\breve\chi_1$ becomes unnecessary. We give a presentation valid in any representation because it generalises to the case of gauged supergravity and ExFT.}

Equation \eqref{eq:twsd ungauged} is invariant under $\widehat{\mathrm{E}}_8\rtimes\Virm$ and independent of the choice of parametrisation of the coset representative \eqref{eq:sugra coset rep}.
In order to make contact with the original system \eqref{eq:ungauged physical eom}, we must choose the $K(\mathrm{E}_9)$  gauge~\eqref{GaugeNegative} where $V_{\scalebox{0.6}{loop}}$ is generated by $T^A_m$ with $m\le0$, such that the component $P_{\mu\,A}^{\hspace{2.3mm}0}$ equals the original E$_8$ current $\mathring{P}_{\mu\,A}$.
The physical equations of motion \eqref{eq:ungauged physical eom} are then identified with the integrability conditions for \eqref{eq:twsd ungauged}, which is best understood by mapping the latter to the Breitenlohner--Maison linear system \cite{Breitenlohner:1986um}, as explained in Section~2~of~\cite{Bossard:2021jix}.

\subsection{Gaugings and the embedding tensor formalism}

\label{sec:gaugings}

Gauging a supergravity theory amounts to promoting  a subgroup of the rigid symmetries of the original model to a local invariance.
In order to do so in $D=2$ maximal supergravity, it is necessary to introduce vector fields, which should transform in an appropriate representation of E$_9$.
Arguments based on the structure of tensor hierarchies \cite{Samtleben:2007an,deWit:2008ta,Cederwall:2021ymp} indicate that this representation should be $R(\Lambda_0)_{-1}$, hence we shall denote the vectors fields $\ket{A_\mu}$.
The selection of a gauge subalgebra of the rigid symmetries and its coupling to the vector fields is most conveniently captured in terms of an \emph{embedding tensor}~\cite{Nicolai:2000sc}.
It allows us to treat all possible gaugings on an equal footing and formally maintain covariance with respect to the duality group of the original (ungauged) theory.
The idea is to introduce the constant object $\bra{\Theta_\alpha}$, called the embedding tensor, to couple $\ket{A_\mu}$ to the symmetry generators.
Partial spacetime derivatives are then covariantised to
\begin{equation}\label{eq:gsugra covd}
D_\mu\Phi = \partial_\mu\Phi -\braket{\Theta_\alpha}{A_\mu}\,\bbdelta^\alpha\Phi\,,
\end{equation}
where $\Phi$ is any field of the theory and $\delta_X \Phi = X_\alpha \bbdelta^\alpha\Phi$ denotes its infinitesimal variation with respect to~$X \in \hevirm$.
The representation content of the embedding tensor in generic dimension is constrained by supersymmetry. 
Currently, we lack a $K(\mathrm{E}_9)$ covariant formulation of the fermionic sector of $D=2$ maximal supergravity, which would be necessary to determine the representation content of the $D=2$ embedding tensor.
However, it can be argued \cite{Samtleben:2007an} that (Lagrangian) gaugings should be parametrised by an element $\bra\theta\in \overline{R(\Lambda_0)_{-2}}$ such that
\begin{equation}\label{eq:gsugra covD only theta}
D_\mu\Phi = \partial_\mu\Phi +\eta_{-1\,\alpha\beta}\bra{\theta}T^\beta\ket{A_\mu}\,\bbdelta^\alpha\Phi\,,
\qquad\qquad
\bra{\Theta_\alpha} = -\eta_{-1\,\alpha\beta} \bra\theta T^\beta\,.
\end{equation}
Note that the appearance of $\eta_{-1\,\alpha\beta}$ here is consistent with the different weights of the representations of $\bra{\theta}$ and $\ket{A_\mu}$ that are being paired.
The simplified ansatz~\eqref{eq:gsugra covD only theta} is further confirmed by matching the formalism with specific models \cite{Ortiz:2012ib}.
With reference to the definition \eqref{eq:etak}, notice that such gaugings always involve $L_{-1}$ and no other Virasoro generator. We will briefly discuss how this formalism extends to gaugings of generic $\virm$ elements at the end of this section. However, we will find in Section~\ref{sec:kinematics} that all gaugings admitting a higher-dimenisonal origin can be cast in the form \eqref{eq:gsugra covD only theta}, or a slight generalisation that is necessary to include non-Lagrangian theories involving the gauging of the $L_0$ scaling symmetry.

Consistency of the gauging requires that the embedding tensor should be invariant under the gauge algebra it defines.
This translates into a quadratic constraint that reads \cite{Samtleben:2007an}
\begin{equation}\label{eq:QC theta only}
\eta_{-1\,\alpha\beta}\bra\theta T^\alpha\otimes\bra\theta T^\beta = 0\,.
\end{equation}
Inequivalent gauged maximal supergravities are determined by duality orbits of solutions of the quadratic constraint.

The dynamics of the bosonic sector of $D=2$ gauged supergravities were partially formulated in \cite{Samtleben:2007an}. 
The approach followed there is based on the gauge-covariantisation of the Breitenlohner--Maison linear system \cite{Breitenlohner:1986um} and leads to a $K(\mathrm{E}_9)$ gauge-fixed formulation of the theory in terms of a (physical) Lagrangian that is gauge invariant but not duality covariant.
All gauged supergravities involve a potential for the scalar fields. 
In higher dimensions, the scalar potential is calculated by requiring invariance under supersymmetry.
However, this is not possible in $D=2$ due to the lack of a $K(\mathrm{E}_9)$ covariant formulation of the fermionic sector of the theory. 
As a result, \cite{Samtleben:2007an} only formulates the dynamics of $D=2$ gauged supergravities up to the construction of the scalar potential, which is left as an unknown.\footnote{%
An exception is the formulation of SO(9) gauged maximal supergravity \cite{Ortiz:2012ib}. This theory could be constructed in full by foregoing E$_9$ covariance and the formulation in terms of a linear system to exploit instead the fact that $D=2$ fermions transform naturally in irreducible representations of Spin(9) $\subset$ SO(16). We will analyse the SO(9) gauging and its higher-dimensional origin in the companion paper~\cite{SO9}.}

In this paper we do not present the formulation of the dynamics of \cite{Samtleben:2007an}, but rather choose to pursue a manifestly duality covariant and $K(\mathrm{E}_9)$ invariant formulation based on a pseudo-Lagrangian and twisted self-duality constraint, derived from E$_9$ ExFT.
To this end, we simply need to describe the covariantisation of the currents introduced in \eqref{eq:ungauged P}.
The scalar fields still parametrise the coset space \eqref{eq:hevirm coset} in terms of a coset representative $V$.
Then, we have
\begin{align}\label{eq: gauged sugra P and Q}
P_\mu &= \frac12 D_\mu V V^{-1} +\mathrm{h.c.}  
= \frac12\Big(\partial_\mu V V^{-1} -\braket{\Theta_\alpha}{A_\mu}  VT^\alpha V^{-1}   \Big)  +\mathrm{h.c.}  \\\nonumber
Q_\mu &= \frac12 D_\mu V V^{-1} -\mathrm{h.c.} 
= \frac12\Big(\partial_\mu V V^{-1} -\braket{\Theta_\alpha}{A_\mu}  VT^\alpha V^{-1}   \Big)  -\mathrm{h.c.}  
-\braket{\Theta_\alpha}{A_\mu} \breve{k}^\alpha
\,,
\end{align}
where we have also defined the anti-Hermitian composite connection $Q_\mu$, so that $D_\mu VV^{-1} = P_\mu + Q_\mu$.
The compensating $K(\mathfrak{e}_9)$ transformation $\breve{k}^\alpha$ was introduced in \eqref{eq:ungauged coset trf}.
Note that with these definitions both $P_\mu$ and $Q_\mu$ include positive Virasoro generators, but the symmetry of the theory is only $\widehat{\mathrm{E}}_8\rtimes\Virm$.

With these definitions, the twisted self-duality constraint is still given by \eqref{eq:twsd ungauged}, by introducing the shift operators \eqref{eq:shift op def} and the auxiliary one-form $\breve\chi_{1\,\mu}$.
In fact, it is convenient to introduce some further objects which will parallel the structure of E$_9$ ExFT in the following sections.
We define the gauged supergravity shifted currents as 
\begin{equation}\label{eq:gsugra shifted P}
P^{(m)}_\mu = \cS_m(P_\mu)+\breve\chi_{\mu\,m}\,\dK\,,\qquad m\in\ints\,,
\end{equation}
where $\mathcal{S}_m$ is defined in \eqref{eq:shift op def} and $\breve\chi_{\mu\,m}$ for $m>0$ are independent auxiliary fields, introduced to achieve $K(\mathfrak{e}_9)$ covariance of the shifted currents.
We also set $\breve\chi_{\mu\,0}=P_{\mu\,\dK}$, so that $P^{(0)}_\mu=P_\mu$, and $\breve\chi_{\mu\,m}=\breve\chi_{\mu\,-m}$. Then, by hermiticity of $P_\mu$, we get $P^{(m)\,\dagger}_\mu=P^{(-m)}_\mu$.
Under $\hevirm$ these auxiliary forms transform only with respect to the $K(\mathfrak{e}_9)$ compensator:
\begin{equation}
\bbdelta^\alpha \breve\chi_{\mu\,m} = -(\breve{k}^\alpha){}_\gamma \,\omega^{\gamma\beta} P^{(m)}_{\mu\,\,\beta}\,,
\end{equation}
where $(\breve{k}^\alpha){}_\gamma$ are the components of the $K(\mathfrak{e}_9)$ element appearing in~\eqref{eq:ungauged coset trf} such that  
\be (\breve{k}^\alpha){}_\gamma T^\gamma = \breve{k}^\alpha\; , \label{KcompoSugra} \ee 
and  
\begin{equation}\label{eq:algebra cocycle}
\omega^{\alpha\beta}=-f^{\alpha\beta}{}_\dK    
\end{equation}
is the Lie algebra cocycle of $\hevir$. We have also introduced the current components such that $ P^{(m)}_{\mu\,\,\alpha}\,T^\alpha = P^{(m)}_\mu$.  

The one-forms $\breve\chi_{\mu\,m}$ are covariant under gauge transformations.
These definitions generalise the right-hand side of \eqref{eq:twsd ungauged}, which we can now rewrite as
\begin{equation}\label{eq:twsd gsugra}
\star P = P^{(1)}\,,\qquad \star^{|m|}P=P^{(m)}\,,\ m\in\ints\,,
\end{equation}
where $\star$ is again the flat space Hodge duality operator.\footnote{Note that the duality equation sets to zero all components in $P$ along the $\mf{so}(16)$ loop generators projected to the coset since they would be related to the $\mf{so}(16) \subset \mf{e}_8$ part that vanishes by construction.}
The second expression is a consequence of the first one except for its $\dK$ component, which gives a relation between different $\breve\chi_{\mu\,m}$.
The Virasoro constraint \eqref{eq:vir cstr ungauged} is also straightforwardly covariantised. 
A more convenient expression is 
\begin{equation}\label{eq:vir cstr gauged sugra}
\delta \tilde{g}^{\mu\nu}\Big(2 P_{\mu\,\dK} P_{\nu\,0} 
- P_{\mu\,0} P_{\nu\,0}
+ D_\mu P_{\nu\,0} -\eta^{AB} P_{\mu\, A}^{\hspace{2.3mm}0} P_{\nu\, B}^{\hspace{2.3mm}0} \Big) = 0\,,
\end{equation}
with $\delta\tilde{g}^{\mu\nu}$ symmetric traceless.
The rest of the dynamics of gauged maximal supergravity will be derived in Sections~\ref{sec:top} and \ref{sec:pot} from the consistent truncation of the E$_9$ ExFT pseudo-Lagrangian.

\medskip

Let us now come back to how \eqref{eq:gsugra covD only theta} is extended beyond the construction of \cite{Samtleben:2007an} to include the gauging of other Virasoro generators.
A first extension is obtained by considering gaugings that do not admit a Lagrangian description. 
These are gaugings that involve $L_0$, since this generator always induces a rescaling of the Lagrangian and is only a symmetry of the equations of motion. Gaugings of $L_0$ are analogous to gaugings of the trombone symmetry in higher-dimensional supergravity theories \cite{LeDiffon:2008sh,LeDiffon:2011wt}.
To describe such non-Lagrangian gaugings we introduce a second bra-vector $\bra{\vartheta}\in \overline{R(\Lambda_0)_{-1}}$ and extend the embedding tensor to
\begin{equation}\label{eq:Theta with vartheta}
\bra{\Theta_\alpha} = -\eta_{0\,\alpha\beta} \bra\vartheta T^\beta -\eta_{-1\,\alpha\beta} \bra\theta T^\beta +\delta_\alpha^\dK \bra\vartheta \,.
\end{equation}
We can still use the definitions \eqref{eq:gsugra covd}, \eqref{eq: gauged sugra P and Q} and simply replace the expression for $\bra{\Theta_\alpha}$.
The shift by $\bra{\vartheta}$ in the last term (where $\delta^\dK_\alpha$ is a Kronecker delta) is introduced such that a vector in the $R(\Lambda_0)_{-1}$ representation transforms according to (see~\eqref{eq:trmV})
\begin{equation}
\big(\bbdelta^\alpha\ket{V}\big)\bra{\Theta_\alpha} 
\ =\ -T^\alpha\ket{V}\bra{\Theta_\alpha}\ +\ \ket{V}\bra{\vartheta}
\ =\ -T^\alpha\ket{V} \big(\eta_{0\,\alpha\beta} \bra{\vartheta} T^\beta+\eta_{-1\,\alpha\beta} \bra{\theta} T^\beta\big)\,.
\end{equation}
This extension of the embedding tensor was derived in \cite{Bossard:2017aae} where a first analysis of generalised Scherk--Schwarz reductions for E$_9$ was carried out.
We will indeed find that all gaugings arising from consistent truncation of higher-dimensional supergravities can be cast in this form.

Covariance under the $\virm$ symmetries introduced in Section~\ref{sec:alg} requires us to treat all non-positive Virasoro generators on the same footing.
This entails a further extension of the embedding tensor, completing the components $\bra\vartheta$, $\bra\theta$ to a representation of $\virm$.
This is obtained by introducing a full series of components 
$\bra{\Theta_{-k}}$ of weights $k+1$ (in the sense of~\eqref{eq:trmV}) associated to gaugings of arbitrary $L_{-k}$ generators, so that the complete embedding tensor can transform covariantly under $\virm$:
\begin{equation}\label{eq:Theta virm exp}
\bra{\Theta_\alpha} = 
-\sum_{k=0}^\infty \eta_{-k\,\alpha\beta} \bra{\Theta_{-k}} T^\beta
+\delta_\alpha^\dK\bra{\Theta_{0}}\,,
\end{equation}
where each $\bra{\Theta_{-k}}$ equals the $L_{-k}$ component of $\bra{\Theta_\alpha}$.
The `original' $L_0$ and $L_{-1}$ components are identified with the first two terms:
\begin{equation}\label{eq:extended theta to unextended}
\bra{\Theta_{0}}=\bra\vartheta\,,\qquad
\bra{\Theta_{-1}}=\bra\theta\,.
\end{equation}
Again, the definitions \eqref{eq:gsugra covd}, \eqref{eq: gauged sugra P and Q} are also valid in this extended setting.
Under $\hevirm$, the $\eta_{k\,\alpha\beta}$ pairings transform according to \eqref{eq:eta trf}.
As a consequence, the components  $\bra{\Theta_{-k}}$ must transform indecomposably with respect to each other so that $\bra{\Theta_\alpha}$ transforms as a tensor:
\begin{equation}\label{eq:embtens rigid trf}
\bbdelta^\alpha\bra{\Theta_{-k}} = \bra{\Theta_{-k}}\,T^\alpha - \delta^\alpha_{L_0}\,\bra{\Theta_{-k}}
+ \sum_{0\le p\le k} (2p-k) \delta^\alpha_{L_{-p}} \bra{\Theta_{p-k}} \,,
\end{equation}
where $\delta^\alpha_{L_{-p}} $ is a Kronecker delta which selects a $\virm$ component of $T^\alpha$.
The original quadratic constraint \eqref{eq:QC theta only} can be always written as the general expression
\begin{equation}\label{eq:Theta QC}
\bra{\Theta_\beta}\otimes\bbdelta^\beta\!\bra{\Theta_\alpha} = -f^{\beta\gamma}{}_\alpha \bra{\Theta_\beta}\otimes\bra{\Theta_\gamma}+\bra{\Theta_\beta}\otimes \bra{\Theta_\alpha} T^\beta 
- \bra{\Theta_{0}}\otimes \bra{\Theta_\alpha} 
= 0\,,
\end{equation}
which simply reflects the statement that the embedding tensor should be a gauge singlet.
In terms of its irreducible expansion \eqref{eq:Theta virm exp} the constraint becomes
\begin{equation}\label{eq:component QC}
\sum_{p=0}^\infty \eta_{-p\,\alpha\beta}\bra{\Theta_{-p}}T^\alpha\otimes\bra{\Theta_{-k}}T^\beta
-\sum_{0\le p\le k} (2p-k) \bra{\Theta_{-p}}\otimes\bra{\Theta_{p-k}} 
= 0 \,,\qquad \forall\,k\ge0\,.
\end{equation}
The quadratic constraint for the case of unextended gaugings involving only $L_{-1}$ and/or $L_0$ is readily reproduced by setting $\bra{\Theta_{-k}}=0$ for $k\ge2$.
The quadratic constraint restricted to $\bra{\Theta_{0}}=\bra{\vartheta}$ and $\bra{\Theta_{-1}}=\bra{\theta}$ was given in~\cite{Bossard:2017aae}:
\begin{align}\label{eq:QCvarthetatheta}
\eta_{0\,\alpha\beta}\bra\vartheta T^\alpha \otimes\bra\vartheta T^\beta + 
\eta_{-1\,\alpha\beta}\bra\theta T^\alpha \otimes\bra\vartheta T^\beta
= 0\,,&\\[1ex]\nonumber
\eta_{0\,\alpha\beta}\bra\vartheta T^\alpha \otimes\bra\theta T^\beta + 
\eta_{-1\,\alpha\beta}\bra\theta T^\alpha \otimes\bra\theta T^\beta +\bra\vartheta\otimes\bra\theta - \bra\theta\otimes\bra\vartheta =0\,.&
\end{align}

\medskip

In Section~\ref{sec:kinematics} we will find that gaugings arising from a geometric generalised Scherk--Schwarz reduction of a higher-dimensional supergravity can always be cast in a form where $\bra{\Theta_{-k}}=0$ for $k\ge2$.\footnote{More precisely, if a gauging does admit a gSS uplift, any non-zero $\langle\Theta_{-k}|$ components with $k\ge2$ can be removed by rigid Vir$^-$ transformations or other field redefinitions. As discussed in Appendix~\ref{app:ha}, this is associated to setting to zero a certain object in the gSS ansatz, called $\bra{\tilde{h}_\alpha}$ and introduced in equation~\eqref{eq:halpha redef} below.}
Therefore, the definition \eqref{eq:Theta with vartheta} and especially the one \eqref{eq:gsugra covD only theta} for Lagrangian gaugings are sufficient to capture such models, at the cost of breaking formal $\Virm$ covariance of the gauged theory.
Nonetheless, it would be interesting to investigate whether gaugings that necessarily involve lower Virasoro generators have a physical interpretation and whether they are  compatible with maximal supersymmetry.

\section{\texorpdfstring{Generalised Scherk--Schwarz for E$_9$}{Generalised Scherk--Schwarz for E9}}
\label{sec:kinematics}

In this section we move to exceptional field theory and present the  Scherk--Schwarz ansatz for the generalised diffeomorphisms. We first review the $\hevirm$ extended generalised diffeomorphisms of \cite{Bossard:2021jix} to determine the most general ansatz. One finds that up to terms that do not affect the fields appearing in the gauged supergravity Lagrangian, one can always restrict to the Scherk--Schwarz ansatz described in \cite{Bossard:2022wvi}, which extends \cite{Bossard:2017aae} by a single additional constrained bra vector that is necessary for covariance under E$_9$. In particular, we prove that the condition that the embedding tensor is constant implies automatically the gauged supergravity quadratic constraint \eqref{eq:QC theta only}. 

\subsection{\texorpdfstring{$\hevirm$ generalised diffeomorphisms}{e8+vir generalised diffeomorphisms}}

Generalised diffeomorphisms act on fields through a generalised Lie derivative, defined in terms of internal derivatives $\bra\partial$ transforming in the infinite-dimensional $\overline{R(\Lambda_0)_{-1}}$ `bra' representation of E$_9$ and subject to an $\widehat{\mathrm{E}}_8\rtimes \Virm$ covariant, algebraic section constraint reviewed below. 
We indicate with a subscript the field the derivative acts on, while the position of the derivative bra is fixed to manifest the E$_9$ contraction, for example for a field  $\Phi$ we can write
\be 
\Phi \; \langle \partial_\Phi | = \left(\partial_M \Phi\right)  \; \langle e^M | \; ,
\ee
where $\bra{e^M}$ are the basis elements of $\overline{R(\Lambda_0)_{-1}}$ introduced in Section~\ref{sec:alg}.
The internal coordinates associated to $\partial_M$ are denoted by $y^M$. We will, however, only consider objects satisfying the section constraint reviewed below, that then depend only on a finite set of physical coordinates denoted~$y^I$.

The gauge parameters are given by a generalised vector $\ket\Lambda$, transforming in the $R(\Lambda_0)_{-1}$ `ket' representation of the rigid E$_9$ invariance, and by a set of ancillary parameters $\Sigma^{(k)}$, $k=1,\,2,\,\ldots$ transforming in the tensor product \mbox{$R(\Lambda_0)_{0}\otimes \overline{R(\Lambda_0)_{-1}}$}, where the `bra' components are required to satisfy the same algebraic section constraint as the internal derivatives $\bra\partial$.
To capture such gauge parameters we use the shorthand notation
\begin{equation}\label{eq:bbLambda1}
\bbLambda=\big(\,\ket\Lambda\,,\ \Sigma^{(k)}\,\big)\,.
\end{equation}
The generalised Lie derivative, acting on a field $\Phi$ with rigid $\hevirm$ variation denoted by $\bbdelta^\alpha\Phi$, reads
\begin{equation}\label{eq:genLie}
\cL_{\bbLambda} \Phi = \braket{\partial_\Phi}{\Lambda}\,\Phi + [\bbLambda]_\alpha \bbdelta^\alpha\Phi\,,
\end{equation}
where the parameter of the rigid transformation in the last term is defined as
\begin{equation}\label{eq:squareProj}
[\bbLambda]_\alpha = \eta_{0\,\alpha\beta}\bra{\partial_\Lambda}T^\beta\ket{\Lambda} +\sum_{k=1}^\infty \eta_{-k\,\alpha\beta} \Tr(\Sigma^{(k)}T^\beta)\,.
\end{equation}
Closure of the generalised Lie derivative requires us to impose the section constraint
\begin{equation}\label{eq:SC}
\eta_{0\,\alpha\beta}\bra{\partial_1} T^\alpha\otimes\bra{\partial_2} T^\beta = \bra{\partial_2} \otimes\bra{\partial_1} - \bra{\partial_1} \otimes\bra{\partial_2} 
\end{equation}
which further implies
\begin{equation}
\eta_{-n\,\alpha\beta}\bra{\partial_1} T^\alpha\otimes\bra{\partial_2} T^\beta = 0\,,\qquad n\ge1\,,
\end{equation}
as well as  
\begin{equation}
\eta_{1\,\alpha\beta}\bra{\partial_1} T^\alpha\otimes\bra{\partial_2} T^\beta
+\eta_{1\,\alpha\beta}\bra{\partial_2} T^\alpha\otimes\bra{\partial_1} T^\beta = 0\,.
\end{equation}
Here the two derivatives are independent and can act on any fields, gauge parameters, or products thereof.
The ancillary parameters are algebraically constrained so that~\cite{Bossard:2017aae}\footnote{The ancillary parameters $\Sigma^{(k)}$ are linear combinations of ket-bra tensors and the ket-part is understood to remain on the same side of the tensor product in the equation below, such that the section constraint only involves the bra vectors.}
\begin{equation}\label{eq:SCsigma}
\eta_{0\,\alpha\beta}\bra{\partial_*} T^\alpha\otimes\Sigma^{(k)} T^\beta = \Sigma^{(k)} \otimes\bra{\partial_*} - \bra{\partial_*} \otimes\Sigma^{(k)}\,,
\end{equation}
with $\bra{\partial_*}$ acting on any field or parameter.
The same section constraint also holds with both derivatives traded for independent ancillary parameters.

The rigid $\virm$ action on the generalised vectors $\ket\Lambda$ is induced by the Sugawara construction for Virasoro generators in the $R(\Lambda_0)_{-1}$ representation.
The ancillary gauge parameters, on the other hand, transform both tensorially under the full $\hevirm$ algebra, according to the $R(\Lambda_0)_{0}\otimes\overline{R(\Lambda_0)_{-1}}$ representation, but also indecomposably with respect to each other and $\ket\Lambda\bra{\partial_\Lambda}$ under $\virm$. 
This is to ensure that the projection $[\bbLambda]_\alpha$ in \eqref{eq:squareProj} transforms as a tensor in the coadjoint representation.
Explicitly, given an element $X\in \hevirm$,
\begin{equation}
X_\alpha \, \bbdelta^\alpha\,\Sigma^{(k)} = -X_\alpha \big[T^\alpha,\,\Sigma^{(k)}\big] + \sum_{0\le p<k} (2p-k) X_{-p} \Sigma^{(k-p)} + k X_{-k} \ket\Lambda\bra{\partial_\Lambda}\,.
\end{equation}

When acting on the field content of E$_9$ ExFT, the role of the ancillary parameters is to gauge some parabolic subgroup of the rigid $\widehat{E}_8$, as well as all strictly negative Virasoro generators.
In particular, this means that $\hevirm$ extended ExFT is invariant under finite, local transformations of the form
\begin{equation}\label{eq:local Virm element}
e^{-\alpha_1(x,y) L_{-1}}e^{-\alpha_2(x,y) L_{-2}}e^{-\alpha_3(x,y) L_{-3}}\cdots
\end{equation}
with local parameters $\alpha_{k}(x,y)$ corresponding to gauge parameters $\Tr\big[\Sigma^{(k)}(x,y)\big]$.
This local invariance can be used to partially gauge-fix the scalar field content of the theory.

A characteristic of the generalised Lie derivative defined above is that there exist trivial parameters of `$\Sigma$-only' type, namely parameters $\widetilde{\bbLambda}$ with vanishing $\ket\Lambda$ component, such that the generalised Lie derivative on any field $\Phi$ vanishes:
\begin{equation}
\widetilde{\bbLambda}=\big(\, 0\,,\ \widetilde\Sigma^{(k)}  \,\big)\,,\qquad
\sum_{k=1}^\infty \eta_{-n\,\alpha\beta} \Tr(\widetilde\Sigma^{(k)}T^\beta) = 0\quad\Leftrightarrow\quad \cL_{\widetilde{\bbLambda}}\Phi=0
\,. 
\end{equation}
Such parameters are always projected out algebraically in the E$_9$ ExFT dynamics.
We will now see how such parameters arise in the closure of the gauge algebra and construct a projected space where they are manifestly absent.

Closure of the algebra of generalised diffeomorphisms as well as the gauge transformation of vector fields are better encoded into a generalised Dorfman product \cite{Hohm:2017wtr,Bossard:2021jix} 
\begin{equation}\label{eq:Dorf1}
\bbLambda_1 \circ \bbLambda_2 = \big(\,\cL_{\bbLambda_1}\ket{\Lambda_2}\,,\ \Sigma^{(k)}_{12}\,\big)\,,
\end{equation}
where
\begin{align}
\Sigma^{(k)}_{12} =\ & 
  \cL_{\bbLambda_1}\Sigma^{(k)}_{2} 
+ \delta^k_1 \eta_{1\,\alpha\beta}\bra{\partial_{\Lambda_1}}T^\beta\ket{\Lambda_1}\,T^\alpha\ket{\Lambda_2}\bra{\partial_{\Lambda_1}}
+ \eta_{0\,\alpha\beta}\Tr(T^\alpha\Sigma^{(k)}) \, T^\beta\ket{\Lambda_2}\bra{\partial_{\Sigma_1}}  \CR&
- \ket{\Lambda_2}\bra{\partial_{\Sigma_1}}\Sigma^{(k)}_1 
- (k-1) \Big( \Tr(\Sigma^{(k)}_1)\,\ket{\Lambda_2}\bra{\partial_{\Sigma_1}} - \Sigma^{(k)}_1\braket{\partial_{\Sigma_1}}{\Lambda_2}  \Big)\,.
\end{align}
The Dorfman product is useful to determine the closure of the generalised Lie derivative, namely one finds
\begin{equation}
[\cL_{\bbLambda_1}\,,\,\cL_{\bbLambda_2}]= \cL_{\bbLambda_1 \circ \bbLambda_2}\,.
\end{equation}

The notion of a generalised Dorfman product was introduced in \cite{Hohm:2017wtr} for O$(8,8)$ and E$_8$ ExFT, such that it would satisfy the Leibniz identity.
This is not achieved for the $\hevirm$ extended product defined above.
One finds instead that \eqref{eq:Dorf1} satisfies the Leibniz identity only up to a $\Sigma$-only trivial parameter~\cite{Bossard:2021jix}:
\begin{equation}\label{eq:DorfLeib1}
  {\bbLambda}_1\circ (\,{\bbLambda}_2 \circ {\bbLambda}_3\,)
- {\bbLambda}_2\circ (\,{\bbLambda}_1 \circ {\bbLambda}_3\,)
- (\,{\bbLambda}_1\circ {\bbLambda}_2\,) \circ {\bbLambda}_3
= \big( \,0\,,\ \widetilde\Sigma^{(k)} \,\big)\,.
\end{equation}
Since $\Sigma$-only trivial parameters are always algebraically projected out in physically relevant expressions, the definitions given above are sufficient to describe the gauge structure of E$_9$ ExFT.
However, it can be desirable to work in a setup where the Leibniz identity holds exactly.
This happens if we restrict ancillary transformations to $\Sigma^{(1)}=\Sigma$ and $\Sigma^{(k)}=0$ for $k\ge2$.
One can then check that such a set of gauge parameters closes onto itself and the Leibniz identity holds:
\begin{equation}\label{eq:DorfLeib2}
  {\bbLambda}_1\circ (\,{\bbLambda}_2 \circ {\bbLambda}_3\,)
- {\bbLambda}_2\circ (\,{\bbLambda}_1 \circ {\bbLambda}_3\,)
- (\,{\bbLambda}_1\circ {\bbLambda}_2\,) \circ {\bbLambda}_3
= 0\,,\qquad \Sigma_i^{(k)} =0\,,\ k\ge2\,.
\end{equation}
This setup reflects the structure of \emph{minimal} E$_9$ ExFT \cite{Bossard:2021jix}, where the $\virm$ scalar fields $\phi_n$ for $n\ge 1$ are gauge-fixed to vanish.

Another option is to simply redefine generalised diffeomorphisms in terms of a space of projected ancillary parameters, where all $\Sigma$-only trivial parameters are automatically removed. This amounts to defining
\begin{equation}\label{sigmahat}
\hat\Sigma_\alpha  = \sum_{k=1}^\infty \eta_{-k\,\alpha\beta} \Tr\big( \Sigma^{(k)} T^\beta \big)
\end{equation}
and treating $\hat\Sigma_\alpha$ as the fundamental object, algebraically constrained to parametrise the same subalgebra of $\hevirm\!$ as its expansion in terms of the covariantly constrained $\Sigma^{(k)}$.
In particular, using the section constraint one proves that $\hat\Sigma_\alpha\bra\partial T^\alpha = 0$  and that $\hat\Sigma_\alpha\bra\partial \cS_{+1}(T^\alpha) $ is itself on section.
We will denote gauge parameters in the projected ancillary space as follows: 
\begin{equation}
\prepi\bbLambda = \big(\,\ket\Lambda\,,\ \hat\Sigma_\alpha  \,\big)\,,
\end{equation}
and the generalised Lie derivative will still be given by \eqref{eq:genLie}, in terms of the $\hevirm$ parameter
\begin{equation}\label{eq:squareProjSmall}
[\prepi\bbLambda]_\alpha = [\bbLambda]_\alpha = \eta_{0\,\alpha\beta}\bra{\partial_\Lambda}T^\beta\ket\Lambda +\hat\Sigma_\alpha\,.
\end{equation}
To guarantee $\hevirm$ covariance of this expression, the variation of $\hat\Sigma_\alpha$ must be
\begin{equation}\label{sigmahat rigid trf}
X_\beta\,\bbdelta^\beta \hat\Sigma_\alpha = 
-X_\beta \hat\Sigma_\gamma f^{\beta\gamma}{}_\alpha +\Big(\sum_{k=1}^\infty\,k\, X_{-k} \, \eta_{-k\,\alpha\beta} \bra{\partial_\Lambda} T^\beta \ket\Lambda\Big)\,.
\end{equation}

The generalised Dorfman product for the projected gauge parameters is still denoted by $\circ$ and is computed to be
\begin{equation}\label{eq:DorfSmallParam}
\prepi\bbLambda_1\circ\prepi\bbLambda_2 = \big(\, \cL_{\bbLambda_1}\ket{\Lambda_2}\,,\ \hat\Sigma_{12\,\alpha} \,\big)
\end{equation}
with
\begin{align}
\hat\Sigma_{12\,\alpha} =\ &
\mathcal L_{\bbLambda_1} \hat\Sigma_{2\,\alpha}
- [\bbLambda_1]_\alpha \braket{\partial_{\bbLambda_1}}{\Lambda_2}
+ f^{\gamma\delta}{}_\alpha [\bbLambda_1]_\gamma \, \eta_{0\,\delta\beta} 
  \bra{\partial_{\bbLambda_1}}T^\beta\ket{\Lambda_2}
\CR&
- [\bbLambda_1]_0 \, \eta_{0\,\alpha\beta}  
  \bra{\partial_{\bbLambda_1}}T^\beta\ket{\Lambda_2}
+ \sum_{k=1}^\infty \,k\, [\bbLambda_1]_{-k} \, \eta_{-k\,\alpha\beta}
  \bra{\partial_{\bbLambda_1}}T^\beta\ket{\Lambda_2}
\end{align}
and the Leibniz identity is satisfied:
\begin{equation}\label{eq:dorf leib proj}
  \prepi{\bbLambda}_1\circ (\,\prepi{\bbLambda}_2 \circ \prepi{\bbLambda}_3\,)
- \prepi{\bbLambda}_2\circ (\,\prepi{\bbLambda}_1 \circ \prepi{\bbLambda}_3\,)
- (\,\prepi{\bbLambda}_1\circ \prepi{\bbLambda}_2\,) \circ \prepi{\bbLambda}_3
= 0\,.
\end{equation}

\subsection{The generalised Scherk--Schwarz condition}
\label{sec:gSScond}

Consistent truncations to gauged maximal supergravities are based on generalised Scherk--Schwarz (gSS) reductions \cite{Grana:2008yw,Aldazabal:2011nj,Geissbuhler:2011mx,Grana:2012rr,Berman:2012uy,Musaev:2013rq,Aldazabal:2013mya,Berman:2013cli,Aldazabal:2013via,Lee:2014mla,Hohm:2014qga,Hohm:2017wtr}.
For E$_{n}$ expectional field theories with $n\le7$, the gSS ansatz is encoded entirely in a twist matrix $\cU\in\mathrm{E}_{n}$ and weight factor $r\in\reals^+$, dependent only on the internal coordinates and subject to the section constraint.

We first review the relevant structures for the case of E$_n$ ExFT with $n\leq 7$.
Using index notation, the internal coordinate dependence of generalised vectors such as the gauge parameter~$\Lambda^M$ is assumed to factorise in terms of the twist matrix:
\begin{equation}\label{eq:gss factorisation low rank}
\Lambda^M(x,y) = r^{-1}(y) \big(\cU^{-1}(y)\big){}^M{}_{\underline{N}} \ \lambda^{\underline N}(x)\ .
\end{equation}
The product $r^{-1}\cU^{-1}$ defines a global frame for the generalised tangent bundle and the condition for existence of a generalised Scherk--Schwarz reduction is that the associated generalised torsion is constant:
\begin{equation}\label{eq:gss cond lowrank}
\cL_{\Lambda_1}\Lambda_2^M = 
- \lambda_1^{\underline{P}} \lambda_2^{\underline{Q}} \  X_{\underline{PQ}}{}^{\underline{N}} \ 
r^{-1}\big(\cU^{-1}\big){}^M{}_{\underline{N}}\,,\qquad \text{$X_{\underline{PQ}}{}^{\underline{N}}$ constant\,.}
\end{equation}
Here both $\Lambda_1^M$ and $\Lambda_2^M$ are assumed to factorise as in \eqref{eq:gss factorisation low rank}.
The constant $X_{\underline{MN}}{}^{\underline{P}}$ is identified with the embedding tensor of the gauged supergravity theory obtained after truncation.
It can always be written as $X_{\underline{MN}}{}^{ \underline{P}} = \Theta_{\underline{M}\,\alpha} T^{\alpha\,\underline{P}}{}_{\underline{N}}$ where $T^{\alpha\,\underline{M}}{}_{{\underline{N}}}$ generate E$_{n}\times\reals^+$.
Importantly, $X_{\underline{MN}}{}^{\underline{P}}$ defined from \eqref{eq:gss cond lowrank} automatically satisfies the linear (representation) constraints for a gauged supergravity embedding tensor. Such linear constraints are imposed for $n\leq 8$ by requiring consistency of the fermion supersymmetry variations~
\cite{Nicolai:2000sc,Nicolai:2001sv,deWit:2004nw,Samtleben:2005bp,deWit:2007kvg,Bergshoeff:2007ef}.

The underlined indices in \eqref{eq:gss factorisation low rank} are not acted upon by the generalised Lie derivative.
On the other hand, we need to specify how rigid $\mathrm{E}_{n}\times\reals^+$ should act on
the right-hand side of \eqref{eq:gss factorisation low rank}.
We distinguish two different actions.
The first acts on $\cU^{-1}$ from the left and reproduces the standard transformation of the generalised vector but leaves the gauged supergravity coefficient $X_{\underline{PQ}}{}^{\underline{N}}$ invariant. 
We refer to this as the `ExFT' rigid transformations.
The second set of rigid transformations leaves the generalised vector invariant but acts on $\lambda^{\underline{M}}$ as well as on $\cU^{-1}$ from the right. This will be the `gauged supergravity' rigid transformation.
The factor $r$ has opposite weights under the ExFT and gauged supergravity $\reals^+$ rigid transformations.
The same distinctions will apply in the case of $\hevirm$ rigid transformations for E$_9$ ExFT, with $L_0$ playing the role analogous to the $\reals^+$ generator here.
We have so far used underlined indices in order to distinguish the two actions, but will henceforth drop them as the distinction can always be inferred by the nature of the objects in play.

\medskip

We now apply a similar logic to E$_9$ exceptional field theory.
A first difference compared to lower-rank exceptional groups, but also present for E$_8$, is the presence of ancillary parameters in the generalised diffeomorphisms.
In the following steps we will investigate in full generality how the ansatz for generalised vectors $\ket\Lambda$ can be extended to the ancillary parameters, extending the analysis carried out in \cite{Bossard:2017aae}.
Following the structure proposed for constrained $p$-forms in lower-rank ExFTs \cite{Hohm:2017wtr} and observing that ancillary gauge transformations are not present in gauged supergravity, we impose that both the generalised vectors $\ket\Lambda$ and ancillary parameters $\hat\Sigma_\alpha$ reduce to the same gauge parameters $\ket\lambda$ for the resulting gauged supergravity, namely
\begin{equation}\label{eq:gssgenvec}
\ket{\Lambda(x,y)} = r^{-1}(y)\,\cU^{-1}(y)\ket{\lambda(x)}\,,\qquad \hat\Sigma_\alpha(x,y) = \braket{H_\alpha(y)}{\lambda(x)}\,,
\end{equation}
where the first expression is the E$_9$ analogue of \eqref{eq:gss factorisation low rank}, except that now we take $\cU\in \hat{\mathrm{E}}_8\rtimes\Virm$ written in the $R(\Lambda_0)_0$ representation, the weight $r$ being its $L_0$ component, such that 
\begin{equation}\label{eq:twistU}
\cU = \cU_{\scalebox{0.6}{loop}} \,r^{-L_0} \,e^{-\psi_1L_{-1}}e^{-\psi_2L_{-2}}e^{-\psi_3L_{-3}}\cdots
\end{equation}
with $\cU_{\scalebox{0.6}{loop}}\in\hat{\mathrm E}_8$. The explicit appearance of $r$ in \eqref{eq:gssgenvec} is due to the $L_0$ weight of generalised vectors (namely, they transform in the $R(\Lambda_0)_{-1}$ representation).
Notice that we can always gauge-fix the twist matrix to sit in E$_9$ only, by exploiting the invariance of ExFT under ancillary transformations gauging the negative part of the Virasoro algebra.
This amounts to performing a finite, local $\Virm$ transformation such as \eqref{eq:local Virm element}, which acts on the twist matrix from the right and sets $\psi_n\to0$ for all $n\ge1$.
The component $\cU_{\scalebox{0.6}{loop}}r^{-L_0}\in\mathrm{E}_9$ is unaffected. 
For the moment we will not perform this gauge-fixing and make our presentation valid for any $\cU$ in \eqref{eq:twistU}.

The object $\bra{H_\alpha}$ in \eqref{eq:gssgenvec} captures the reduction ansatz for ancillary parameters and is a priori independent of $\cU$.
Notice that we only require the ancillary parameters to satisfy a reduction ansatz up to $\Sigma$-only trivial parameters, hence we work in terms of the projected $\hat\Sigma_\alpha$ as defined in \eqref{sigmahat}.
Still, the gSS ansatz for the projected $\hat\Sigma_\alpha$ must be compatible with the section constraint \eqref{eq:SCsigma}.
This means that we must be able to write an expansion analogous to~\eqref{sigmahat}:
\begin{equation}\label{eq:Halpha expansion}
\bra{H_\alpha} = \sum_{k=1}^\infty \eta_{-k\,\alpha\beta} \Tr\big(\Xi^{(k)}_M T^\beta\big)\bra{e^M}\,,
\end{equation}
where $\bra{e^M}$ form a basis of $\overline{R(\Lambda_0)_{-1}}$ and the $\Xi^{(k)}_M$ satisfy the section constraint just as $\Sigma^{(k)}$ in~\eqref{eq:SCsigma}.

The generalised Scherk--Schwarz condition analogous to \eqref{eq:gss cond lowrank} is then cast in terms of the generalised Dorfman product \eqref{eq:DorfSmallParam} and reads
\begin{equation}\label{eq:gss cond e9}
\prepi\bbLambda_1 \circ \prepi\bbLambda_2 = - \Braket{\Theta_\alpha{-}\delta_\alpha^\dK\Theta_{0}}{\lambda_1} \big(\,r^{-1}\cU^{-1} T^\alpha \ket{\lambda_2}\,,\, \bra{H_\beta}T^\alpha\ket{\lambda_2}  \,\big)\,,
\end{equation}
with constant $\bra{\Theta_\alpha}$.
We have introduced the shorthand notation $\Bra{\Theta_\alpha{-}\delta_\alpha^\dK\Theta_{0}} =\bra{\Theta_\alpha} -  \delta_\alpha^\dK\bra{\Theta_{0}}$.
The $\bra{\Theta_{0}}$ term is defined as the $L_0$ component of $\bra{\Theta_\alpha}$.  Its appearance here is analogous to the shift of the central charge component in \eqref{eq:Theta virm exp} and reflects the fact that $\ket{\lambda_i}$ has $L_0$ weight  equal to $-1$, so that $\Braket{\Theta_\alpha{-}\delta_\alpha^\dK\Theta_{0}}{\lambda_1}\ T^\alpha\ket{\lambda_2}=-\braket{\Theta_\alpha}{\lambda_1}\,\bbdelta^\alpha\ket{\lambda_2}$.
Condition \eqref{eq:gss cond e9} is motivated by the requirement that the ExFT covariant derivative $\cD_\mu$, involving vector gauge fields undergoing the same reduction ansatz as \eqref{eq:gssgenvec}, must reduce to the covariant derivative of gauged supergravity \eqref{eq:gsugra covd} when acting on fields undergoing a covariant factorisation ansatz.
The Leibniz identity \eqref{eq:dorf leib proj} will then also guarantee that the field strengths of the ExFT vector fields reduce to those of gauged supergravity.

The gSS condition \eqref{eq:gss cond e9} is made of two components, corresponding to the generalised vector and the ancillary gauge parameter.
The former component is equivalent to \eqref{eq:gss cond lowrank} of lower-rank ExFTs, although it is important to stress that ancillary parameters do enter this condition through the generalised Lie derivative.
Nonetheless, just as for lower-rank ExFTs invertibility of $\cU$ guarantees that one can always find a non-constant $\bra{\Theta_\alpha}$ satisfying the equality, hence the only non-trivial requirement is that $\bra{\Theta_\alpha}$ must be constant in order to define a consistent truncation.
On the other hand, the ancillary component of \eqref{eq:gss cond e9} is a priori more restrictive, as there is no reason in principle why its right-hand side should display the same structure of contractions, nor involve the  $\bra{\Theta_\alpha}$ tensor appearing in the generalised vector component.
In the following we will study the structure of these conditions and derive the expression of the embedding tensor.

We begin by defining the Weitzenb\"ock connection associated with $\cU$, which we denote \mbox{$\bra{W_\alpha}\otimes T^\alpha$} and write
\begin{equation}\label{eq:weitz}
\bra{\partial_\cU}\otimes\cU = \bra{W_\alpha}\,r \cU\otimes T^\alpha \cU\,.
\end{equation}
In index notation and temporarily restoring underlined indices, this definition reads
\begin{equation}
W_{\underline M\,\alpha} T^{\alpha\,\underline N}{}_{\underline P}
= r^{-1} \big(\cU^{-1}\big)^{M}{}_{\underline M} \big(\cU^{-1}\big)^{N}{}_{\underline P}\partial_M \cU^{\underline N}{}_N \,.
\end{equation}
The indecomposability of the transformation \eqref{sigmahat rigid trf} between the ancillary parameters $\hat\Sigma_\alpha$ and the associated $\ket\Lambda$ translates into an indecomposability between $\bra{H_\alpha}$ and the Weitzenb\"ock connection:
\begin{equation}\label{eq:Halpha trf}
X_\beta \,\bbdelta^\beta\braket{H_\alpha}{\lambda} 
= -f^{\gamma\delta}{}_\alpha X_\gamma \braket{H_\delta}{\lambda}
  - \sum_{k=1}^\infty k \, X_{-k}\,\eta_{-k\,\alpha\beta}\bra{W_\gamma{-}\delta_\gamma^\dK W_0}\, \cU T^\beta \cU^{-1}T^\gamma\ket{\lambda}\,.
\end{equation}

In the following, we will analyse the ansatz and its consistency in more detail. Compared to ExFT in higher dimensions, the discussion is slightly more technical due to the indecomposable structure of the representations. The final result for the components of the embedding tensor is summarised  in~\eqref{eq:theta summary} at the end of this section.

The transformation~\eqref{eq:Halpha trf} can be exponentiated to any element of $\widehat{\mathrm{E}}_8\rtimes\Virm$ and guarantees that the projection \eqref{eq:squareProjSmall} transforms covariantly.\footnote{One cannot exponentiate Lie algebra  elements with a positive mode number, but one can generate the group action by exponentiating elements for well chosen $X_\alpha$, see Appendix~F of~\cite{Bossard:2021jix}.}
This means that we can more efficiently encode the information in $\bra{H_\alpha}$ in terms of the projection \eqref{eq:squareProjSmall} conjugated by the twist matrix itself:
\begin{equation}\label{eq:LambdaProjGSS}
[\bbLambda]_\alpha \, \cU T^\alpha \cU^{-1} =
\Big(-\eta_{\alpha\beta}\Bra{W_\gamma{-}\delta_\gamma^\dK W_0} T^\beta T^\gamma\Ket\lambda + \braket{h_\alpha}{\lambda}\Big)\, T^\alpha\,,
\end{equation}
where $\bra{h_\alpha}$ is obtained by exponentiation of \eqref{eq:Halpha trf} with coefficients determined by \eqref{eq:twistU}.
It will be easier to work with $\bra{h_\alpha}$ in the following steps.\footnote{If we gauge-fix $\cU\in\mathrm{E}_9$, $\bra{H_\alpha}$ and $\bra{h_\alpha}$ are simply related by the action of $\cU$ on the coadjoint index.}
The shift of the central charge component of the Weitzenb\"ock connection is due to the explicit $r^{-1}$ weight in \eqref{eq:gssgenvec}.

We can now identify $\bra{\Theta_\alpha}$ from the generalised vector component of \eqref{eq:gss cond e9}. Writing
\begin{align}
T^\alpha\ket{\lambda_2}\, \Braket{\Theta_\alpha{-}\delta^\dK_\alpha\Theta_0}{\lambda_1}
&= r\cU\, \cL_{\bbLambda_1} \ket{\Lambda_2} 
\CR
&= 
\ket{\lambda_2}\bra{W_\alpha}T^\alpha\ket{\lambda_1}
-T^\alpha\ket{\lambda_2}\braket{W_\alpha}{\lambda_1}
-\cU T^\alpha\cU^{-1}\ket{\lambda_2}[\bbLambda_1]_\alpha
\end{align}
and comparing with \eqref{eq:LambdaProjGSS} we find
\begin{equation}\label{eq:Theta1}
\bra{\Theta_\alpha} = \bra{W_\alpha} 
-\eta_{\alpha\beta}\bra{W_\gamma}T^\beta T^\gamma
+\eta_{\alpha\beta}\bra{W_0}T^\beta + \bra{h_\alpha}\,.
\end{equation}
Notice that similar manipulations also lead us to the simple relation
\begin{equation}\label{eq:bbLambda red}
[\bbLambda]_\alpha \, \cU T^\alpha \cU^{-1} =
\Braket{\Theta_\alpha{-}W_\alpha}{\lambda}\,T^\alpha\,,
\end{equation}
which will be useful later on.
Looking at \eqref{eq:Theta1}, we stress that the coadjoint index of $\bra{h_\alpha}$ is constrained in a way similar to $\hat\Sigma_\alpha$ and $\bra{H_\alpha}$, meaning that it must be possible to find an expansion analogous to \eqref{sigmahat}:
\begin{equation}\label{eq:halpha expansion}
\bra{h_\alpha} = \sum_{k=1}^\infty \eta_{-k\,\alpha\beta} \Tr\big(\xi^{(k)}_M T^\beta\big)\bra{e^M}\,,
\end{equation}
where $\xi^{(k)}_M$ satisfy a `flat' version of the section constraint~\eqref{eq:SCsigma}
\begin{equation}
\eta_{\alpha\beta} \,\xi^{(k)}_M T^\alpha\otimes \bra{\partial_*}\cU^{-1} T^\beta 
= \bra{\partial_*}\cU^{-1}\otimes\xi^{(k)}_M-\xi^{(k)}_M\otimes\bra{\partial_*}\cU^{-1}\,,
\end{equation}
and the derivative $\bra{\partial_*}$ is on section and can act on any field or parameter of the theory.
This algebraic constraint on $\bra{h_\alpha}$ means that requiring constancy of the expression \eqref{eq:Theta1} is non-trivial, because the dependence on $\bra{W_\alpha}$ cannot entirely be compensated by $\bra{h_\alpha}$.

Notice that the expression analogous to \eqref{eq:Theta1} for lower rank $E_n$ ExFT does not include the additional term in $\langle h_\alpha |$ for $n\le 7$. This term is associated to the ancillary gauge parameter $\Sigma$ and appears starting from $n=8$. It follows that for $n\le 7$ the only allowed $E_n$  irreducible representations in the embedding tensor descending from the gSS ansatz are the ones known to be compatible with supersymmetry. For $E_9$ we do not know how to use supersymmetry to constrain the embedding tensor, but we have argued that $ \bra{\Theta_\alpha}$ must take the form \eqref{eq:Theta virm exp} in gauged supergravity. In order to recover this formula, we first compute the $L_0$ component of \eqref{eq:Theta1} to identify $\bra{\Theta_{0}}=\bra\vartheta$ and find
\begin{equation}\label{eq:vartheta}
\bra{\Theta_{0}} = \bra{\vartheta} = \bra{W_\alpha}T^\alpha\,.
\end{equation}
Then, we shift $\bra{h_\alpha}$ by terms compatible with its algebraic expansion \eqref{eq:halpha expansion}:
\begin{equation}\label{eq:halpha redef}
\bra{h_\alpha} = 
  \eta_{-1\,\alpha\beta}\bra{W_\gamma}T^\beta \cS_{+1}(T^\gamma) 
+ \eta_{-1\,\alpha\beta}\bra{w^+}T^\beta
+ \sum_{k=1}^\infty \eta_{-k\,\alpha\beta}\bra{W_{-k}}T^\beta
+ \bra{\tilde{h}_\alpha}\,.
\end{equation}
The term $\bra{\tilde{h}_\alpha}$ is constrained analogously to $\bra{{h}_\alpha}$ to admit an expansion such as \eqref{eq:halpha expansion}.
 We will show in Appendix~\ref{app:ha} that there is no loss of generality in setting $\bra{\tilde{h}_\alpha}$ to zero. 
In this expression $\bra{w^+}$ is a-priori independent from $\cU$ and is introduced to be the $\dK$ completion of $\bra{W_\alpha}\otimes \cS_{+1}(T^\alpha)$, so that we can define\footnote{More precisely, $\bra{W_\alpha}$ is invariant under the `ExFT' rigid transformations as described below \eqref{eq:gss factorisation low rank}, while it is a tensor under the gauged supergravity transformations. This is the completion needed for $\bra{W^+_\alpha}$ to also transform as a tensor under the latter.}
\begin{equation}\label{eq:shiftweitz}
\bra{W^+_\alpha} \otimes T^\alpha = \bra{W_\alpha}\otimes \cS_{+1}(T^\alpha) + \bra{w^+}\otimes \dK\,.
\end{equation}
The object $\bra{w^+}$ satisfies the `flat' version of the section constraint, namely it is on section with respect to $\bra{\partial_*}\cU$, analogously to $\xi^{(k)}_M$ in \eqref{eq:halpha expansion}, with $\bra{\partial_*}$ an internal derivative acting on any field or parameter.

Using \eqref{eq:halpha redef} and expressions \eqref{eq:commutator shifts} and \eqref{eq:commutator shifts for W}, we can rewrite \eqref{eq:Theta1} as
\begin{align}\label{eq:Theta2}
\bra{\Theta_\alpha} -\delta_\alpha^\dK\bra\vartheta = -\eta_{\alpha\beta}\bra{\vartheta}T^\beta
-\eta_{-1\,\alpha\beta}\bra{\theta}T^\beta 
+\bra{\tilde h_\alpha}\,,
\end{align}
with $\bra\vartheta$ given by \eqref{eq:vartheta} and
\begin{equation}\label{eq:theta}
\bra\theta = -\bra{W_\alpha}\cS_{+1}(T^\alpha)-\bra{w^+} = -\bra{W^+_\alpha}T^\alpha\,.
\end{equation}

We argue in Appendix~\ref{app:ha} that we can set $\bra{\tilde h_\alpha}=0$ in~\eqref{eq:Theta2}  without loss of generality. This further simplifies the setup of gSS reductions.
As a first step, we use the fact that ancillary gauge transformations in ExFT include the gauging of all $L_{-k}$ generators for $k\ge1$.
We can therefore gauge-fix the twist matrix in \eqref{eq:gssgenvec} to $\cU\in\mathrm{E}_9$, i.e. set $\psi_k=0$ in \eqref{eq:twistU}.
The expression for $\bra{H_\alpha}$ is then readily deduced from \eqref{eq:halpha expansion} to be 
\begin{equation}
\langle H_\alpha |  = r\,\eta_{-1\,\alpha\beta}\bra{W^+_\gamma}\cU T^\beta\cU^{-1} T^\gamma \,, \qquad \cU\in\mathrm{E}_9\,.
\end{equation}
This expression can be written in terms of a single (unprojected) ancillary parameter $\Sigma^{(1)}\equiv\Sigma$, 
so that the gSS ansatz \eqref{eq:gssgenvec} can be rewritten in unprojected form\footnote{This form of the gSS ansatz is also the one compatible with the minimal formulation of E$_9$ ExFT \cite{Bossard:2021jix}.}
\begin{equation}\label{eq:gssgenvec minimal}
\ket{\Lambda} = r^{-1}\,\cU^{-1}\ket{\lambda}\,,\quad \cU\in \mathrm{E}_9\,,\qquad
\Sigma^{(1)} = \Sigma = \cU^{-1}T^\alpha\ket{\lambda}\bra{W^+_\alpha}r\cU\,,\quad
\Sigma^{(k)} = 0\,.
\end{equation}
The expression for $\Sigma$ displayed here agrees with the gSS ansatz first described in \cite{Bossard:2017aae} and completes it by introducing the independent component $\bra{w^+}$ not considered there but required for covariance under rigid $\hevirm$ transformations. This additional object might be necessesary to write the most general generalised Scherk--Schwarz ansatz, but we do not know of an example where it cannot be set to zero. 
The same completed expression \eqref{eq:gssgenvec minimal} was presented in \cite{Bossard:2022wvi}.
The gSS condition \eqref{eq:gss cond e9} now reads
\begin{equation}
\label{eq:gss cond e9 unproj}
\bbLambda_1 \circ \bbLambda_2 = - \Braket{\Theta_\alpha{-}\delta_\alpha^\dK\vartheta}{\lambda_1} \Big(\,r^{-1}\cU^{-1} T^\alpha \ket{\lambda_2}\,,\, 
\cU^{-1}T^\beta T^\alpha \ket{\lambda_2}\bra{W^+_\beta}r\cU  \,\Big)\,,
\end{equation}
where it is now understood that only the $\Sigma^{(1)}$ ancillary parameters are non-vanishing on both sides of the equation.

Looking now at the ancillary component of the gSS condition,
we find that the gSS condition is automatically satisfied if $\bra\vartheta$ and $\bra\theta$ defined in \eqref{eq:vartheta} and \eqref{eq:theta} are constant.
More precisely, in Appendix~\ref{ClosureDorf} we compute
\begin{align}\label{eq:bbGamma minimal}
\bbLambda_1 \circ \bbLambda_2 =\ & - \Braket{\Theta_\alpha{-}\delta_\alpha^\dK\vartheta}{\lambda_1} \Big(\,r^{-1}\cU^{-1} T^\alpha \ket{\lambda_2}\,,\, 
\cU^{-1}T^\beta T^\alpha \ket{\lambda_2}\bra{W^+_\beta}r\cU  \,\Big)\\\nonumber&
-\big(\eta_{+1\,\alpha\beta}\bra\vartheta T^\alpha \ket{\lambda_1}+\eta_{0\,\alpha\beta}\bra\theta T^\alpha\ket{\lambda_1}\big)\ 
\Big(\,0\ ,\,  \cU^{-1} T^\beta \ket{\lambda_2} \bra{\partial_{\vartheta,\theta}}\, \Big)\,.
\end{align}
Combining this result with the Leibniz identity~\eqref{eq:DorfLeib2} for the generalised Dorfman product, we deduce the integrability conditions for equation~\eqref{eq:bbGamma minimal}
\bea 
   0   \hspace{-2mm} &=& \hspace{-2mm}  \eta_{-1\, \alpha\beta} \langle \theta | T^\alpha \otimes \langle \vartheta | T^\beta  + \eta_{0\, \alpha\beta} \langle \vartheta | T^\alpha \otimes \langle \vartheta | T^\beta  \\
& & \hspace{-2mm} - \Bigl( \eta_{-1\, \alpha\beta} \langle \theta | T^\alpha +  \eta_{0\, \alpha\beta} \langle \vartheta | T^\alpha\Bigr) \otimes \langle {\partial}_{\theta,\vartheta} | r^{-1} \mathcal{U}^{-1}   T^\beta + \langle \vartheta | \otimes \langle {\partial}_{\vartheta} | r^{-1} \mathcal{U}^{-1}   - \langle {\partial}_{\vartheta} | r^{-1} \mathcal{U}^{-1}   \otimes \langle \vartheta |\,, \CR
0 \hspace{-2mm} &=& \hspace{-2mm}  \eta_{-1\, \alpha\beta} \langle \theta | T^\alpha \otimes \langle \theta | T^\beta  + \eta_{0\, \alpha\beta} \langle \vartheta | T^\alpha \otimes \langle \theta | T^\beta + \langle \vartheta | \otimes \langle \theta | -  \langle \theta | \otimes \langle \vartheta | \CR
& & \hspace{-2mm} + \Bigl( \eta_{0\,\alpha\beta} \langle \theta | T^\alpha +  \eta_{1\, \alpha\beta} \langle \vartheta | T^\alpha\Bigr) \otimes \langle {\partial}_{\theta,\vartheta} | r^{-1} \mathcal{U}^{-1}   T^\beta + \langle \theta | \otimes \langle {\partial}_{\theta} | r^{-1} \mathcal{U}^{-1}   -  \langle {\partial}_{\theta} | r^{-1} \mathcal{U}^{-1}   \otimes \langle \theta |\; . \nonumber \eea
For constant $\bra\vartheta$ and $\bra\theta$ these expressions reduce to the quadratic constraints~\eqref{eq:QCvarthetatheta}. This is proved in Appendix~\ref{ClosureDorf}.

\medskip

To summarise, we have proved that all $D=2$ gauged supergravities arising from generalised Scherk--Schwarz reduction of E$_9$ ExFT are determined by an embedding tensor \eqref{eq:Theta virm exp} with components
\begin{equation}\label{eq:theta summary}
\bra{\Theta_{0}}=\bra\vartheta=\bra{W_\alpha}T^\alpha\,,\qquad
\bra{\Theta_{-1}}=\bra\theta= -\bra{W^+_\alpha}T^\alpha\,,\qquad
\bra{\Theta_{-k}}=0\,,\ k\ge2\,.
\end{equation}
The Weitzenb\"ock connection $\bra{W_\alpha}$ and its shifted version are defined in \eqref{eq:weitz} and \eqref{eq:shiftweitz} in terms of a twist matrix $\cU$ and an independent constrained bra $\bra{w^+}$.
The twist matrix can be gauge-fixed to take values only in E$_9$, but \eqref{eq:theta summary} holds regardless.
Analogously to the lower-rank case, the quadratic constraint is automatically satisfied if $\bra\vartheta$ and $\bra\theta$ in \eqref{eq:theta summary} are constant.

\subsection{Identities and uplift conditions for Lagrangian gaugings}\label{sec:gauging id}

In this paper we only consider geometric gaugings, which are the ones for which one can find a generalised Scherk--Schwarz ansatz with both the twist matrix $\mathcal{U}$ and the ancillary field $\langle w^+|$ satisfying the section constraint. One can then always choose a solution to the section constraint that defines a vector space of dimension $d\le 9$ inside the basic module $\overline{R(\Lambda_0)_{-1}}$.
These vector spaces are classified by the associated subgroups ${\rm GL}(d) \times {\rm E}_{9-d} \subset {\rm E}_9$ such that ${\rm GL}(d)$ is identified with the geometric linear group acting on the local coordinates $y^I$ of the internal manifold. Any constrained bra can be written in the corresponding basis as a dimension $d$ vector in the degenerate ground state
\be \langle \partial | = \langle 0 |^I \partial_I \; . \ee
All these vector spaces $\mathds{R}^d$ embed either inside $\mathds{R}^9$ of ${\rm GL}(9)$ or $\mathds{R}^8$ of ${\rm GL}(8) \times {\rm SL}(2)$, which correspond respectively to eleven-dimensional and type IIB supergravity. We define the subgroup ${\rm GL}(d) \times {\rm E}_{9-d} $ as the subgroup $G_0^{(\ell)} \subset {\rm E}_9$ commuting with the generator 
\be {\sf L}_0 = L_0 + \ell_A T_0^A + \tfrac12 \eta^{AB} \ell_A \ell_B \dK \; , \label{eq:flowL0}\ee
with $\ell_A T_0^A $ an appropriately normalised semi-simple element of $\mf{e}_8$ commuting with ${\rm GL}(d{-}1) \times {\rm E}_{9-d}\subset {\rm E}_8 $. This determines $\ell_A$ as a rational element in the  Cartan subalgebra up to E$_9$ conjugation.\footnote{In our conventions we can fix $\ell_A$ such that $\eta^{AB} \ell_A \ell_B = \frac{d-1}{d}$ takes the smallest possible value in the $E_9$ Weyl orbit. It is then a fundamental weight of $\mf{e}_8$, that is $0$ for $d=1$, $\Lambda_{d-1}/d$ for $2\le d\le  6$, $\Lambda_7/4$ for $d=8$ type IIB and $\Lambda_8/3$ for $d=9$.}
This formula extends to the Virasoro algebra of generators 
\be {\sf L}_n =  L_n + \ell_A T_n^A + \tfrac12 \eta^{AB} \ell_A \ell_B  \delta_{n,0} \dK \; , \label{Lnl}  \ee
and one can redefine the mode number in $\hevir$ according to ${\sf L}_0$ such that 
\be \hat{{\mf{e}}}_8 = \bigoplus_q \mf{g}_q^{(\ell)} \; , \ee
\vskip -4mm
\noindent where any generator ${\sf T}^{A_{q}}_q \in \mf{g}_q^{(\ell)}$  of mode number $q$  satisfies  
\be [ {\sf T}^{A_{q}}_q , {\sf L}_0  ] = q  {\sf T}^{A_{q}}_q \; . \ee
Here the range of the index $A_q$ depends on $q$ mod $1$ and corresponds to the representations in the branching of $\mf{e}_8$ under ${\rm SL}(d) \times {\rm E}_{9-d}\subset {{\rm E}}_8 $. In particular $q\in \frac{\mathds{Z}}{d}$ for $d\le 7$, $q\in \frac{\mathds{Z}}{4}$ in the type IIB $d=8$ case  and $q \in \frac{\mathds{Z}}{3}$  for $d=9$. The component with $q = \frac{p}{d}$ corresponds to the $p$-form fields in $d+2$ dimensions. For $q\notin \mathds{Z}$ all the vector spaces 
\be \mf{g}_{q}^{(\ell)} \cong \mf{g}_{q+n}^{(\ell)} \; , \qquad \forall n \in \mathds{Z} \; , \ee
are therefore isomorphic representations of $G_0^{(\ell)}$, while 
\be  \mf{g}_{0}^{(\ell)} \cong \mf{g}_{n}^{(\ell)} \oplus {\mathds{R}}  \; , \qquad \forall n \in \mathds{Z} \smallsetminus \{ 0\} \; , \ee
where $\mathds{R}$ is the centre. 
This gives a definition of the shift operators 
\be \mathsf{S}_m^{(\ell)}({\sf T}_q^{A_q}) = {\sf T}_{q+n}^{A_q}  \; , \qquad \mathsf{S}_m^{(\ell)}({\sf L}_n)= {\sf L}_{m+n}\; ,   \ee
that only differs from \eqref{eq:shift op def} by central terms
\be  \mathsf{S}_m^{(\ell)}(T_n^A)  =  \mathcal{S}_m(T_n^A) + \delta_{m+n} \eta^{AB} \ell_B \dK\; , \qquad   \mathsf{S}_m^{(\ell)}(L_n)  =  \mathcal{S}_m(L_n) -  \delta_{m+n} \tfrac12 \eta^{AB} \ell_A \ell_B  \dK   \; .\label{Shiftl}  \ee
E$_9$ exceptional field theory can be defined with the shift operators $\mathsf{S}_m^{(\ell)}$ for an arbitrary choice of $\ell$, such that these constants are simply absorbed in a redefinition of the constrained fields $\tilde{\chi}_n$ and $\langle\tilde\chi_1|$ that will be introduced in Section~\ref{sec:ExFTfields}. In particular, one then defines the twist matrix ansatz with this set of Virasoro operators
\begin{align}
\mathcal{U} = \mathsf{U}_{\scalebox{0.6}{loop}} r^{- \mathsf{L}_0} \prod_{n\ge 1} e^{- \psi_n \mathsf{L}_{-n}} \; . 
\end{align}
Note that one  can obtain this expression from \eqref{eq:twistU} by a redefinition of the loop component, while the Virasoro components $r$ and $\psi_n$ are unchanged. 

The basic module also decomposes accordingly into eigenspaces of $\mathsf{L}_0$ as (valid for any $s$)
\be \overline{R(\Lambda_0)_s} = \bigoplus_{q\ge 0} \overline{R_{\frac{d-1}{2d}+q}^{(\ell)}} \; 
.
\ee 
\vskip -3mm
\noindent The ground state $\langle 0|^I \in \overline{R_{\frac{d-1}{2d}}^{(\ell)}} $  transforms as a vector under GL$(d)$, and one computes that 
\be \langle 0|^I   \mathsf{L}_{-n} = \delta_{n,0} \frac{d-1}{2d} \langle 0|^I  \; , \qquad \forall n\ge 0\; . \ee
By construction any constrained bra $\langle 0|^I  \partial_I$ belongs to $ \overline{R_{\frac{d-1}{2d}}^{(\ell)}}$, and is therefore annihilated by all the negative mode generators. 

One can now use this property to compute that  for all $n\ge 1$
\be   \langle W_\alpha | T^\alpha  \mathsf{U}_{\scalebox{0.6}{loop}}  \mathsf{L}_{-n} \bigl( \mathsf{L}_0 + n - \tfrac{d-1}{2d}\bigr)^{-1} =  \langle W_\alpha | \mathsf{U}_{\scalebox{0.6}{loop}} \mathsf{S}^{({\ell})}_{-n}( \mathsf{U}_{\scalebox{0.6}{loop}}^{-1} T^\alpha \mathsf{U}_{\scalebox{0.6}{loop}})  \label{FlatLnShift}   \; . \ee
In the above formula, the inverse  operator is well-defined due to the spectrum of $\mathsf{L}_0$.

To prove this formula, one  rewrites   
\be \langle W_\alpha | T^\alpha \mathsf{U}_{\scalebox{0.6}{loop}}   = r^{- \frac{1+d}{2d}} W_{i \alpha}  \langle 0|^I   \mathsf{U}^{-1}_{\scalebox{0.6}{loop}}T^\alpha \mathsf{U}_{\scalebox{0.6}{loop}}  \ee
and uses that for each generator $\mathsf{T}_{q}^{A_q}$ and $n\ge 1$
 \bea &&  \langle 0|^I {\sf T}_{q}^{A_q} \mathsf{L}_{-n} \bigl( \mathsf{L}_0 + n - \tfrac{d-1}{2d}\bigr)^{-1} = \langle 0|^I \bigl[ {\sf T}_{q}^{A_q} , \mathsf{L}_{-n}\bigr] \bigl( \mathsf{L}_0+ n - \tfrac{d-1}{2d}\bigr)^{-1} \CR
  &=& \langle 0|^I q {\sf T}_{q-n}^{A_q} \bigl( \mathsf{L}_0+ n - \tfrac{d-1}{2d}\bigr)^{-1} =   \langle 0|^I  {\sf T}_{q-n}^{A_q}  =  \langle 0|^I \mathsf{S}^{({\ell})}_{-n}({\sf T}_{q-n}^{A_q})    \; . \eea
Expanding $W_{i \alpha}  \mathsf{U}^{-1}_{\scalebox{0.6}{loop}}T^\alpha \mathsf{U}_{\scalebox{0.6}{loop}}  $ in the  ${\sf T}_{q}^{A_q}$ and the Virasoro generators ${\sf L}_{-m}$, one obtains that the Virasoro components cancel and the loop-algebra ones recombine into the right-hand side of \eqref{FlatLnShift}. It follows that if the trombone $\langle \vartheta|=0$,  one gets  
\be \langle W_\alpha | \mathsf{S}^{({\ell})}_{-n}(T^\alpha) =  \omega^{(\ell)  \alpha}_n(\mathsf{U}_{\scalebox{0.6}{loop}}^{-1}  )  \langle W_\alpha   |  \label{tauflatzero}  \; ,\ee
 for all $n\ge 1$, where $\omega^{(\ell)  \alpha}_n(\mathsf{U}_{\scalebox{0.6}{loop}}^{-1}  ) $ is the cocycle defined in \cite{Bossard:2018utw} for the shift operator $\mathsf{S}^{({\ell})}_{-n}$
\be \mathsf{U}_{\scalebox{0.6}{loop}} \mathsf{S}^{({\ell})}_{-n}( \mathsf{U}_{\scalebox{0.6}{loop}}^{-1} T^\alpha \mathsf{U}_{\scalebox{0.6}{loop}})  \mathsf{U}_{\scalebox{0.6}{loop}}^{-1} =  \mathsf{S}^{({\ell})}_{-n}( T^\alpha )+  \omega^{(\ell)  \alpha}_n(\mathsf{U}_{\scalebox{0.6}{loop}}^{-1}  ) \dK \; . \ee
We therefore conclude that $\langle W_\alpha | \mathcal{S}_{-n}(T^\alpha)$ is a constrained bra.\footnote{The conversion from $\cS_{m}$ to $\mathsf{S}^{(\ell)}_m$ only involves central terms, which do not spoil the conclusion.} It will be important in the following that $\langle W_\alpha | \mathcal{S}_{-n}(T^\alpha)$ can consistently appear in the ansatz for $\langle \tilde{\chi}|$ when $\langle \vartheta|=0$. 
  
 \medskip

   In the same way one computes that 
 \be \langle \theta  |  \mathsf{U}_{\scalebox{0.6}{loop}} {\sf L}_{-n-1} \bigl( {\sf L}_0 + n+1 - \tfrac{d-1}{2d}\bigr)^{-1} = -  \langle W_\alpha | \mathsf{U}_{\scalebox{0.6}{loop}} \mathsf{S}^{({\ell})}_{-n}( \mathsf{U}_{\scalebox{0.6}{loop}}^{-1} T^\alpha \mathsf{U}_{\scalebox{0.6}{loop}})  = 0 \; , \ee
 for all $n\ge 1$, and therefore 
 \be \langle \theta  | \mathsf{U}_{\scalebox{0.6}{loop}} {\sf L}_{-n} = 0  \; , \qquad \forall n \ge 2\; .  \ee
 Note that this equation only depends on the internal coordinates through $ \mathsf{U}_{\scalebox{0.6}{loop}} $, and is true point-wise, giving therefore more than one equation for each $n$ through the span of $ \mathsf{U}_{\scalebox{0.6}{loop}} {\sf L}_{-n} \mathsf{U}_{\scalebox{0.6}{loop}}^{-1}$. 
This allows us to prove that 
 \be \langle \theta | \otimes \langle \theta | \eta_{-1\, \alpha\beta} T^\alpha \otimes T^\beta \bigl( \dK \otimes \mathsf{U}_{\scalebox{0.6}{loop}} {\sf L}_{-n} \mathsf{U}_{\scalebox{0.6}{loop}}^{-1}  +\mathsf{U}_{\scalebox{0.6}{loop}} {\sf L}_{-n} \mathsf{U}_{\scalebox{0.6}{loop}}^{-1}  \otimes \dK\bigr) = (n-1) \langle \theta | \otimes \langle \theta | \eta_{{-}1{-}n \, \alpha\beta} T^\alpha \otimes T^\beta \ee
 for all $n\ge 2$. Once again this formula is obtained using a commutator 
 \bea && \bigl[ \,  \eta_{-1\, \alpha\beta} T^\alpha \otimes T^\beta \, ,\,   \dK \otimes \mathsf{U}_{\scalebox{0.6}{loop}} {\sf L}_{-n} \mathsf{U}_{\scalebox{0.6}{loop}}^{-1}  +\mathsf{U}_{\scalebox{0.6}{loop}} {\sf L}_{-n} \mathsf{U}_{\scalebox{0.6}{loop}}^{-1}  \otimes \dK\, \bigr] \CR
 &=&  \bigl[ \,  \eta_{-1\, \alpha\beta} \, \mathsf{U}_{\scalebox{0.6}{loop}}  T^\alpha  \mathsf{U}_{\scalebox{0.6}{loop}}^{-1}   \otimes  \mathsf{U}_{\scalebox{0.6}{loop}}  T^\beta  \mathsf{U}_{\scalebox{0.6}{loop}}^{-1}  \,   ,\,   \dK \otimes \mathsf{U}_{\scalebox{0.6}{loop}} {\sf L}_{-n} \mathsf{U}_{\scalebox{0.6}{loop}}^{-1}  +\mathsf{U}_{\scalebox{0.6}{loop}} {\sf L}_{-n} \mathsf{U}_{\scalebox{0.6}{loop}}^{-1}  \otimes \dK\, \bigr] \CR
&=&  \mathsf{U}_{\scalebox{0.6}{loop}} \! \otimes \mathsf{U}_{\scalebox{0.6}{loop}}  \bigl[ \,  \eta_{-1\, \alpha\beta} T^\alpha \otimes T^\beta\,  ,\,   \dK \otimes  {\sf L}_{-n}   + {\sf L}_{-n}  \otimes \dK\, \bigr] \mathsf{U}_{\scalebox{0.6}{loop}}^{-1} \otimes \mathsf{U}_{\scalebox{0.6}{loop}}^{-1}  \CR
&=& (n-1) \mathsf{U}_{\scalebox{0.6}{loop}}\! \otimes \mathsf{U}_{\scalebox{0.6}{loop}}  \bigl( \eta_{-1{-}n\, \alpha\beta} T^\alpha \otimes T^\beta \bigr)  \mathsf{U}_{\scalebox{0.6}{loop}}^{-1} \otimes \mathsf{U}_{\scalebox{0.6}{loop}}^{-1}  \CR
&=&  (n-1)  \eta_{-1{-}n\, \alpha\beta} T^\alpha \otimes T^\beta  \; .   \eea 
 This equation together with the quadratic constraint \eqref{eq:QC theta only} gives that 
 \be \eta_{-n\, \alpha\beta}   \langle \theta | T^\alpha \otimes \langle \theta |  T^\beta  = 0 \; , \qquad \forall n \ge 3\; ,\label{eq:extra cons} \ee
 whenever $\langle \vartheta|=0$. In the same way one derives that
 \be \eta_{-n\, \alpha\beta} \eta_{-m\, \gamma\delta}   \langle \theta | T^\alpha  T^\gamma \otimes \langle \theta |  T^\beta  T^\delta = 0 \; , \qquad \forall m , n \ge 1 \, \; ,\label{eq:extra cons 2} \ee
and similarly for any higher number of $\eta_{-n}$. 

The E$_9$ invariant conditions~\eqref{eq:extra cons} and~\eqref{eq:extra cons 2} must be satisfied by any $\bra\theta$ admitting a higher-dimen\-sional origin through a gSS ansatz.
In the companion paper~\cite{SO9} we will prove that any Lagrangian embedding tensor $\bra\theta$ admitting an uplift is only parametrised by \emph{finitely} many components, which are identified explicitly at the cost of breaking E$_9$ covariance.\footnote{The statement that there are only finitely many components is E$_9$ invariant.
Namely, for any $\bra\theta$ admitting an uplift there exists an E$_9$ transformation mapping it to the set of components displayed in~\cite{SO9}.
The fact that there are finitely many components also does not depend on the choice of grading with respect to which $\overline{R(\Lambda_0)_{-2}}$ is decomposed.
However, a bad choice may require to expand $\bra\theta$ up to an extremely large (albeit finite) grade in order to identify all the linearly independent entries with uplift. A convenient choice of grading is~\eqref{eq:flowL0}.}

\section{The topological term}
\label{sec:top}

There is no kinetic term in E$_9$ exceptional field theory. In particular the gauge fields only appear in the topological term in the Virasoro extended formulation of exceptional field theory \cite{Bossard:2021jix}. In this section we carry out the generalised Scherk--Schwarz reduction of the topological term, leaving the potential term to Section~\ref{sec:pot}.

\subsection{Exceptional field theory topological term}

We begin with a quick review of the field content and structures appearing in $\hevirm$ extended ExFT. 

\subsubsection{Field content}
\label{sec:ExFTfields}

The gauge connection for generalised diffeomorphisms is given by a set of vector fields $\bbA_\mu$ denoted as follows, in analogy with \eqref{eq:bbLambda1}:
\begin{equation}
\mathbbm A_\mu = \Big(\,\ket{\cA_\mu} \,,\ \cB^{(k)}_\mu \,\Big)\,,
\end{equation}
where $\cB^{(k)}_\mu$ determine the gauge connection for the ancillary transformations parametrised by $\Sigma^{(k)}$.
They sit in the $R(\Lambda_0)_{0}\otimes\overline{R(\Lambda_0)_{-1}}$ representation, transform indecomposably with respect to $\ket{\cA_\mu}$ under rigid $\virm$ and are subject to the section constraint in their $\overline{R(\Lambda_0)_{-1}}$ `bra' component analogously to \eqref{eq:SCsigma}.
In particular, $\Tr\big[\cB^{(k)}\big]$ act as gauge fields for the local $\Virm$ transformations introduced in \eqref{eq:local Virm element}.
External derivatives covariant under generalised diffeomorphisms are then defined as
\begin{equation}\label{eq:exft covd}
\cD_\mu = \partial_\mu - \cL_{\bbA_\mu}\,.
\end{equation}
Field strengths for the gauge fields are obtained from the generalised Dorfman product.
For our purposes we can use for instance \eqref{eq:DorfSmallParam} and write in form notation
\begin{equation}
(\dd-\prepi\bbA\circ)\wedge(\dd-\prepi\bbA\circ) = -{}^\pi\bbF\circ\,,\qquad
\bbF=\big(\,\ket\cF\,,\ \cG^{(k)}\,\big)
\end{equation}
acting on any gauge parameter. The expression $(\dd-\prepi\bbA\circ)$ is analogous to \eqref{eq:exft covd} but uses the Dorfman product in place of the Lie derivative in order to act correctly on ancillary parameters.
This definition of the field strengths only holds up to trivial parameters, which are compensated for by introducing two-form fields. 
Using the Leibniz identity \eqref{eq:dorf leib proj} we have in particular
\begin{equation}
{}^\pi\bbF = \prepi\dd\bbA -\frac12 \prepi\bbA\circ\prepi\bbA + {}^\pi\hspace{-.1em} \mathbbm{C}\label{eq:DorfFS}
\end{equation}
where the wedge product is understood in the second term and ${}^\pi\hspace{-.1em}\mathbbm{C}$ is a suitable trivial parameter combination of two-form fields. Explicit expressions were given in \cite{Bossard:2021jix}.%
\footnote{%
\label{fn:18}%
The expression for the two-forms is given by equation (A.25) of \cite{Bossard:2021jix}, that translates to the present convention by the  identification $^\pi\mathbbm{C} = {^\pi}\varpi(\mathbf{C}+\tfrac12\upiota(\bbA,\bbA))$.
}

The scalar fields parametrise the coset space
\begin{equation}\label{eq:hevirm coset}
\frac{\widehat{\mathrm{E}}_8\rtimes\Virm}{K(\mathrm{E}_9)}
\end{equation}
in terms of a coset representative 
\begin{equation}\label{eq:exft coset rep}
\cV = \Gamma\,\cV_{\scalebox{0.6}{loop}} = \rho^{-L_0} e^{-\phi_1 L_{-1}} e^{-\phi_2 L_{-2}}\cdots \,\cV_{\scalebox{0.6}{loop}}\,,
\end{equation}
where $\Gamma\in\Virm$ has been expanded explicitly on the right-hand side and $\cV_{\scalebox{0.6}{loop}}\in \widehat{\mathrm{E}}_8$ also includes the exponential of the central element.
We work in the conformal gauge for the $D=2$ external metric:
\begin{equation}
g_{\mu\nu} = e^{2\upsigma}\eta_{\mu\nu}
\end{equation}
and the conformal factor contributes as $e^{-\upsigma\dK}$ to the central element of $\cV_{\scalebox{0.6}{loop}}$.
Under a rigid $\hevirm$ transformation we have
\begin{equation}\label{eq:coset trf}
\bbdelta^\alpha \cV = k^\alpha \cV + \cV T^\alpha\,,\qquad k^\alpha\in K(\mathfrak e_9)\,,
\end{equation}
where $k^\alpha=-(k^\alpha)^\dagger$ is the local compensating transformation needed to preserve a choice of $K(E_9)$ gauge for $\cV$.
We can also construct a generalised metric as
\begin{equation}
\cM=\cV^\dagger\cV
\end{equation}
invariant under the local action of $K(\mathrm E_9)$ and transforming in the symmetric tensor product $R(\Lambda_0)_0 \vee R(\Lambda_0)_0$ under $\hevirm$.
Under generalised diffeomorphisms, $\cV$ and equivalently $\cM$ transform covariantly according to the generalised Lie derivative \eqref{eq:genLie}.

External currents for the scalar fields are constructed by projecting the covariantised Maurer--Cartan derivative on its Hermitian part, while the anti-Hermitian part gives a composite connection as usual for coset space scalar manifolds:
\begin{align}
\label{eq:external exft currents}
\cP_{\mu} &= \frac12 \cD_\mu \cV\cV^{-1} + \mathrm{h.c.} 
= \frac12\Big( \partial_\mu \cV\cV^{-1} 
               - \big(\braket{\partial_\cV}{\cA_\mu}\cV\big)\cV^{-1}
               -[\bbA_\mu]_\alpha\,\cV T^\alpha\cV^{-1}
  \Big) +\text{h.c.}
\,,\\\nonumber
\cQ_{\mu} &= \frac12 \cD_\mu \cV\cV^{-1} - \mathrm{h.c.} 
= \frac12\Big( \partial_\mu \cV\cV^{-1} 
               - \big(\braket{\partial_\cV}{\cA_\mu}\cV\big)\cV^{-1}
               -[\bbA_\mu]_\alpha\,\cV T^\alpha\cV^{-1}
  \Big) -\text{h.c.}
  -[\bbA_\mu]_\alpha k^\alpha\,.
\end{align}
The internal currents are defined as
\begin{equation}\label{eq:P internal}
\bra{\cP_\alpha}\otimes T^\alpha = \frac12\bra{e^M}\otimes\partial_M\cV\cV^{-1}+\mathrm{h.c.}
\end{equation}
Note that $\bra{\cP_\alpha}$ do not transform covariantly under generalised diffeomorphims.
We also need to introduce shifted currents, defined in term of the shift operators \eqref{eq:shift op def}.
The external shifted currents are\footnote{By hermiticity of $\cP_\mu$, positive and negative shifts are related and we shall take $\tilde\chi_{\mu\,m}=\tilde\chi_{\mu\,-m}$. Also, for convenience we define $\tilde\chi_{\mu\,0}=\cP_{\mu\,\dK}$.}
\begin{equation}\label{eq:shift P external}
\cP^{(m)}_\mu = \cS_{m}(\cP_\mu) + \tilde\chi_{\mu\,m}\dK\,,
\end{equation}
where the one-forms $\tilde\chi_{\mu\,m}$ are introduced to restore covariance under rigid $\hevirm$ as well as generalised diffeomorphisms.
We define the components $(k^\alpha{})_\beta $ of the $K(\mathrm{E}_9)$ compensating transformation from \eqref{eq:coset trf} as follows:
\begin{equation}\label{eq;k components}
k^\alpha = (k^\alpha{})_\beta T^\beta = -(k^\alpha{})_\beta T^\beta{}^\dagger\,,
\end{equation}
where we avoid introducing a separate index for the $K(\mathfrak{e}_9)$ algebra.
Then, the auxiliary form rigid transformation reads
\begin{equation}
\bbdelta^\alpha \tilde\chi_{\mu\,m} = -(k^\alpha){}_\gamma \,\omega^{\gamma\beta}\cP^{(m)}_{\mu\,\,\beta}\,,
\end{equation}
where $\omega^{\alpha\beta}=-f^{\alpha\beta}{}_\dK$ is the Lie algebra cocycle of $\hevir$ and  $\cP^{(m)}_\mu = \cP^{(m)}_{\mu\,\,\alpha}\,T^\alpha$.
The one-forms $\tilde\chi_{\mu\,m}$ transform covariantly under generalised diffeomorphisms.
A similar approach applies to shifted internal currents, where only a shift by a single unit is required:
\begin{equation}\label{eq: P internal}
\bra{\cP_\alpha^{(1)}} \otimes T^\alpha =  \bra{\cP_\alpha} \otimes \cS_{1}(T^\alpha) + \bra{\tilde\chi_1}\otimes\dK\,.
\end{equation}
The scalar field $\bra{\tilde\chi_1}$ is algebraically constrained to satisfy the section constraint~\eqref{eq:SC}, analogously to an internal derivative $\bra\partial$, so that only finitely many components are non-vanishing.
Under rigid $\hevirm$ transformations, $\bra{\tilde\chi_1}$ transforms analogously to its one-form siblings:
\begin{equation}
\bbdelta^\alpha \bra{\tilde\chi_{1}} = 
\bra{\tilde\chi_{1}} \big(T^\alpha -\delta^\alpha_{L_0}\dK  \big)
-(k^\alpha){}_\gamma \,\omega^{\gamma\beta}\,\big[\cS_1(T^\delta)\big]_\beta\,\bra{\cP^{(m)}_\delta}\,,\label{eq:internal chi var}
\end{equation}
where the first term is the covariant action on the `bra' components.
In contrast with its one-form siblings, $\bra{\tilde\chi_1}$ transforms non-covariantly under generalised diffeomorphisms. Its transformation can be deduced by requiring invariance of the ExFT Lagrangian and is displayed in \eqref{eq: gauge var internal chi}.

\subsubsection{The dynamics}

The dynamics of $\hevirm$ extended ExFT are captured by a pseudo-Lagrangian~\cite{Bossard:2018utw,Bossard:2021jix} 
\begin{equation}\label{eq:exft pseudoL}
\cL^{\rm pseudo}_{\rm ExFT} = \cL^{\rm top}_{\rm ExFT} - V_{\rm ExFT}^{\phantom{top}}
\end{equation}
where $\cL^{\rm top}_{\rm ExFT}$ is a topological term to be discussed momentarily, while $V_{\rm ExFT}$ is the scalar potential, which only depends on internal derivatives.
It will be presented in Section~\ref{sec:pot}.

The pseudo-Lagrangian is supplemented by a twisted self-duality constraint as well as a Virasoro constraint.
The latter is required because we use the conformal gauge for the external metric.
The twisted self-duality constraint reads, in form notation ($\cP = \cP_\mu \dd x^\mu$)
\begin{equation}\label{eq:twsd exft}
\star\cP = \cP^{(1)}\,,\qquad \star^{|m|}\cP=\cP^{(m)}\,,\ m\in\ints\,,
\end{equation}
which is the ExFT extension of~\eqref{eq:twsd gsugra}.
The second expression is a consequence of the first one except for its $\dK$ component, which gives a relation between the auxiliary one-forms $\tilde\chi_{m\,\mu}$.
The Virasoro constraint reads
\begin{equation}\label{eq:vir cstr exft}
\delta \tilde{g}^{\mu\nu}\Big(2 \cP_{\mu\,\dK} \cP_{\nu\,0} 
- \cP_{\mu\,0} \cP_{\nu\,0}
+ \cD_\mu \cP_{\nu\,0} -\eta^{AB} P_{\mu\, A}^{\hspace{2.3mm}0}P_{\nu\, B}^{\hspace{2.3mm}0} \Big) = 0\,,
\end{equation}
with $\delta \tilde{g}^{\mu\nu}$ symmetric traceless.
This expression is not manifestly covariant under local $K(\mathfrak e_9)$ transformations, but it is invariant by virtue of twisted self-duality.

To describe the topological term $\cL^{\rm top}_{\rm ExFT}$ we first need to introduce covariantised, shifted versions of the projector \eqref{eq:squareProj} or equivalently \eqref{eq:squareProjSmall}.
To do so one first introduces $\Virm$-invariant versions of the shift operators \eqref{eq:shift op def} by dressing them with the coset representative
\begin{equation}\label{eq:Sgamma def}
\cS_m^{\upgamma}(X) 
= \cS_0\Big(\cV^{-1}\cS_m\big(\cV \,X\, \cV^{-1}\big)\cV\Big)
= \cS_0\Big(\Gamma^{-1}\cS_m\big(\Gamma \,X\, \Gamma^{-1}\big)\Gamma\Big)\,,\qquad X\in\hat{\mathfrak e}_8\oleft\vir
\,.
\end{equation}
The $\upgamma$ superscript indicates that such operators depend on the Virasoro fields $\rho$, $\phi_k$ contained in the $\Gamma\in\Virm$ element of \eqref{eq:exft coset rep}.
The notation reflects the relation between this dressing and the variable spectral parameter of the linear system for $D=2$ gravity \cite{Bossard:2021jix}.
For later use, we also recall that such field-dependent shift operators admit an expansion in terms of the constant ones:
\begin{equation}\label{eq:Sgamma expansion}
\cS^\upgamma_{m} = \rho^{-m}\big(\cS_m - m \phi_1 \cS_{m-1}+\ldots \big)\,.
\end{equation}
Using \eqref{eq:Sgamma def} we can now define
\begin{equation}\label{eq:shift F}
[\bbF]^{(m)}_\alpha \, T^\alpha = [\bbF]_\alpha \,\cS^\upgamma_m(T^\alpha) 
+ \widehat{\bbF}^\upgamma_m \,\dK\,,
\end{equation}
where we introduced the last term to reinstate tensoriality under rigid $\hevirm$ transformations, in analogy with the introduction of $\tilde\chi_{m\,\mu}$ in \eqref{eq:shift P external}.
However, in this case the $\dK$ completion can be constructed from the existing fields and no new auxiliaries are required:
\begin{equation}\label{eq:hat F}
\widehat{\bbF}^\upgamma_{m} = 
-\bra{\partial_\cF}\cS^\upgamma_m(L_0)\ket\cF
-\sum_{k=1}^\infty \Tr\big(\, \cG^{(k)} \cS^\upgamma_m(L_{-k}) \, \big)\,.
\end{equation}
For $m\le1$ this expression does not involve two-form potentials, namely trivial parameters are projected out.

We can now present the topological term as an external two-form~\cite{Bossard:2021jix}:
\begin{align}\label{eq:exft Ltop}
\rho^{-1}\cL^{\rm top}_{\rm ExFT}\,\dd x^0\wedge\dd x^1\ =\ \ &
  2\cD\tilde\chi_{1} 
- 2\eta^{AB}\sum_{m\in\ints}m\,\cQ^m_A\wedge\cP^{-m-1}_B
+ 2\sum_{k=1}^{\infty}\cP_k\wedge \big( \tilde\chi_{k+1}-\tilde\chi_{k-1} \big)
\CR&
+ 2\braket{\tilde\chi_1}{\cF}
+ \widehat{\bbF}^\upgamma_{1} + \widehat{\bbF}^\upgamma_{-1}
+ \omega^\alpha(\cV)\big(\, [\bbF]^{(1)}_\alpha + [\bbF]^{(-1)}_\alpha \,\big)\,.
\end{align}
In the last term we introduced the group cocycle $\omega^\alpha(\,\cdot\,)$. It is the Virasoro extension of the standard group cocycle  
associated with the central extension $\widehat{\mathrm E}_8$ of the loop-group over E$_8$, and is defined as:
\begin{equation}\label{eq:group cocycle def}
\omega^\alpha(g) \dK 
\ =\ \big( g T^\alpha  g^{-1}\big|_\dK -\delta^\alpha_\dK\big)\dK
\ =\ \cS_0(T^\alpha) -  g^{-1}\cS_0( g T^\alpha g^{-1}) g
\,,
\qquad
g\in\widehat{\rm E}_8\rtimes\Virm\,,
\end{equation}
with $T^\alpha\in\hevir$.
Another useful relation is 
\begin{equation}\label{eq:group cocycle vs Sgamma}
\cV^{-1}\cS_m\big(\cV T^\alpha \cV^{-1} \big)\cV 
=
\cS^\upgamma_m(T^\alpha) - \omega^\beta(\cV) \big[\cS^\upgamma_m(T^\alpha) \big]_\beta\,\dK\,.
\end{equation}
Notice that the constant shift operator appears on the left-hand side, and the field-dependent one on the right-hand side.
It is also useful to display explicitly the covariant derivative of the auxiliary one forms:
\begin{equation}\label{eq:tildechi covd}
\cD\tilde\chi_1 = 
\dd\tilde\chi_1 -\braket{\partial_{\tilde\chi_1}}{\cA}\wedge\tilde\chi_1
- [\bbA]_\alpha \wedge (k^\alpha)_\beta \, \omega^{\beta\gamma} \cP^{(1)}_\gamma\,,
\end{equation}
where $(k^\alpha){}_\beta$ is defined as in \eqref{eq;k components}.
Importantly, the compensating contributions of $k^\alpha$ to $\cD\tilde\chi_1$ and to $\cQ^m_A$ (see~\eqref{eq;k components})  in the first two terms of \eqref{eq:exft Ltop} cancel out, so that the topological term is independent of the $K(\mathrm E_9)$ compensator.

\subsection{Ansatz for the fields}
\label{sec:gSSans}

We can now study the generalised Scherk--Schwarz ansatz for the ExFT fields.
The gSS ansatz \eqref{eq:gssgenvec} for generalised diffeomorphism parameters applies directly to the vector fields:
\begin{equation}\label{eq:bbA gss}
\prepi\mathbbm{A}_\mu(x,y) = \Big(\,r^{-1}(y)\,\cU^{-1}(y)\ket{A_\mu(x)}\,,\ 
\braket{H_\alpha(y)}{A_\mu(x)}
\,\Big)\,,
\end{equation}
where $\ket{A_\mu(x)}=\ket{A_\mu}$ are the gauged supergravity vector fields.
As usual, we have projected out $\Sigma$-only trivial components of the ancillary fields.
Notice that we do not introduce an analogue of the ancillary $\cB_\mu^{(k)}$ fields in gauged supergravity: the ExFT ancillary vectors reduce only to the standard gauge supergravity connection $\ket{A_\mu}$.

Assuming the gSS condition \eqref{eq:gss cond e9} is satisfied, the Leibniz identity implies that the field strengths obey themselves the gSS ansatz\footnote{More precisely, any na\"ive violation of the gSS ansatz is a trivial parameter and can be removed by appropriately choosing the reduction ansatz for the ExFT two-forms.}
\begin{equation}
{}^\pi\bbF_{\mu\nu}(x,y) = \Big(\,r^{-1}(y)\,\cU^{-1}(y)\ket{F_{\mu\nu}(x)}\,,\ \braket{H_\alpha(y)}{F_{\mu\nu}(x)}  \,\Big)\,+\,\ldots\,,\label{eq:truncFSproj}
\end{equation}
where the dots correspond to extra terms in the reduction ansatz for the ExFT two-forms, which do not affect the reduction of the pseudo-Lagrangian but are important for the non-Lagrangian duality equation~\eqref{eq:dualityeqExFT} of the field strengths. They will be discussed in Section~\ref{sec:nonLag}.
The fields $\ket{F_{\mu\nu}}$ are the field strengths of the gauged supergravity vectors, which are straightforward to compute, and which read (moving now to form notation)
\begin{equation}
\ket{F} = \ket{\dd A} +\frac12 T^\alpha\ket{A} \wedge \big( 
\eta_{\alpha\beta}\bra\vartheta T^\beta\ket{A} + \eta_{-1\,\alpha\beta}\bra\theta T^\beta\ket{A}
 \big)+\ldots\label{eq:FSgs}
\end{equation}
where we have substituted the final expression \eqref{eq:theta summary} for the embedding tensor with \mbox{$\bra{\tilde h_\alpha}=0$}.
The dots correspond to gauged supergravity two-form contributions which vanish under contraction with $\bra{\Theta_{\alpha}}$ and will drop out from the gauged supergravity pseudo-Lagrangian.
They are displayed in equation~\eqref{eq:sugraFS} for vanishing $\bra\vartheta$.
Notice that if we gauge-fix $\cU\in\mathrm{E}_9$, the expressions above can also be rewritten in terms of the minimal ansatz \eqref{eq:gssgenvec minimal}.
The gauged supergravity field strength is unaffected.
The following computations are valid for arbitrary $\cU \in \widehat{E}_8\rtimes \text{Vir}^-$.

The reduction ansatz for the scalar fields in \eqref{eq:exft coset rep} is simply given by
\begin{equation}\label{eq:coset gss}
\cV(x,y) = V(x)\,\cU(y)\,,
\end{equation}
which partially fixes the local $K(\mathrm E_9)$ gauge, such that $V(x)$ is the coset representative for the gauged supergravity $\widehat{\mathrm E}_8\rtimes\Virm$ scalar fields.
In particular we shall write in analogy with \eqref{eq:exft coset rep}
\begin{equation}\label{eq:sugra coset rep}
V(x) = \sugraGamma(x) V_{\scalebox{0.6}{loop}}(x) = 
\sugrarho(x)^{-L_0} e^{-\varphi_1(x)L_{-1}} e^{-\varphi_2(x)L_{-2}}\cdots V_{\scalebox{0.6}{loop}}(x)
\end{equation}
and the loop component includes the $D=2$ external conformal factor $e^{-\sugrasigma(x)\dK}$.
It is also convenient to explicitly write the factorisation ansatz for the dilaton:
\begin{equation}\label{eq:dilaton gss}
\rho(x,y) = \sugrarho(x) r(y)\,.
\end{equation}
It is worth stressing that, while invariance under ExFT ancillary transformations allows us to gauge-fix $\cU\in\mathrm E_9$, the gauged supergravity Virasoro fields contained in $V(x)$ cannot be gauge-fixed to vanish without violating the factorisation ansatz.%
\footnote{One may be tempted to first gauge-fix $\phi_k(x,y)=0$ in ExFT and then impose a gSS factorisation ansatz such as \eqref{eq:coset gss}. However, for $\phi_k(x,y)=0$ the ExFT twisted self-duality relations impose $\Tr[\cB^{(k)}]\neq0$, which is incompatible with \eqref{eq:bbA gss}. In order to properly gauge-fix $\varphi_k(x)=0$ one must introduce $\cB$-like fields in $D=2$ supergravity, which would also allow the switch to a `minimal' formulation analogous to the one of E$_9$ ExFT \cite{Bossard:2021jix}. We refrain from doing so in this paper.}
As already noted under \eqref{eq:bbA gss}, we do not introduce an analogue of $\Tr\big[\cB^{(k)}_\mu\big]$ in gauged supergravity to play the role of gauge connection for all $L_{-k}$, $k\ge1$. 
Hence the fields $\varphi_{k}(x)$ are not pure gauge.
The only exception is $\varphi_1(x)$, which can be gauge-fixed to vanish whenever $\bra\theta\neq0$, since then the $L_{-1}$ generator is gauged by a linear combination of the vectors $\ket{A_\mu(x)}$.

Combining \eqref{eq:external exft currents} with \eqref{eq:coset gss} and \eqref{eq:gss cond e9} we find that under the gSS ansatz the external currents reduce directly to their gauged supergravity counterparts:
\begin{equation}\label{eq:ext currents gss}
\cP_\mu(x,y) = P_\mu(x)\,,
\end{equation}
where $P_\mu(x)$ is defined in \eqref{eq: gauged sugra P and Q} in terms of the gauged supergravity covariant derivative \eqref{eq:gsugra covd}.
We can therefore apply the same ansatz to the auxiliary one-forms $\tilde\chi_{m\,\mu}$:
\begin{equation}\label{eq:ext shift currents gss}
\tilde\chi_{\mu\,m}(x,y) = \sugrachi_{\mu\,m}(x)\,,\qquad
\cP^{(m)}_\mu(x,y) = P^{(m)}_\mu(x) = \cS_{m}\big(P_\mu(x)\big)+\sugrachi_{\mu\,m}(x)\,\dK\,,
\end{equation}
where $P^{(m)}_\mu$ and $\sugrachi_{\mu\,m}$ were defined in \eqref{eq:gsugra shifted P} and are the gauged supergravity versions of $\cP^{(m)}_\mu$ and $\tilde\chi_{\mu\,m}$ respectively.
The reduction ansatz for the internal currents will be discussed in Section~\ref{sec:GSS pot}.
The expression for the reduction of the constrained scalar $\bra{\tilde\chi_1}$ is fixed by requiring consistency of the truncation of the scalar potential, as will be shown in Section~\ref{sec:GSS pot}, or from similar consistency of the topological term, as shown in the next few pages.
It is a non-trival cross-check that the two expressions agree.

With this information we immediately conclude that both the twisted self-duality \eqref{eq:twsd exft} and the Virasoro constraint \eqref{eq:vir cstr exft} reduce to their gauged supergravity counterparts, simply obtained by trading the ExFT currents $\cP$, $\cP^{(m)}$ for $P$, $P^{(m)}$, reproducing \eqref{eq:twsd gsugra} as well as \eqref{eq:vir cstr gauged sugra}.
The non-trivial task is to compute the reduction of the topological and potential terms.

\subsection{Reduction of the topological term}\label{sec:ExFTtop}

We are now set to derive the gSS reduction of the topological term \eqref{eq:exft Ltop}.
We shall henceforth focus on gSS reductions giving rise to Lagrangian gaugings, hence we will assume
\begin{equation}\label{eq:vartheta=0}
\bra\vartheta=0\,.
\end{equation}
To our knowledge, there is no consistent compactification leading to trombone gaugings. Indeed, consistent truncations leading to non-Lagrangian gaugings must be performed at the level of the equations of motion, because the truncation ansatz violates the boundary conditions required for integration by parts along the internal space in the variation of the higher-dimensional action~\cite{Hohm:2014qga}.
Furthermore, including $\bra\vartheta\neq0$ complicates the structure of the reduced theory. In particular several couplings, while consistent, do not appear to be expressible directly in terms of the bra vectors $\bra\vartheta$ and $\bra\theta$. These terms only disappear for $\bra\vartheta=0$ by virtue of relation \eqref{tauflatzero}.

The overall dilaton factor in $\cL^{\rm top}_{\rm ExFT}$ factorises as in \eqref{eq:dilaton gss}, hence we expect to reproduce a topological term for gauged supergravity, multiplied by an overall $r(Y)$ factor (the same will hold for the scalar potential).
The first line in \eqref{eq:exft Ltop} indeed reduces immediately to the same expression in terms of gauged supergravity quantities. 
To see that this is the case, we simply combine \eqref{eq:ext currents gss}, \eqref{eq:ext shift currents gss} with the expressions \eqref{eq:external exft currents} and \eqref{eq:tildechi covd} for $\cQ$ and $\cD\tilde\chi_1$, recalling that the compensating transformation $k^\alpha$ contained in these two objects cancels out.
All other terms in such expressions map directly to their gauged supergravity counterparts, so we conclude
\begin{align}\label{eq:Ltop first line gss}
&\quad  2\cD\tilde\chi_{1} 
- 2\eta^{AB}\sum_{m\in\ints}m\,\cQ^m_A\wedge\cP^{-m-1}_B
+ 2\sum_{k=1}^{\infty}\cP_k\wedge \big( \tilde\chi_{k+1}-\tilde\chi_{k-1} \big)\\\nonumber
&=
  2D\sugrachi_{1} 
- 2\eta^{AB}\sum_{m\in\ints}m\, Q^m_A\wedge P^{-m-1}_B
+ 2\sum_{k=1}^{\infty} P_k\wedge \big( \sugrachi_{k+1}-\sugrachi_{k-1} \big)\,,
\end{align}
where explicitly
\begin{equation}
D\tilde\chi_{1} = \dd\sugrachi_1 
-\braket{\Theta_\alpha}{A}\wedge (\breve{k}^\alpha)_\beta\,\omega^{\beta\gamma}P^{(1)}_\gamma
\end{equation}
with $(\breve{k}^\alpha)_\beta$ the $x$-dependent $K(\mathfrak{e}_9)$ compensating transformation in gauged supergravity, introduced in \eqref{eq:ungauged coset trf},\eqref{KcompoSugra}, which cancels out in $\cL_{\rm top}$ with the analogous term within $Q^m_A$.

The reduction of the second line of \eqref{eq:exft Ltop} is more laborious.
We begin by noticing that we can dress the shifted field strengths \eqref{eq:shift F} with the twist matrix to deduce expressions analogous to \eqref{eq:LambdaProjGSS}.
The $\pm1$ shifts of $[\bbF]_\alpha$ are given by
\begin{align}\label{eq:Fhat red intermediate step}
[\bbF]^{(\pm1)}_\alpha \cU T^\alpha \cU^{-1}
\ =\ \ &
[\bbF]_\alpha \, \cU \cS^\upgamma_{\pm1}\big(T^\alpha\big) \cU^{-1}
+\widehat\bbF^\upgamma_{\pm1}\,\dK
\CR=\ \ &
\Braket{\Theta_\alpha{-}W_\alpha}{F} \,
\cU \cS^\upgamma_{\pm1}\big(\, \cU^{-1} T^\alpha \cU \,\big) \cU^{-1}
+\widehat\bbF^\upgamma_{\pm1}\,\dK\,.
\intertext{%
The $\cU$ conjugation of the shift operator can be treated on the same footing as a rigid $\widehat{\rm E}_8\rtimes\Virm$ transformation, since there are no derivatives acting on these twist matrices.
Bringing $\cU$ through the shift operator generates a group cocycle $\sim\omega^\alpha(\cU^{-1})$ that is reabsorbed in $\widehat\bbF^\upgamma_{\pm1}$, reproducing its rigid transformation properties.
Using the expansions \eqref{eq:halpha expansion} as well as \eqref{eq:bbLambda red} we can also deduce the expression
}
[\bbF]^{(\pm1)}_\alpha \cU T^\alpha \cU^{-1}
\ =\ \ &
\Braket{\Theta_\alpha{-}W_\alpha}{F}
\,\cS^{\sugraupgamma}_{\pm1}\big( T^\alpha \,\big) 
\label{eq:topterm step1}
\\\nn&
+ \Big( 
   \bra{W_\alpha} \cS^{\sugraupgamma}_{\pm1}(L_0)T^\alpha\ket{F}
   -\sum_{k=1}^\infty  \Tr\big( \xi^{(k)}_M \cS^{\sugraupgamma}_{\pm1}(L_{-k}) \big) \braket{e^M}{F}
  \Big)\,\dK
\end{align}
where in analogy with \eqref{eq:Sgamma def} we defined the field-dependent shift operators of gauged supergravity
\begin{equation}\label{eq:Sbargamma def}
\cS_m^{\sugraupgamma}(X) = \cS_0\Big(V^{-1}\cS_m\big(V \,X\, V^{-1}\big)V\Big)
= \cS_0\Big(\sugraGamma{}^{-1}\cS_m\big(\sugraGamma \,X\, \sugraGamma{}^{-1}\big)\sugraGamma\Big)
\,,\qquad X\in\hat{\mathfrak e}_8\oleft\vir\,.
\end{equation}
Notice that this admits an expansion analogous to \eqref{eq:Sgamma expansion}, with the supergravity fields $\varrho$, $\varphi_n$ in place of $\rho$, $\phi_n$.
The $\dK$ component of \eqref{eq:topterm step1} can be computed directly, taking into account the indecomposable transformation \eqref{eq:Halpha trf} of $\bra{H_\alpha}$ and its map to $\bra{h_\alpha}$, or simply deduced as it is the natural $\dK$ completion of the first line in that same equation.
So far we have concluded that $\widehat\bbF^\upgamma_{\pm1}$ will reduce to the second line of \eqref{eq:topterm step1}, plus a cocycle term proportional to $\omega^\alpha(\cU^{-1})$.
We can easily compute the cocycle term by combining \eqref{eq:coset gss} with \eqref{eq:Sgamma def} and \eqref{eq:group cocycle def}:
\begin{align}\label{eq:omegaU}
\cU \cS^{\upgamma}_{\pm1}\big(\cU^{-1}T^\alpha\cU\big)\cU^{-1} &=
\cU \cS_0\Big[\cU^{-1}V^{-1}\cS_{\pm1}\big(V T^\alpha V^{-1}\big)V\cU\Big]\cU^{-1}\CR
&=\cS^{\sugraupgamma}_{\pm1}\big(T^\alpha\big)
- \omega^\beta(\cU^{-1})\big[ \cS^{\sugraupgamma}_{\pm1}(T^\alpha) \big]_\beta\,\dK\,.
\end{align}
This expression is contracted with $\braket{\Theta_\alpha{-}W_\alpha}{F}$ in \eqref{eq:Fhat red intermediate step}.
By $\widehat{\rm E}_8\rtimes\Virm$ invariance of the topological term, the $\omega^\alpha(\cU^{-1})$ cocycle contributions are guaranteed to cancel out and we will see that this is the case.
\footnote{One way to motivate this expectation is to recall from \cite{Bossard:2021jix} that the topological term is obtained as the $\dK$ completion of a shifted Maurer--Cartan equation for the currents. On one hand, dressing such covariantised equation by $\cU$ leaves the topological term invariant because all non-$\dK$ components are identically zero. On the other hand, the $\omega^\alpha(\cU^{-1})$ terms correspond exactly to the transformation of the $\dK$ component, hence they must cancel out.}

Explicit expressions for the $\xi^{(k)}_M$ appearing in~\eqref{eq:topterm step1} can be deduced from \eqref{eq:halpha redef} after setting $\bra{\tilde h_\alpha}=0$:\footnote{With a slight abuse of notation, the left and right-hand sides both have a single `ket' component, which is identified despite appearing on opposite sides of the tensor product.}
\begin{align}\label{eq:xi gss}
\xi^{(1)}_M\otimes\bra{e^M} &= 
\bra{W_\alpha}\otimes\cS_{1}(T^\alpha)+ (\bra{w^+}+\bra{W_{-1}})\otimes\dK\,,\\
\xi^{(k)}_M\otimes\bra{e^M} &= \bra{W_{-k}}\otimes\dK\,,\quad k\ge2\,.
\end{align}
The next step is to combine the first term in the $\dK$ component of \eqref{eq:topterm step1} with a similar contribution coming from $\xi^{(1)}$.
By expanding the field-dependent shift operators in analogy with \eqref{eq:Sgamma expansion} and adding and subtracting terms in order to find commutators of the form $[L_m,T^\gamma]-[L_{m-1},\cS_{+1}(T^\gamma)]$, we find the identity
\begin{align}\label{eq:W xi1 identity}
&\bra{W_\gamma}\cS_{\pm1}^{\sugraupgamma}(L_0)T^\gamma
-\bra{W_\gamma}\cS_{\pm1}^{\sugraupgamma}(L_{-1})\cS_{+1}(T^\gamma)
\\\nonumber&\qquad
=
-\bra{W_\alpha} \cS_{+1}(T^\alpha) \cS^{\sugraupgamma}_{\pm1}(L_{-1})
+\sum_{k=1}^\infty \bra{W_{-k}}\cS^{\sugraupgamma}_{\pm1}(L_{-k}) \,,
\end{align}
where we also used $\bra\vartheta=0$.
We use this identity to plug \eqref{eq:xi gss} into the second line of \eqref{eq:topterm step1} and obtain
\begin{align}
&\bra{W_\alpha} \cS^{\sugraupgamma}_{\pm1}(L_0)T^\alpha\ket{F}
   -\sum_{k=1}^\infty  \Tr\big( \xi^{(k)}_M \cS^{\sugraupgamma}_{\pm1}(L_{-k}) \big) \braket{e^M}{F}
\\\nonumber&\qquad\qquad\qquad\qquad\qquad\qquad
= \bra\theta \cS^{\sugraupgamma}_{\pm1}(L_{-1})\ket{F} 
  + \bra{W_\alpha}\cS^{\sugraupgamma}_{\pm1}(T^\alpha)\ket{F}\,,
\end{align}
We conclude that the gSS reduction of the $\widehat\bbF^\upgamma_{\pm1}$ terms in $\cL^{\rm top}_{\rm ExFT}$ gives
\begin{equation}\label{eq:Fhat gss}
\widehat\bbF^\upgamma_{\pm1} =
\bra\theta \cS^{\sugraupgamma}_{\pm1}(L_{-1})\ket{F} 
+ \bra{W_\alpha}\cS^{\sugraupgamma}_{\pm1}(T^\alpha)\ket{F} 
+\omega^\alpha(\cU^{-1}) \big[\cS^{\sugraupgamma}_{\pm1}(T^\beta)\big]_\alpha \Braket{\Theta_\beta{-}W_\beta}{F}\,,
\end{equation}
The first term in this expression is the $\dK$ completion of $\braket{\Theta_\alpha}{F}\cS^{\sugraupgamma}_{\pm1}(T^\alpha)$, in full analogy with the definition of $\widehat\bbF^\upgamma_{\pm1}$ in ExFT in \eqref{eq:shift F}, \eqref{eq:hat F}.

To compute the reduction of the last term in \eqref{eq:exft Ltop}, we use the first definition in \eqref{eq:group cocycle def} and recall that $\omega^\dK(\,\cdot\,)=0$ to write
\begin{align}\label{eq:Ltop cocycle gss}
\omega^\alpha(\cV) [\bbF]^{(\pm1)}_\alpha &=
[\bbF]_\alpha \cV \cS_0\Big( \cV^{-1} \cS_{\pm1}(\cV T^\alpha \cV^{-1}) \cV \Big) \cV^{-1}\Big|_\dK   \\\nonumber
&=
\braket{\Theta_\alpha-W_\alpha}{F} \cV \cS_0\Big( \cV^{-1} \cS_{\pm1}(V T^\alpha V^{-1}) \cV \Big) \cV^{-1}\Big|_\dK\\\nonumber
&=
  \omega^\alpha(V)\big[\cS^{\sugraupgamma}_{\pm1}(T^\beta)\big]_\alpha \Braket{\Theta_\beta{-}W_\beta}{F}
- \omega^\alpha(\cU^{-1}) \big[\cS^{\sugraupgamma}_{\pm1}(T^\beta)\big]_\alpha \Braket{\Theta_\beta{-}W_\beta}{F}\,,
\end{align}
and we see that the last term will cancel out with the reduction of $\widehat\bbF^\upgamma_{\pm1}$.

Adding together \eqref{eq:Fhat gss} and \eqref{eq:Ltop cocycle gss}, we see that there are still some terms that depend explicitly on $\bra{W_\alpha}$ rather than just its projection to $\bra\theta$.
These terms must be reabsorbed into the reduction ansatz for the constrained scalar field $\bra{\tilde\chi_1}$.
One then deduces that under gSS reduction,
\begin{equation}\label{eq:internal chi gss from Ltop}
2r^{-1}\bra{\tilde\chi_1}\cU^{-1} = 
\bra{W_\alpha}\big( \cS^{\sugraupgamma}_{+1}(T^\alpha) {+} \cS^{\sugraupgamma}_{-1}(T^\alpha) \big)
+\omega^\alpha(V) \big[\cS^{\sugraupgamma}_{+1}(T^\beta){+}\cS^{\sugraupgamma}_{-1}(T^\beta)\big]_\alpha \bra{W_\beta}
-\varrho^{-1}\bra{\theta}\,.
\end{equation}
Recall that $\bra{\tilde\chi_1}$ is algebraically constrained to be on section.
The last term in \eqref{eq:internal chi gss from Ltop} is introduced to respect this requirement.
Indeed, from \eqref{tauflatzero} we know that, since we set $\bra\vartheta=0$, all combinations of the form $\bra{W_\alpha}\cS_{-k}(T^\alpha)\cU$ are on section for $k\ge1$.
Looking then at the first term in \eqref{eq:internal chi gss from Ltop} and expanding $\cS^{\sugraupgamma}_{\pm1}$ analogously to \eqref{eq:Sgamma expansion} and using again $\bra\vartheta=0$ we find that the only term that is not on section is $\varrho^{-1}\bra{W_\alpha}\cS_{+1}(T^\alpha)$, which is eliminated by the last term in \eqref{eq:internal chi gss from Ltop}.
The reason for introducing the full embedding tensor \eqref{eq:theta} in order to remove $\varrho^{-1}\bra{W_\alpha}\cS_{+1}(T^\alpha)$ is that its contribution to the topological term reduces to a total derivative $\braket{\theta}{F}=\braket{\theta}{\dd A}$, where one uses the quadratic constraint \eqref{eq:QC theta only} to prove that $\braket{\theta}{F}$ is abelian.

Combining \eqref{eq:Ltop first line gss}, \eqref{eq:Fhat gss}, \eqref{eq:Ltop cocycle gss} and \eqref{eq:internal chi gss from Ltop} we finally find
\begin{equation}
\cL^{\rm top}_{\rm ExFT} =
r\,\cL^{\rm top}_{\rm gsugra}
\end{equation}
with 
\begin{align}\label{eq:sugra Ltop}
&\hspace{-3em}\varrho^{-1}\,\cL^{\rm top}_{\rm gsugra}\ \dd x^0\wedge\dd x^1\ =\CR 
&
  2D\sugrachi_{1} 
- 2\eta^{AB}\sum_{m\in\ints}m\, Q^m_A\wedge P^{-m-1}_B
+ 2\sum_{k=1}^{\infty} P_k\wedge \big( \sugrachi_{k+1}-\sugrachi_{k-1} \big)
\\\nonumber&
+\bra\theta \Big( \cS^{\sugraupgamma}_{+1}(L_{-1})+\cS^{\sugraupgamma}_{-1}(L_{-1})\Big) \ket{F}
-\omega^\alpha(V)\big[\cS^{\sugraupgamma}_{+1}(T^\beta)+\cS^{\sugraupgamma}_{-1}(T^\beta)\big]_\alpha \eta_{-1\beta\gamma}\bra{\theta}T^\gamma\ket{F} 
\,,
\end{align}
up to a total derivative. 
The first and second line in this equation correspond to the two terms $2\mathbf{D}\breve\chi_1$ and $\bra{\theta}\mathbf{O}(M)\ket{F}$ displayed in \eqref{eq:Ltop gsugra intro}, respectively.
This concludes the proof of the reduction of the topological term.

Equation \eqref{eq:sugra Ltop} defines the E$_9$ covariant topological term for $D=2$ gauged maximal supergravity.
The terms in the second line of \eqref{eq:sugra Ltop} play the same role as the $\widehat\bbF^\upgamma_{\pm1}$ and cocycle terms in \eqref{eq:exft Ltop} and the combinations of embedding tensor, scalar field dependent shift operators and field strengths appearing here can be regarded as the gauged supergravity counterparts of \eqref{eq:shift F} and \eqref{eq:hat F}.
Let us stress that despite the fact that the second term in the first line of \eqref{eq:sugra Ltop} is expanded in terms of the $\mathfrak{e}_8$ decomposition of $\mathfrak{e}_9$, it simply equals the $\mathfrak{e}_9$ algebra cocycle, so that the whole first line is in fact the $K(\mathfrak{e}_9)$ covariant differential $\mathbf{D}\breve\chi_1$ and can be expressed straightforwardly in any other graded decomposition.
Invariance of \eqref{eq:sugra Ltop} under gauge transformations follows from the reduction, but can also be proved directly by noticing that \eqref{eq:sugra Ltop} can be constructed from a shifted Maurer--Cartan equation for the scalar field currents, following the very same procedure that was used in \cite{Bossard:2021jix} to construct the ExFT topological term.
We display such an equation in \eqref{eq:shiftedMC gsugra}.

The equations of motion of gauged maximal supergravity can be computed from \eqref{eq:sugra Ltop} and the scalar potential \eqref{eq:VsugraIntro} following the same steps as in Section~4.6 of \cite{Bossard:2021jix} and we expect that they must agree up to the duality equation and field redefinitions with the ones derived in \cite{Samtleben:2007an}.
The formulation of \cite{Samtleben:2007an} is directly based on the linear system \cite{Breitenlohner:1986um} written in the E$_8$ decomposition of the loop algebra. However, the integrability of the equations is lost in gauged supergravity and it is more convenient to write the equations of motion in a duality frame in which the gauged supergravity Lagrangian only involves propagating physical fields. We proved that our Lagrangian is well defined from ExFT and we shall not attempt to check the equivalence with \cite{Samtleben:2007an}. 
Indeed, a more effective approach is to derive the physical Lagrangian for specific gaugings from \eqref{eq:sugra Ltop} and then compute the equations of motion for the finite set of fields that appear there. This procedure will be displayed in the companion paper \cite{SO9}, and in particular we will match the Lagrangian of SO($9$) gauged supergravity constructed in \cite{Ortiz:2012ib}.

\section{The scalar potential}
\label{sec:pot}

An `unextended' form of the ExFT scalar potential, for which $\mathcal{V}\in\widehat{\mathrm{E}}_8\rtimes (\mathbb{R}^+_{L_0}\ltimes \mathbb{R}_{L_{-1}})$, was already constructed in \cite{Bossard:2018utw}. Here we introduce the Virasoro extension of this potential, where $\mathcal V\in \widehat{\mathrm{E}}_8\rtimes \text{Vir}^-$, which reduces to the expression in \cite{Bossard:2018utw} upon gauge-fixing $\phi_n=0$ for $n\ge2$.
Subsequently, we perform its reduction to two dimensions using the ans\"atze for the scalars fields presented in the previous section. This leads to the expression of the gauged supergravity scalar potential, which only depends on the supergravity scalars $V(x)$, and which is quadratic in the embedding tensor $\bra{\theta}$. The fact that the dependence on the internal coordinates $y$ consistently factorises out of the potential \text{can} be seen as additional check of the truncation ansatz \eqref{eq:internal chi gss from Ltop} for the constrained scalar $\bra{\tilde\chi_1}$.  

\subsection{Virasoro-extended scalar potential}

The extended scalar potential is constructed, in analogy with the unextended case, by requiring invariance under both rigid $\hevirm$ transformations and extended generalised diffeomorphisms. Some of the computational details are relegated to Appendix~\ref{app:gaugeextpot} and we only summarise here the main results. The extended potential appearing in the ExFT Lagrangian \eqref{eq:exft pseudoL} can be conveniently decomposed into
\begin{equation}\label{eq:extpot}
\rho\,  V_{\text{ExFT}}=V_1-2\,V_2+2\,V_3+2\,V_4+\frac12 V_5\,,
\end{equation}
with
\begin{subequations}
\begin{align}
&V_1=\eta_0^{\alpha\beta}\bra{\cP_\alpha}\cM^{-1}\ket{\cP_\beta}+4\sum\limits_{q=1}^\infty\bra{\cP_\alpha}\mathcal{V}^{-1}\cS_q\,(T^\alpha)\,\mathcal{V}^{-\dagger}\ket{\cP_q}\,,\label{eq:extpot1}\\
&V_2=\bra{\cP_\alpha}\mathcal{V}^{-1}\,T^\beta T^\alpha\,\mathcal{V}^{-\dagger}\ket{\cP_\beta}\,,\\[2mm]
&V_3=\bra{\cP^{(1)}_\alpha}\mathcal{V}^{-1}\,T^{\beta\dagger} T^\alpha\,\mathcal{V}^{-\dagger}\ket{\cP^{(1)}_\beta}\,,\\[2mm]
&V_4=\bra{\cP_0}\mathcal{V}^{-1}\,T^\alpha\,\mathcal{V}^{-\dagger}\ket{\cP_\alpha}\,,\\
&V_5=c_\mathfrak{vir}\sum\limits_{q=2}^\infty q(q-1)\bra{\cP_q}\mathcal{M}^{-1}\ket{\cP_q}\,.
\end{align}
\end{subequations}
We comment further below on how this expression reduces to the one given in~\cite{Bossard:2018utw} after explaining the notation. The (shifted) internal scalar currents appearing in the above expressions were defined in \eqref{eq:P internal} and \eqref{eq: P internal} and take value in $\mathfrak{\hat e}_8\oleft \mathfrak{vir}$. The Virasoro components of the current are denoted by $\bra{\cP_q}$. In addition, we simply denote $(\cV^{-1})^\dagger$ by $\cV^{-\dagger}$, and we use the notation for Hermitian conjugation of bra-ket vectors  $\ket{\mathcal P_\alpha}=(\bra{\mathcal P_\alpha})^\dagger$  to simplify the expressions.\footnote{Note that $\ket{\cP_\alpha}$ is \textit{not} in the E$_9$ representation $R(\Lambda_0)_{-1}$. This notation allows the following rewriting of generic expressions $\bra{A} \mathcal{V}^{-1} T^\alpha \mathcal{V}^{-\dagger} \ket{B}=\bra{B}\mathcal{V}^{-1} T^{\alpha\dagger}\mathcal{V}^{-\dagger} \ket{A}$ with $\bra{A},\,\bra{B}$ in $\overline{R(\Lambda_0)_{-1}}$ and $\ket{A}=(\bra{A})^\dagger$, $\ket{B}=(\bra{B})^\dagger$. This pairing descends from the $K(\mathfrak{e}_9)$-invariant pairing on $\overline{R(\Lambda_0)_{-1}}$ since $\bra{A} \cV^{-1}$ transforms under $K(\mathfrak{e}_9)$.} 
Note also the presence of the Hermitian conjugate on the generator in $V_3$ which is due to the fact that, unlike $\bra{\cP_\alpha}\otimes T^\alpha$, the shifted current $\bra{\cP^{(1)}_\alpha}\otimes T^\alpha$ is not Hermitian as an $\hevir$ element. Finally, $c_\mathfrak{vir}$ in $V_5$ denotes the Virasoro central charge in the basic representation, which we recall for $\mathfrak{e}_9$ takes the value $c_\mathfrak{vir}=8$. We keep the symbol explicit here, as it allows us to  generalise the results directly to exceptional field theories based on other affine algebras. We will also show explicitly in Section~\ref{sec:GSS pot} that, after performing the reduction to two dimensions, $c_\mathfrak{vir}$ drops out of the supergravity potential.

From the transformation properties \eqref{eq:coset trf} and \eqref{eq:internal chi var} of the scalars $\mathcal{V}$ and $\bra{\tilde \chi_1}$, it follows that both currents transform covariantly under rigid $\hevirm$ as
\begin{align}
\bbdelta^\alpha \bra{\mathcal P_\beta}\,\otimes\,T^\beta=&\,\bra{\mathcal P_\beta}\big(T^\alpha{-}\delta^\alpha_{L_0}\mathsf K\big)\,\otimes \,T^\beta+\bra{\mathcal P_\beta}\,\otimes\, [k^\alpha,T^\beta]\,,\label{eq:hevirmP}\\
\bbdelta^\alpha \bra{\mathcal P^{(1)}_\beta}\,\otimes\,T^\beta=&\,\bra{\mathcal P^{(1)}_\beta}\big(T^\alpha{-}\delta^\alpha_{L_0}\mathsf K\big)\,\otimes \,T^\beta+\bra{\mathcal P^{( 1)}_\beta}\,\otimes\, [k^\alpha,T^\beta]\,,\label{eq:hevirmP1}
\end{align}
with $k^\alpha\in K(\mathfrak e_9)$. With the above formulae, it is then straightforward to verify that the terms $V_2$, $V_3$, $V_4$ and $V_5$ of the potential are manifestly invariant under rigid $\hevirm$, up to a uniform scaling under the action of $L_0$. The term $V_1$ requires more care due to the presence of the shift operators, and of the bilinear form $\eta^{\alpha\beta}$ which is only invariant over $\mathfrak e_9$. Let us then look at the variation of the first term in \eqref{eq:extpot1}. Because the loop components $\bra{\mathcal P_A^n}$ in \eqref{eq:hevirmP} are partially rotated into the Virasoro components $\bra{\cP_q}$ under the compensating $K(\mathfrak{e}_9)$ transformation $k^\alpha$, one is left with the following contribution
\begin{align}
\bbdelta^\alpha\big(\eta_0^{\beta\delta}\bra{\cP_\beta}\cM^{-1}\ket{\cP_\delta}\big)=&\,2\sum\limits_{n\in\mathbb{Z}}\sum_{q=1}\limits^\infty n\,\eta^{AB}(k^\alpha)_A^{n}\,\big(\bra{\cP_B^{-n-q}}\cM^{-1}\ket{\cP_q}+\bra{\cP_{B}^{-n+q}}\cM^{-1}\ket{\cP_{-q}}\big)\label{eq:varV11}\\
&-2\,\delta^\alpha_{L_0}\,\eta^{\beta\delta}\bra{\cP_\beta}\cM^{-1}\ket{\cP_\delta}\nonumber\\
=&\,4\sum\limits_{n\in\mathbb{Z}}\sum\limits_{q=1}^\infty n\,\eta^{AB}(k^\alpha)_A^{n}\,\bra{\cP_B^{-n-q}}\cM^{-1}\ket{\cP_q}-2\,\delta^\alpha_{L_0}\,\eta^{\beta\delta}\bra{\cP_\beta}\cM^{-1}\ket{\cP_\delta}\,.\nonumber
\end{align}
In going from the first to the second line we simply flipped $n\rightarrow -n$, and subsequently used that the current is Hermitian while $k^\alpha$ is anti-Hermitian. The variation of the second term in $V_1$, can be easily computed using \eqref{eq:hevirmP} and 
\begin{equation}
\bra{\cP_\beta}\otimes \cS_q([k^\alpha,T^\beta])=\bra{\cP_\beta}\otimes [k^\alpha,\cS_q(T^\beta)]-\sum\limits_{n\in\mathbb{Z}}n\,\eta^{AB}(k^\alpha)^n_A\bra{\cP_B^{-n-q}}\otimes \dK\,.
\end{equation}
The resulting loop variation of the second term, which is proportional to $(k^\alpha)^A_n$, precisely cancels against that of \eqref{eq:varV11}. This implies that $V_1$ is also invariant, up to the $L_0$ scaling. Taking into account that $\bbdelta^\alpha \rho=-\delta^\alpha_{L_0}\rho$, with a Kronecker delta on the right-hand side, one then finds that under rigid $\hevirm$ the extended potential \eqref{eq:extpot} simply scales as  
\begin{equation}\label{eq:hevirmextpot}
\bbdelta^\alpha V_{\text{ExFT}}=-\delta^\alpha_{L_0}V_{\text{ExFT}}\,.
\end{equation}

At this point, the relative coefficients between the five terms appearing in the potential still require a justification. As is usual in exceptional field theories, these coefficients are fixed by requiring gauge invariance. Under extended generalised diffeomorphisms one can indeed show that the potential \eqref{eq:extpot} transforms into total internal derivatives. We prove  explicitly in Appendix~\ref{app:gaugeextpot} that this fixes all the coefficients of the $V_i$ in \eqref{eq:extpot}. We have not proved that this is the most general potential invariant under extended generalised diffeomorphisms, but we know from~\cite{Bossard:2018utw} that it correctly reproduces the dynamics of eleven-dimensional and type IIB supergravity. Let us finally point out that by gauge fixing $\phi_n=0$ for $n\ge2$ in \eqref{eq:exft coset rep} so that $\cV \in\widehat{\mathrm{E}}_8\rtimes (\mathbb R^+_{L_0}\ltimes \mathbb{R}_{L_{-1}})$, the Virasoro components of the currents $\bra{\cP_q}$ vanish for $|q|\ge2$. Out of the sum in $V_1$, only the $q=1$ term survives and the whole term $V_5$ vanishes. In this case, one then correctly recovers the expression of the unextended scalar potential derived in Section~4.3 of \cite{Bossard:2018utw}.\footnote{The field $\bra{\tilde\chi}$ used in \cite{Bossard:2018utw} is related to the one appearing here in the shifted currents by $\langle\tilde\chi|=\,2\langle \tilde \chi_1|+2\,\langle P_1| $.} 

\subsection{Reduction of the scalar potential}
\label{sec:GSS pot}

We will now compute the reduction of the ExFT scalar potential \eqref{eq:extpot}, using the generalised Scherk--Schwarz ans\"atze \eqref{eq:coset gss} and \eqref{eq:internal chi gss from Ltop} for the scalar fields $\cV$ and $\bra
{\tilde\chi_1}$. As for the reduction of the topological term, we will set the trombone $\bra{\vartheta}=0$ as we are only interested in Lagrangian gaugings. However, here we will keep the twist matrix $\cU\in\widehat{\mathrm{E}}_8\rtimes \text{Vir}^-$. One could optionally set $\bra{\tilde h_\alpha}=0$ without loss of generality, but it will not affect the following computations as the ans\"atze for the scalar fields do not depend on it. 
Our final result for the supergravity potential will then only depend on the embedding tensor $\bra{\theta}$.

The substitution of the ans\"atze in the ExFT potential, and the subsequent necessary manipulations  generate numerous cocycles due to the presence of the shift operators. 
One can circumvent this difficulty by choosing to work with a version of the Weitzenb\"ock connection \eqref{eq:weitz} that is dressed by the supergravity scalars \eqref{eq:sugra coset rep}. We therefore define
\begin{equation}
\bra{\underline{W}_{\alpha}}\otimes T^\alpha= \,\frac12\bra{W_{\alpha}}\otimes V T^\alpha V^{-1}\,,\label{eq:VconjW}
\end{equation}
which still takes values in $\hevirm$. 
For instance, $\bra{\underline{W}_0}\cU= \bra{\partial_r} r^{-1}$ from the ansatz in Section~\ref{sec:gSSans}.
It is also useful to define shifted versions of the dressed Weitzenb\"ock connection by
\begin{equation}
\bra{\underline{W}_\alpha^{(\pm1)}}\otimes T^\alpha= \bra{\underline{W}_\alpha}\otimes \cS_{\pm 1}(T^\alpha)+\bra{\underline{\widetilde W}_{\pm}}\otimes \dK\,,\label{eq:weitzuns}
\end{equation} 
which are not Hermitian conjugate to each other. Importantly, their $\dK$ completions $\bra{\underline{\widetilde W}_{\pm}}$ are not independent fields and their expression read
\begin{align}
2\,\bra{\underline{\widetilde{W}}_{+}}=&\,\omega^\alpha(V)\,\big[\cS^{\sugraupgamma}_1\,(T^\beta)\big]_\alpha\,\bra{W_\beta}-\bra{W_\alpha}\cS^{\sugraupgamma}_1\,(T^\alpha)
-\varrho^{-1}\bra{\theta}\,,\label{eq:tildeW+}\\
2\,\bra{\underline{\widetilde{W}}_{-}}=&\,\omega^\alpha(V)\,\big[\cS^{\sugraupgamma}_{-1}(T^\beta)\big]_\alpha\,\bra{W_\beta}-\bra{W_\alpha}\cS^{\sugraupgamma}_{-1}\,(T^\alpha)\,,\label{eq:tildeW-}
\end{align}
in terms of the embedding tensor $\bra{\theta}$ defined in \eqref{eq:theta}, as well as the series of shift operators \eqref{eq:Sbargamma def}. It is important to note that $\bra{\underline{\widetilde{W}}_{\pm}}$ both satisfy the `flat' section constraint, which implies that $\bra{\underline{W}_\alpha^{(\pm1)}}$ do as well. This relies on the results of Section~\ref{sec:gauging id}, where we showed that $\bra{W_\alpha}\cS_{-n}(T^\alpha)$ is on section for all $n\geq 1$ when $\bra{\vartheta}=0$.  For $\bra{\underline{\widetilde{W}}_{+}}$, this also relies on the observation that the contribution proportional to $\bra{W_\alpha}\cS_0(T^\alpha)$ coming from the second term of \eqref{eq:tildeW+} is on section due to the condition $\bra{\vartheta}=0$. Note finally that the contribution proportional to $\bra{W_\alpha}\cS_1(T^\alpha)$ is not on section, but that the combination $\bra{W_\alpha}\cS_1(T^\alpha)+ \bra{\theta}= -\bra{w^+}$ simplifies and is on section.
This is analogous to what is discussed under \eqref{eq:internal chi gss from Ltop}.

Making use of the dressed versions of the (shifted) Weitzenb\"ock connections \eqref{eq:VconjW} and \eqref{eq:weitzuns}, will simplify the upcoming computations. As mentioned previously, this will especially allow us to avoid dealing with cocycles. These connections also satisfy several useful simple relations. In particular, the condition $\bra{\vartheta}=0$ directly implies 
\begin{equation}
\bra{\underline{W}_\alpha}V^{-1} T^\alpha V=\,0\,.\label{eq:W0vanish}
\end{equation}
It is important to note the absence of tensor product in the above formula, when compared \eqref{eq:VconjW}. Here the generators act on the constrained `bra' (up to the conjugation by $V$). Furthermore, pulling the implicit conjugation by $V$ out of the shift operators in \eqref{eq:weitzuns} by using \eqref{eq:group cocycle vs Sgamma}, leads to
\begin{equation}
\bra{\underline{W}_\alpha^{(-1)}}V^{-1} T^\alpha V=\, 0\,,\label{eq:W-vanish}    
\end{equation}
as well as 
\begin{align}
\bra{\underline{W}_\alpha^{(1)}}V^{-1} T^\alpha V=\,-\frac{1}{2\varrho}\bra{\theta}\,,\label{eq:W+theta}
\end{align}

The reduction ansatz \eqref{eq:internal chi gss from Ltop} for the scalar $\bra{\tilde\chi_1}$ can be written straightforwardly using the definitions \eqref{eq:tildeW+} and \eqref{eq:tildeW-},  
\begin{equation}
2\,\bra{\tilde \chi_1}=\big(\bra{\underline{\widetilde{W}}_{+}}+\bra{\underline{\widetilde{W}}_{-}}\big)r\,\cU\,.
\end{equation}
Together with the ansatz \eqref{eq:coset gss} for the scalars $\cV$, this implies that the currents \eqref{eq:P internal} and \eqref{eq: P internal} decompose into
\begin{align}
\bra{\cP_\alpha}\otimes T^\alpha= &\,\bra{\underline{W}_{\alpha}}r\,\cU\otimes (T^\alpha+T^{\alpha\dagger})\,,\label{eq:Pan}\\
\bra{\cP^{(1)}_\alpha}\otimes T^\alpha=&\,\bra{\underline{W}_\alpha^{(1)}}r\,\cU\otimes T^\alpha+\bra{\underline{W}_\alpha^{(-1)}}r\,\cU\otimes T^{\alpha\,\dagger}\,.\label{eq:sPan}
\end{align}
Let us now substitute directly the above ans\"atze for the currents in the various terms of the scalar potential \eqref{eq:extpot}. We find 
\begin{subequations}
\begin{align}
r^{-2}\,V_1=&\,2\,r^{-1}\sum\limits_{n\in\mathbb{Z}}\eta^{AB}\langle\underline{W}_A^n|V^{-1}\cV^{-\dagger}|\cP_B^{-n}\rangle-8\,\langle\underline{W}_0|M^{-1}|\underline{W}_{\textrm K}\rangle\nonumber\\
&+4\sum\limits_{q=1}^\infty\langle\underline{W}_\alpha|V^{-1}\big(\cS_q(T^\alpha)+\cS_{-q}(T^\alpha)^\dagger\big)V^{-\dagger}|\underline{W}_{-q}\rangle\,,\\[2mm]
r^{-2}\,V_2=&\,r^{-1}\langle\underline{W}_\alpha|V^{-1}[T^\beta,T^\alpha]\cV^{-\dagger}|\cP_\beta\rangle\nonumber\\[.5mm]
&+\langle\underline{W}_\alpha|V^{-1}T^\beta T^\alpha V^{-\dagger}|\underline{W}_\beta\rangle+\langle\underline{W}_\alpha|V^{-1}T^{\beta}T^{\alpha\,\dagger}V^{-\dagger}|\underline{W}_\beta\rangle\,,\\[2mm]
r^{-2}\,V_3=&\,r^{-1}\langle\underline{W}_\alpha|V^{-1}[\cS_{-1}(T^\beta),\cS_{1}(T^\alpha)]\cV^{-\dagger}|\cP_\beta\rangle+\langle\underline{W}_\alpha^{(1)}|V^{-1}T^\alpha T^{\beta\,\dagger}V^{-\dagger}|\underline{W}^{(1)}_\beta\rangle\nonumber\\[.5mm]
&+\langle\underline{W}^{(1)}_\alpha|V^{-1}\{T^\alpha ,T^\beta\} V^{-\dagger}|\underline{W}^{(-1)}_\beta\rangle+\langle\underline{W}^{(-1)}_\alpha|V^{-1}T^\beta\,T^{\alpha\dagger} V^{-\dagger}|\underline{W}^{(-1)}_\beta\rangle\,,\\[2mm]
r^{-2}\,V_4=&\,2\,\langle\underline{W}_0|V^{-1}T^\alpha V^{-\dagger}|\underline{W}_\alpha\rangle\,,\\[2mm]
r^{-2}\,V_5=&\,c_\mathfrak{vir}\,\sum\limits_{q=2}^\infty q(q-1)\bra{\underline{W}_{-q}}M^{-1}\ket{\underline{W}_{-q}}\,,\label{eq:GSSV5}
\end{align}
\end{subequations}
where we denote by $M(x)=V^\dagger V$ the Hermitian matrix describing the supergravity scalar fields. For $V_2$ and $V_3$ we followed the same method, which is to take only the commutator of the $\bra{\underline{W}_\alpha}\otimes T^\alpha$ (or $\bra{\underline{W}^{(1)}_\alpha}\otimes T^\alpha$) term that arises from opening up the $\bra{\cP_\alpha}\otimes T^\alpha$ (or $\bra{\cP^{(1)}_\alpha}\otimes T^\alpha$) on the left. In $V_2$ and $V_4$ we dropped terms that vanish according to \eqref{eq:W0vanish}. The commutators that appear in $V_3$ and $V_2$ can be shown to cancel the first term of $V_1$ by using the identity \eqref{eq:commutator shifts}. In the process, a term proportional to the Virasoro central charge in the basic representation $c_\mathfrak{vir}$ is generated, and precisely cancels against the contribution coming from $V_5$. 
Combining the various results and using the ansatz \eqref{eq:dilaton gss} for the dilaton, one obtains rather easily 
\begin{align}
\varrho\,r^{-1}\,V_{\text{ExFT}}=&-2\,\langle\underline{W}_\alpha|V^{-1}T^\beta T^\alpha V^{-\dagger}|\underline{W}_\beta\rangle-2\,\langle\underline{W}_\alpha|V^{-1}T^{\beta}T^{\alpha\,\dagger}V^{-\dagger}|\underline{W}_\beta\rangle\nonumber\\[1mm]
&+2\,\langle\underline{W}_\alpha^{(1)}|V^{-1}T^\alpha T^{\beta\,\dagger}V^{-\dagger}|\underline{W}^{(1)}_\beta\rangle+2\,\langle\underline{W}^{(1)}_\alpha|V^{-1}\{T^\alpha, T^\beta\} V^{-\dagger}|\underline{W}^{(-1)}_\beta\rangle\nonumber\\[1mm]
&+2\,\langle\underline{W}^{(-1)}_\alpha|V^{-1}T^\beta\,T^{\alpha\dagger} V^{-\dagger}|\underline{W}^{(-1)}_\beta\rangle+2\,\langle\underline{W}_0|V^{-1}T^\alpha V^{-\dagger}|\underline{W}_\alpha\rangle\nonumber\\
&+2\sum\limits_{q=1}^\infty\langle\underline{W}_{-q}|V^{-1}\big(\cS_{-q}(T^\alpha)-\cS_{-q}(T^\alpha)^\dagger\big)V^{-\dagger}|\underline{W}_\alpha\rangle \,.\label{eq:VExFTstep}
\end{align}

The goal is now to express the right-hand side of \eqref{eq:VExFTstep} purely in terms of the embedding tensor $\bra{\theta}$ and the supergravity scalars $V(x)$, such that the only remaining dependence on the internal coordinates is correctly factorised in $r^{-1}(Y)$ on the left-hand side. Using \eqref{eq:W+theta}, one can immediately see that the first term of the second line can be rewritten as $\tfrac{1}{2\varrho^2}\bra{\theta}M^{-1}\ket{\theta}$. For the rest of the terms however, identifying the correct rewriting in terms of the embedding tensor is much less obvious. It is possible to constrain the possibilities by focusing on rewritings which are manifestly invariant under the rigid $\hevirm$ transformations of supergravity. But even so, one is a priori left with an infinity of terms to consider that take the generic form 
\begin{equation}
\frac{1}{\varrho^2}\,\eta_{k\,\alpha\beta}\,\eta_{p\,\gamma\delta}\ldots\bra{\theta}V^{-1}T^\alpha\,T^\gamma\ldots T^{\delta\dagger} T^{\beta \dagger} V^{-\dagger}\ket{\theta}\,,\;\;\;\;\;\;\;\;\;\;k,p,\ldots\in\mathbb{Z}\,,\label{eq:possibilities}
\end{equation}
where the generators that are contracted with the bilinear forms~\eqref{eq:etak} must be ordered in this specific way (corresponding to the successive action of coset Virasoro generators on $\bra\theta\otimes\bra\theta$) in order for the expression to be well defined. 
To make contact with the terms in \eqref{eq:VExFTstep}, one has to open up the two $\bra{\theta}$ using the expression \eqref{eq:W+theta}, and subsequently move the generators contracted with the bilinear forms towards the `bra' and `ket' of the two Weitzenb\"ock connections. For specific values of $k,p,\ldots$, this ultimately allows us to get rid of all the terms that depend on the bilinear forms by using the `flat' section constraints. This procedure requires taking various commutators which, along the way, generate some of the terms that appear in the potential \eqref{eq:VExFTstep}. We give a concrete example of how the method can be carried out below.
A possible issue with this strategy is that a specific term in \eqref{eq:VExFTstep} can arise from different terms in \eqref{eq:possibilities}, and it therefore looks like we are facing a problem with a high degree of degeneracy. Here we ignore this potential issue, and pragmatically carry on with our analysis by considering the simplest possibilities which are of the form $\eta_{k\,\alpha\beta}\bra{\theta}V^{-1}T^\alpha T^{\beta\dagger}V^{-\dagger}\ket{\theta}$, for $k<0$. We will see that this turns out to be sufficient for our purpose.\footnote{%
One can also apply the teleparallelism approach to ExFT~\cite{Cederwall:2021xqi} and extend it to the affine case. Because the potential then only brings in the Bianchi identity for the teleparallel field strengths (similar to the quadratic constraints on the embedding tensor), this suggests that the gauged supergravity potential should only involve terms with a single $\eta_{k\,\alpha\beta}$.} Let us however already remark that all such quadratic terms in $\bra{\theta}$ vanish, except for $k=-2$. To see this, one must first pass $V\in \widehat{\mathrm{E}}_8\rtimes \text{Vir}^-$ through $\eta_k$ using \eqref{eq:eta trf}, such that the generators directly act on $\bra{\theta}$'s. This generates an infinite series of terms proportional to $\eta_{k+q}$ with $q\leq 0$ and $q\neq k$, which all vanish for $k+q\neq -2$ as a consequence of the quadratic constraint and \eqref{eq:extra cons}.

Following the strategy outlined directly below \eqref{eq:possibilities}, we then compute
\begin{align}
\frac{1}{4\varrho^{2}}\,\eta_{-2\,\alpha\beta}\,\bra{\theta}V^{-1}T^\alpha T^{\beta\dagger}V^{-\dagger}\ket{\theta}=&\,\eta_{-2\,\alpha\beta}\,\bra{\underline{W}_\gamma^{(1)}}V^{-1}T^\gamma \,T^\alpha \,T^{\beta\dagger}\,T^{\delta\dagger}V^{-\dagger}\ket{\underline{W}_\delta^{(1)}}\nonumber\\
=&\,\eta_{-2\,\alpha\beta}\,\bra{\underline{W}_\gamma^{(1)}}V^{-1}T^\alpha \,T^\gamma \,T^{\delta\dagger}\,T^{\beta\dagger}V^{-\dagger}\ket{\underline{W}_\delta^{(1)}}\nonumber\\
&\,+\eta_{-2\,\alpha\beta}\,\bra{\underline{W}^{(1)}_\gamma}V^{-1}[T^\gamma, T^\alpha] \,\{T^{\delta\dagger},T^{\beta\dagger}\}V^{-\dagger}\ket{\underline{W}_\delta^{(1)}}\,.\label{eq:example}
\end{align}
The second line can be shown to vanish by passing $V$ through $\eta_k$ and subsequently using the section constraint. The commutator in the last line can be handled using the identity
\begin{align}
\eta_{k-1\,\alpha\beta}\bra{\underline W^{(1)}_\gamma}V^{-1}\big[ T^\gamma, T^\alpha \big]\otimes T^\beta =&\Big(\eta_{k\,\alpha\beta}\bra{\underline W_\gamma}V^{-1}\big[  T^\gamma, T^\alpha \big] +\sum\limits_{q=0}^\infty\eta_{k-q\,\alpha\beta}\bra{\underline W_{-q}}V^{-1}T^\alpha\Big)\otimes T^\beta\nonumber\\
&\,- \bra{\underline W_\gamma}V^{-1}\cS_k (T^\gamma)\otimes \dK+\bra{\underline W_\gamma}V^{-1}\otimes \cS_k(T^\gamma)\,,\label{eq:identity shift com}
\end{align}
which holds for all $k\in\mathbb Z$, and which follows from \eqref{eq: id A2}. We therefore obtain
\begin{align}
\frac{1}{4\varrho^{2}}\,\eta_{-2\,\alpha\beta}\,\bra{\theta}V^{-1}T^\alpha T^{\beta\dagger}V^{-\dagger}\ket{\theta}=
&\,-\eta_{-1\,\alpha\beta}\,\bra{\underline{W}^{(1)}_\gamma}V^{-1}[T^\gamma,T^\alpha] \,T^{\delta\dagger}\,T^{\beta\dagger}V^{-\dagger}\ket{\underline{W}_\delta}\nonumber\\
&\,+\sum\limits_{q=0}^\infty\eta_{-1-q\,\alpha\beta}\,\bra{\underline W^{(1)}_\gamma}V^{-1}[T^\gamma, T^\alpha]\,T^{\beta\dagger}V^{-\dagger}\ket{\underline{W}_{-q}}\nonumber\\
&\,+\bra{\underline W_\alpha}V^{-1}\{T^{\beta\dagger},\cS_{-1}(T^\alpha)^\dagger\}V^{-\dagger}\ket{\underline W_\beta^{(1)}}\nonumber\\[1mm]
&\,-2\,\bra{\underline W_\alpha}V^{-1}\cS_{-1}(T^\alpha)\,T^{\beta\dagger}V^{-\dagger}\ket{\underline W_\beta^{(1)}}\,,\label{eq:eta -2 int}
\end{align}
where we again dropped terms using the section constraint and \eqref{eq:W0vanish}. The shifted connections in the last two lines can be completed into $\bra{\underline W_\alpha^{(-1)}}\otimes T^\alpha$ at no cost, by noticing that their $\dK$ components $\bra{\underline{\widetilde W}_-}$ cancel out. This allows us to use the relation \eqref{eq:W-vanish}, after which the last two lines simply reduce to $\langle\underline{W}^{(1)}_\alpha|V^{-1}\{T^\alpha,T^\beta\} V^{-\dagger}|\underline{W}^{(-1)}_\beta\rangle$. Note that this is one of the terms that we were trying to reproduce in \eqref{eq:VExFTstep}. We can take care of the remaining commutators by making use once more of \eqref{eq:commutator shifts}. The final expression then reads
\begin{align}
\frac{1}{4\varrho^{2}}\,\eta_{-2\,\alpha\beta}\,\bra{\theta}V^{-1}T^\alpha T^{\beta\dagger}V^{-\dagger}\ket{\theta}=&\,\bra{\underline W_\alpha}V^{-1}T^\beta\, T^{\alpha\dagger}V^{-\dagger}\ket{\underline W_\beta}-\bra{\underline W_\alpha}V^{-1}T^\beta\, T^{\alpha}V^{-\dagger}\ket{\underline W_\beta}\nonumber \\
&\,-\bra{\underline W_0}V^{-1}T^\alpha V^{-\dagger}\ket{\underline W_\alpha}+\langle\underline{W}^{(1)}_\alpha|V^{-1}\{T^\alpha,T^\beta\} V^{-\dagger}|\underline{W}^{(-1)}_\beta\rangle\nonumber\\
&+\sum\limits_{q=1}^\infty\langle\underline{W}_{-q}|V^{-1}\big(\cS_{-q}(T^\alpha)-\cS_{-q}(T^\alpha)^\dagger\big)V^{-\dagger}|\underline{W}_\alpha\rangle\,,
\end{align}
The terms proportional to $\bra{\underline W_\alpha}\otimes \,\cS_0(T^\alpha)$ that arise from the commutators in \eqref{eq:eta -2 int} have been promoted into $\bra{\underline W_\alpha} \otimes T^\alpha$ using the fact that their $\dK$ components $\bra{\underline W_\dK}$ cancel each other. In the process, we used again \eqref{eq:W0vanish}. We can now substitute the above result in the ExFT scalar potential \eqref{eq:VExFTstep} and write
\begin{align}
\varrho\,r^{-1}\,V_{\text{ExFT}}=&\,\frac{1}{2\varrho^2}\Big(\bra{\theta}M^{-1}\ket{\theta}+\eta_{-2\,\alpha\beta}\,\bra{\theta}V^{-1}T^{\alpha}\, T^{\beta\dagger}V^{-\dagger}\ket{\theta}\Big)\nonumber\\
&\,-4\,\langle\underline{W}_\alpha|V^{-1}T^{\beta}T^{\alpha\,\dagger}V^{-\dagger}|\underline{W}_\beta\rangle+4\,\langle\underline{W}_0|V^{-1}T^\alpha V^{-\dagger}|\underline{W}_\alpha\rangle\nonumber\\[1mm]
&\,+2\,\langle\underline{W}^{(-1)}_\alpha|V^{-1}T^\beta\,T^{\alpha\dagger} V^{-\dagger}|\underline{W}^{(-1)}_\beta\rangle\,.\label{eq:VExFTfinal}
\end{align}
By following a similar computational logic as the one used above, one proves that the last line can be rewritten in terms of the embedding tensors as
\begin{align}
\frac{1}{2\varrho^2}\eta_{-4\,\alpha\beta}\,\bra{\theta}V^{-1}T^\alpha T^{\beta\dagger}V^{-\dagger}\ket{\theta}=2\,\langle\underline{W}^{(-1)}_\alpha|V^{-1}T^\beta\,T^{\alpha\dagger} V^{-\dagger}|\underline{W}^{(-1)}_\beta\rangle\,,\label{eq:eta-4}
\end{align}
and therefore vanishes according to \eqref{eq:extra cons}. One furthermore shows that the second line of \eqref{eq:VExFTfinal} reduces to the total internal derivative
\begin{equation}
4\,\langle\underline{W}_0|V^{-1}T^\alpha V^{-\dagger}|\underline{W}_\alpha\rangle-4\,\langle\underline{W}_\alpha|V^{-1}T^{\beta}T^{\alpha\,\dagger}V^{-\dagger}|\underline{W}_\beta\rangle=2\,r^{-1}\,\bra{\partial}\Big(\cV^{-1}\,T^\alpha\,V^{-\dagger}\ket{\underline W_\alpha}\Big)\,,
\end{equation}
up to terms that vanish according to $\eqref{eq:W0vanish}$. As a last step, we can rewrite the second term of the first line of \eqref{eq:VExFTfinal} in terms of the supergravity scalar matrix $M=V^\dagger V$ by passing $V$ through the bilinear form $\eta_{-2}$, and subsequently using \eqref{eq:extra cons}. This leads to the final result
\begin{equation}
V_{\text{ExFT}}=r\,V_{\text{gsugra}}+ \,\bra{\partial}\Big( \sugrarho^{-1}  \cU^{-1} T^\alpha M^{-1} \ket{ W_\alpha}\Big)\,,
\end{equation}
where the scalar potential of two-dimensional gauged supergravity reads
\begin{align}
\label{eq:VsugraFinal}
V_{\text{gsugra}}=\frac{1}{2\varrho^3}\,\bra{\theta}M^{-1}\ket{\theta}+\frac{1}{2\varrho}\eta_{-2\,\alpha\beta}\,\bra{\theta}T^{\alpha}M^{-1} T^{\beta\dagger}\ket{\theta}\,.
\end{align}
In~\cite{Bossard:2022wvi}, the potential $V_{\text{sugra}}$ was given with the  additional term \eqref{eq:eta-4}, which we now find to vanish using equation~\eqref{eq:extra cons}. In reductions where there is a mild violation of the section constraint, such as in massive type IIA~\cite{Hohm:2011cp,Ciceri:2016dmd}, the term \eqref{eq:eta-4}  could in principle contribute.

\section{Duality equation for the gauge field strength}
\label{sec:nonLag}

Although the Euler--Lagrange equations of the pseudo-Lagrangian \eqref{eq:Lintro} together with the duality equations \eqref{eq:twsdintro} determine entirely the dynamics of the gauged supergravity degrees of freedom, they do not determine directly the expressions of some one-form and two-form gauge fields that  appear in the uplift ansatz to eleven or ten dimensions. Only the gauge covariant quantities in higher dimensions are directly determined, but one generally wishes to have the explicit expression of the Kaluza--Klein vectors in the metric ansatz. To fix directly these gauge fields it is important to consider the duality equation for the field strength $ |F\rangle$. In this section we derive the duality equation for this field strength from the gSS ansatz.

The Euler-Lagrange equation of the ExFT pseudo-Lagrangian \eqref{eq:exft pseudoL} for the constrained field $\bra{\tilde\chi_1}$ gives a projection of the following duality equation for the gauge field strength  $\ket{\cF}$
\be  |\mathcal{F}\rangle -2 \,\rho^{-2}\star  \cV^{-1} T^{\alpha\dagger} \cV^{-\dagger} |\cP^{(1)}_\alpha\rangle = 0 \; , \label{eq:dualityeqExFT} \ee
where the shifted current $\bra{\cP^{(1)}_\alpha}\otimes T^\alpha$ was defined in \eqref{eq: P internal}. It was shown in  \cite[Section 5]{Bossard:2021jix} that the duality equation \eqref{eq:twsd exft} transforms into \eqref{eq:dualityeqExFT} under external diffeomorphism, and this equation therefore holds without projection as a non-Lagrangian duality equation. 

Note that the field strength $\ket{\cF}$ in  \eqref{eq:dualityeqExFT} also depends on the two-form gauge fields according to \eqref{eq:DorfFS}. Because the two-forms appear without external derivative, this equation turns out to be tautological for all but finitely many components of  $ |\mathcal{F}\rangle$ depending of the chosen solution to the section constraint. It does nonetheless contain some components that are not Euler--Lagrange equations and that cannot be trivially satisfied by an appropriate choice of the two-form gauge fields.

It is important to point out that the unprojected field strength $ |\mathcal{F}\rangle$ does not transform covariantly under generalised diffeomorphisms. Its non-covariant variation stems from one of the two-forms, and therefore takes the form of (the $\Lambda$ component of) a  trivial parameter. The latter cancels against the non-covariant variation of the scalar current $\langle \cP^{(1)}_\alpha|\otimes T^\alpha$ and in this way ensures the covariance of the unprojected duality equation \eqref{eq:dualityeqExFT} under generalised diffeomorphisms. As we shall see below, this is reflected in the Scherk--Schwarz ans\"atze for the two-forms. Note that this subtlety did not appear in checking the invariance of the topological term \eqref{eq:exft Ltop}, since the latter does not depend on the two-form by construction.

A detailed presentation of two-forms and trivial parameters in ExFT can be found in Appendix~A of \cite{Bossard:2021jix}, and we refer to it for notations.\footnote{Note that the two-forms were denoted by $C$ there. Here this font is reserved for the supergravity two-form, and we instead use $\cC$ for those in ExFT.} Here we simply recall how they appear in the expression of the ExFT field strength,
\begin{align}
\ket{\cF}=&\,\ket{\dd \cA}-\frac12\, \mathcal L_\bbA\wedge|\cA\rangle+\eta_{0\,\alpha\beta}\, \langle \partial_\cC | T^\alpha |\cC_{(1}\rangle\, T^\beta |\cC_{2)}\rangle    \nonumber\\
&\,+  \eta_{0\,\alpha\beta} \,\langle \pi_\cC | T^\alpha |\cC_{[1}\rangle\, T^\beta |\cC_{2]}\rangle + \,2 \,\langle \pi_\cC | \cC_{[1}\,\rangle |\cC_{2]}\rangle
+\eta_{-1\,\alpha\beta}\, \Tr[ T^\alpha \cC^+_2 ]\, T^\beta |\cC^+_1\rangle\,\label{eq:ExFTF}
\end{align}
The derivative in the last term of the first line in \eqref{eq:ExFTF} must be understood as acting on the two kets, which together represent a single two-form field $\cC_{\mu\nu}^{(MN)}$. Note also that the above expression corresponds to the $\Lambda$ component\footnote{Note that the $\Lambda$ component in \eqref{eq:DorfFS} is not affected by the ${}^\pi$ projection.} of \eqref{eq:DorfFS}, in which the two-forms have been redefined by the $\Lambda$ components of trivial parameters that are quadratic in the gauge fields. In the context of footnote~\ref{fn:18}, this redefinition simply amounts to $\mathbbm{C}+\tfrac12\upiota(\bbA,\bbA)\rightarrow  \mathbbm{C}$. 

To simplify the upcoming expressions we now fix, without loss of generality, the twist matrix $\cU\in \textrm E_9$. We also set $\langle \vartheta|=0$. The ans\"atze for the two-forms then read 
\begin{align}
 |\mathcal{C}_{(1}\rangle \otimes |\mathcal{C}_{2)}\rangle &= \frac{1}{2} \,r^{-2} \,\mathcal{U}^{-1} |C_{(1} \rangle \otimes  \mathcal{U}^{-1}   |C_{2)}\rangle\,,\\ 
  |\mathcal{C}_{[1} \rangle \otimes |\mathcal{C}_{2]} \rangle \otimes \langle \pi_\cC|  &=  \frac{1}{2} \,r^{-1} \Big(\mathcal{U}^{-1}|C_{(1}\rangle\otimes  \mathcal{U}^{-1} T^\alpha |C_{2)}\rangle -\mathcal{U}^{-1}T^\alpha |C_{(1}\rangle\otimes  \mathcal{U}^{-1} |C_{2)} \rangle  \Bigr) \otimes \langle W_\alpha |\,\cU\,,\nonumber \\
  |\mathcal{C}^+_{1} \rangle \otimes \mathcal{C}^+_{2} &= \frac{1}{2}\,   \mathcal{U}^{-1} |C_{(1} \rangle \otimes  \mathcal{U}^{-1} T^\alpha |C_{2)} \rangle \,  \langle W^+_\alpha |\,\cU+|\mathcal{\delta\cC}^+_{1} \rangle \otimes \mathcal{\delta\cC}^+_{2}\,,\label{eq:2form3}
\end{align}
where the shifted connection $\langle W_\alpha^+|\otimes T^\alpha$ was defined in \eqref{eq:shiftweitz}. We denote the purely $x$-dependent supergravity two-form $C_{\mu\nu}^{(MN)}$ by $ |C_{(1} \rangle \otimes |C_{2)}\rangle$. Just like the ExFT two-form $|\mathcal{C}_{(1}\rangle \otimes |\mathcal{C}_{2)}\rangle$, it belongs to the symmetric tensor product of two $R(\Lambda_0)$ representation, with the representation $R(2\Lambda_0)$ subtracted. Note that the ansatz \eqref{eq:2form3} involves an extra contribution denoted by $|\mathcal{\delta\cC}^+_{1} \rangle \otimes \mathcal{\delta\cC}^+_{2}$ that depends on both $x$ and $y$ coordinates, and which remains to be specified. This will be discussed below. When substituting the above ans\"atze into \eqref{eq:ExFTF}, the two-form contributions to the field strength~\eqref{eq:ExFTF} reduce to
\begin{subequations}
{\allowdisplaybreaks
\begin{align}
    \eta_{0\,\alpha\beta}\, \langle \partial_\cC | T^\alpha |\cC_{(1}\rangle\, T^\beta |\cC_{2)}\rangle&=-\frac{1}{2}r^{-1}\cU^{-1}\eta_{0\,\alpha\beta}\langle W_\gamma|T^\alpha\, T^\gamma|C_{(1}\rangle\,T^\beta|C_{2)}\rangle\nonumber\\*
&\quad-\frac{1}{2}r^{-1}\cU^{-1}\eta_{0\,\alpha\beta}\langle W_\gamma|T^\alpha|C_{(1}\rangle\,T^\beta\,T^\gamma|C_{2)}\rangle\nonumber\\*
&\quad+r^{-1}\cU^{-1}\eta_{0\,\alpha\beta}\langle W_0|T^\alpha|C_{(1}\rangle\,T^\beta|C_{2)}\rangle\,,\label{2formred1}\\[2mm]
\eta_{0\,\alpha\beta} \,\langle \pi_\cC | T^\alpha |\cC_{[1}\rangle\, T^\beta |\cC_{2]}\rangle + \,2 \,\langle \pi_\cC | \cC_{[1}\,\rangle |\cC_{2]}\rangle&=-\frac{1}{2}r^{-1}\cU^{-1}\eta_{0\,\alpha\beta}\langle W_\gamma|T^\alpha\, T^\gamma|C_{(1}\rangle\,T^\beta|C_{2)}\rangle\nonumber\\*
&\quad+\frac{1}{2}r^{-1}\cU^{-1}\eta_{0\,\alpha\beta}\langle W_\gamma|T^\alpha|C_{(1}\rangle\,T^\beta\,T^\gamma|C_{2)}\rangle\nonumber\\*
&\quad+r^{-1}\cU^{-1}\langle W_\alpha|C_{(1}\rangle\,T^\alpha|C_{2)}\rangle\,,\label{eq:2formred2}\\[2mm]
\eta_{-1\,\alpha\beta}\, \Tr[ T^\alpha \cC^+_2 ]\, T^\beta |\cC^+_1\rangle&=r^{-1}\cU^{-1}\eta_{-1\,\alpha\beta}\langle W_\gamma^+|T^\alpha \,T^\gamma|C_{(1}\rangle\,T^\beta|C_{2)}\rangle\nonumber\\*
&\quad +\eta_{-1\,\alpha\beta}\, \Tr[ T^\alpha \delta\cC^{+}_2 ]\, T^\beta |\delta\cC^{+}_1\rangle\,,\label{eq:2formred3}
\end{align}
}
\end{subequations}
where we dropped terms proportional $\langle\vartheta|$. Taking the commutator of the generators in the first term of \eqref{eq:2formred3}, and subsequently using the identity \eqref{eq:commutator shifts} as well as \eqref{eq:theta summary}, leads to
\begin{align}
|\cF\rangle=r^{-1}\cU^{-1}|F\rangle+\eta_{-1\,\alpha\beta}\, \Tr[ T^\alpha \delta\cC_2^+ ]\, T^\beta |\delta\cC_1^+\rangle\,,\label{eq:trunc FS}
\end{align}
where the gauged supergravity field strength reads
\be |F\rangle = |\dd A\rangle -\frac12  \eta_{-1\alpha\beta} \langle \theta | T^\alpha |A\rangle \wedge T^\beta |A\rangle -  \eta_{-1\alpha\beta} \langle \theta | T^\alpha |C_{(1}\rangle \,T^\beta |C_{2)}\rangle\,, \label{eq:sugraFS}\ee
and correctly reproduces \eqref{eq:FSgs} for $\langle \vartheta|=0$. The induced truncation ansatz for the field strength \eqref{eq:trunc FS} differs by a trivial parameter\footnote{This corresponds to the $\Lambda$ component of the trivial parameter. The ansatz for $\cG^{(k)}$ differs accordingly from the one in \eqref{eq:truncFSproj} by the corresponding $\Sigma$ component of the trivial parameter.} from the one in \eqref{eq:truncFSproj}, which was used throughout Section~\ref{sec:ExFTtop}. Note, however, that in the topological term \eqref{eq:exft Ltop}, $|\cF\rangle$ only appears through an expression in which such trivial parameters are projected out. In other words, the ExFT topological term is independent of the  two-forms, as explained in detail in \cite{Bossard:2021jix}. The ansatz \eqref{eq:truncFSproj} was therefore sufficient to perform the reduction of the pseudo-Lagrangian. Here, the presence of the extra trivial parameter in \eqref{eq:trunc FS} is crucial to reduce consistently the unprojected version of the duality equation \eqref{eq:dualityeqExFT}, as it allows to absorb the unwanted $y$-dependent contributions coming from the scalar current $\langle P_\alpha^{(1)}|\otimes T^\alpha$.

Let us then turn to the reduction of the second term in \eqref{eq:dualityeqExFT}. Substituting the ansatz \eqref{eq:sPan} for (the conjugate of) $\bra{\cP^{(1)}_\alpha}\otimes T^\alpha$ and using \eqref{eq:W+theta} directly leads to 
\begin{align}
 \rho^{-2}\,\cV^{-1} T^{\alpha\dagger} \cV^{-\dagger} |\cP^{(1)}_\alpha\rangle= r^{-1}\cU^{-1}\Big(-\frac{1}{2\varrho^{3}}\,M^{-1}  |\theta\rangle+\frac{1}{\varrho^2}V^{-1}T^\alpha V^{-\dagger}|\underline W_\alpha^{(-1)}\rangle\Big)\,.\label{eq:dual2ndterm}
\end{align}
The last term can be rewritten using the relation 
\begin{align}
 -\frac{1}{2\varrho}\eta_{-2\,\alpha\beta}V^{-1} T^\alpha T^{\beta\dagger} V^{-\dagger}\ket{\theta}
-V^{-1}T^\alpha V^{-\dagger} |\underline W^{(-1)}_\alpha \rangle 
=\ &-\eta_{-1\,\alpha\beta}V^{-1}T^\alpha T^{\gamma\dagger}T^{\beta\dagger}V^{-\dagger}|\underline  W_\gamma\rangle\label{eq:idscal}\\ 
&+\sum\limits_{q=0}^\infty\eta_{-1-q\,\alpha\beta} V^{-1}T^\alpha T^{\beta\dagger}V^{-\dagger}|\underline  W_{-q}\rangle\nonumber\\[1mm]
&+\eta_{-2\,\alpha\beta}V^{-1}T^\alpha T^{\gamma\dagger}T^{\beta\dagger}V^{-\dagger}|\underline  W^{(1)}_\gamma\rangle\,,\nonumber
\end{align}
which can be proven by opening up $\bra{\theta}$ and moving the generators contracted with the $\eta$'s towards the Weitzenb\"ock connection. The same strategy was used in Section~\ref{sec:GSS pot} to perform the reduction of the scalar potential. 
To get to this result, we also used \eqref{eq:W0vanish}, \eqref{eq:W-vanish}, and \eqref{eq:W+theta}. Note that the terms on the right-hand side of \eqref{eq:idscal} are all $y$-dependent. In order for the duality equation \eqref{eq:dualityeqExFT} to consistently reduce to a gauged supergravity duality equation (\textit{i.e.} for the $y$-dependence to factorise out into an overall $r^{-1}\cU^{-1}$ factor), these terms must be absorbed in the Scherk--Schwarz ansatz for $|\delta\cC^+_{1} \rangle \otimes \delta\cC^+_{2}$ in \eqref{eq:2formred3}. This can be done and we explain it in the case of the first term in~\eqref{eq:idscal}. Passing $V$ through the bilinear form $\eta_{-1}$ we first get a series of terms of the form
\begin{align}
    \varrho\,\eta_{-1-k\, \alpha\beta}\, T^\alpha M^{-1}T^{\gamma\dagger} T^{\beta\dagger}|W_\gamma\rangle\,,\;\;\;\forall k\geq 0\,,\label{eq:extraterms2f} 
\end{align} 
which are multiplied by different binomial coefficients depending on the supergravity Virasoro scalars $\varphi_q$,  with $q>0$. Here we also used the definition \eqref{eq:VconjW} of the dressed Weitzenb\"ock connection $\bra{\underline W_\gamma}\otimes T^\gamma$. Since $\bra{W_\gamma}$ satisfies the `flat' section constraint, all the contributions in \eqref{eq:extraterms2f} can ultimately be absorbed in the ansatz for the two-form. For instance, the $k=0$ contribution is absorbed by choosing
\begin{align}
    \braket{e^M}{\delta\cC^+_1}\bra{e^N}\delta\cC^+_2\ket{e_P} = \big(\delta\cC^+\big)^{MN}{}_P=\,\star\,\frac{2}{\varrho\,r}\,\big(\cU^{-1}  T^{\gamma }  M^{-1}  \cU^{-1 \dagger} \big)^{NM} \,\,\big(\bra{W_\gamma}\,\cU\big)_P+\ldots\,,
\end{align}
where we switched back momentarily to an index notation to clarify the structure of the ansatz. The $k>0$ contributions in \eqref{eq:extraterms2f} can be absorbed by completing the above ansatz with similar terms  that also involve the operators $\widehat \cS_{-k}$ introduced in Appendix A.4 of \cite{Bossard:2021jix}. Finally, the second and last line of \eqref{eq:idscal} can also be absorbed in the above ansatz by following the same procedure as outlined above. Note that this hinges on the fact that $\bra{\underline W_{-q}}$ and $\bra{\underline W_\gamma^{(1)}}$ also satisfy the `flat' section constraint.

With the various Scherk--Schwarz ans\"atze presented in this section the ExFT duality equation \eqref{eq:dualityeqExFT} reduces to the following gauged supergravity duality equation 
\be   |{F}\rangle + \star \,\Big( \frac{1}{\varrho^{3}} M^{-1}  |\theta\rangle+\frac{1}{\varrho} \eta_{-2\,\alpha\beta}  T^\alpha M^{-1} T^{\beta\dagger}  \ket{\theta}\Big)= 0\,, \label{FieldStrengthDuality} 
\ee
where the non-abelian field strength $|F\rangle$ is given by \eqref{eq:sugraFS}. As in higher dimensions~\cite{deWit:2008ta,deWit:2008gc}, one expects that this equation should follow from the variation of the gauged supergravity pseudo-Lagrangian with respect to the embedding tensor, considered as a field of the theory, consistently with the interpretation that $|F\rangle$ is dual to the embedding tensor.\footnote{Note that in higher dimensions the $D-1$-form field Euler--Lagrange equation imposes the condition that the embedding tensor be constant. This does not follow so trivially in $D=2$ because the one-form is also the Yang--Mills connection. We nonetheless expect that constancy of $\langle \theta|$ will follow from the one-form Euler--Lagrange equation combined with the scalar field duality equation.} Note that the second term in \eqref{FieldStrengthDuality} is indeed the derivative of the potential \eqref{eq:VsugraFinal} with respect to the embedding tensor. We expect that the variation of the topological term \eqref{eq:sugra Ltop} should reproduce the non-abelian components of $|{F}\rangle $ up to the equations of motion, but we have not checked this. 

Let us finally comment on ambiguities in this equation that are not determined by the gSS reduction of exceptional field theory. Because of the section constraint on the Weitzenb\"ock connection and the condition $\langle \vartheta | =0$, any term in the potential of the form $\eta_{-k\,\alpha\beta}\,\bra{\theta}T^{\alpha}M^{-1} T^{\beta\dagger}\ket{\theta}$ for $k\ge 3$ vanishes. This is consistent with the field strength duality equation because one checks similarly that any term of the form 
$\star\, \eta_{-k\,\alpha\beta}\, T^{\alpha}M^{-1} T^{\beta\dagger}\ket{\theta}$ for $k\ge 3$ in the duality equation can be absorbed in a redefinition of $|\mathcal{\delta\cC}^+_{1} \rangle \otimes \mathcal{\delta\cC}^+_{2}$.

\section{Concluding comments}
\label{sec:conclusions}

In this paper we have constructed the complete bosonic dynamics of gauged maximal supergravities in two dimensions arising as consistent truncations from higher dimensions.

It is natural to ask whether the pseudo-Lagrangian~\eqref{eq:Lintro} may also capture correctly the dynamics of gauged maximal supergravities that do not admit a geometric uplift to higher dimensions, or what kind of modifications should be needed to do so. One observes that the shifted Maurer--Cartan equation \eqref{eq:shiftedMC gsugra} defining $\cL^{\rm top}_{\rm gsugra}$ can be written for any embedding tensor, suggesting that the topological term might be valid for all gaugings. On the contrary, one checks that the potential  $V_{\rm gsugra}$ must be modified by terms that vanish when the additional constraints \eqref{eq:extra cons} and \eqref{eq:extra cons 2} are satisfied.  We have verified that the Kaluza--Klein circle reduction of $D=3$ gauged supergravity  \cite{Nicolai:2000sc},  for a generic embedding tensor $\Theta_{AB} = \theta \eta_{AB} + \acute{\Theta}_{AB}$ in the ${\bf 1}\oplus{\bf 3875}$ satisfying the quadratic constraint, is correctly reproduced by $D=2$ gauged supergravity for $\langle \theta | = ( \frac{1}{62} \theta \eta_{AB} + \frac{1}{14} \acute{\Theta}_{AB} ) \langle 0 | T_1^A T_1^B$, provided we correct the potential \eqref{eq:VsugraIntro} to 
\bea
\label{eq:VsugraE8}
V_{\text{gsugra}}&=&\frac{1}{2\varrho^3}\,\bra{\theta}M^{-1}\ket{\theta}+\frac{1}{2\varrho}\eta_{-2\,\alpha\beta}\,\bra{\theta}T^{\alpha}M^{-1} T^{\beta\dagger}\ket{\theta}\\\nonumber&&- \frac{3}{28}\varrho\,\eta_{-2\,\alpha\beta}\eta_{-2\,\gamma\delta} \,\bra{\theta}T^{\alpha} T^\gamma M^{-1} T^{\delta \dagger} T^{\beta\dagger}\ket{\theta}  \, .
\eea
The additional term would vanish according to \eqref{eq:extra cons 2} if the theory admitted a geometric uplift to ten or eleven dimensions. The fact that only one additional term is sufficient to reproduce the correct $D=3$ gauged supergravity potential follows from the property that the constraints \eqref{eq:extra cons} and most of the constraints \eqref{eq:extra cons 2} still apply in this case. There is no reason to expect that the complete potential of maximal gauged supergravity in two dimensions involves finitely many terms in general. The bilinear $\langle \theta | \otimes \langle \theta|$ can be decomposed into the tensor product of two irreducible modules of E$_9$ and two irreducible modules of the coset Virasoro algebra generated by the $L_n^{\scalebox{0.6}{coset}} = - \frac{1}{32} \eta_{n\, \alpha\beta}  T^\alpha \otimes  T^\beta$ \cite{KacWaki}. A general E$_9$ invariant potential can therefore be written for an arbitrary formal series in the coset Virasoro generators $L_n^{\scalebox{0.6}{coset}} $ with $n\le 0$. For any fixed embedding tensor, $\langle \theta | \otimes \langle \theta|$ admits a maximal finite $L_0^{\scalebox{0.6}{coset}} $ degree so that only finitely many terms can contribute. But there is no bound on this  maximal finite $L_0^{\scalebox{0.6}{coset}} $ degree in gauged supergravity. One could in principle check more terms in the potential using the generalised Scherk--Schwarz reduction of generic gauged supergravity theories in arbitrary dimensions by generalising the formalism introduced in~\cite{Ciceri:2016dmd} to E$_9$. But this could only fix finitely many terms and the complete form of the potential can most likely only be obtained by a careful $K({\rm E}_9)$ covariant supersymmetry analysis.

The construction of $D=2$ gauged maximal supergravities in this paper relies on generalised Scherk--Schwarz reductions of E$_9$ ExFT.
As noted in \cite{Bossard:2021ebg}, such a formalism can be directly applied to theories based on duality groups $\widehat{G}$ that are affine extensions of other finite-dimensional Lie groups $G$. This allows us to straightforwardly determine the bosonic dynamics of many less- or non-supersymmetric $D=2$ gravity theories.
Natural instances would be gauged half-maximal supergravities, or the consistent $D=2$ Kaluza--Klein reductions of pure higher-dimensional general relativity.

Our paper opens up the possibility of exploring the vast landscape of $D=2$ gauged supergravity  theories admitting an uplift and the resulting ten- and eleven-dimensional geometries.
One important question is how to effectively classify and construct all such models, identifying the associated internal spaces and twist matrices.
This is a daunting task even in $D>2$.
Progress has been made in recent years, by identifying necessary and sufficient conditions for the existence of an uplift of a gauged supergravity and by devising a general procedure for constructing the twist matrix \cite{Inverso:2017lrz,Bugden:2021wxg,Bugden:2021nwl,Hulik:2022oyc,Hassler:2022egz}.
In order to carry out an explicit classification of all gauged maximal supergravities admitting a higher-dimensional origin, one would need to classify duality orbits of inequivalent embedding tensors satisfying the uplift conditions found in these works, as well as the standard gauged supergravity quadratic constraints. 
This is a very difficult and yet unsolved problem for all $D\le7$. 
The $D=9,8$ cases are treated in~\cite{Dibitetto:2012rk}. 
It would be highly desirable to extend  the analyses of~\cite{Inverso:2017lrz,Bugden:2021wxg,Bugden:2021nwl,Hulik:2022oyc} to~$D\le3$.
The simple observation \cite{Grana:2008yw} that the internal space must be a coset space $G/H$ with $G$ the gauge group of the $D$ dimensional theory applies also to the lower-dimensional cases, as the presence of ancillary transformations in the gauge structure of E$_8$ and E$_9$ ExFTs does not enter in the argument.
Recently, a duality invariant, necessary condition for a gauging of $D=3$ maximal supergravity to admit a gSS uplift was determined in \cite{Galli:2022idq,Eloy:2023zzh}. In equations~\eqref{eq:extra cons} and \eqref{eq:extra cons 2} we give an analogous set of constraint for $D=2$.
Furthermore, in appendix~E of the companion paper~\cite{SO9} we show that any Lagrangian embedding tensor $\bra\theta$ admitting an uplift to eleven-dimensional or type IIB supergravity is only parametrised by \emph{finitely} many components, which we identify explicitly.
A full analysis of necessary and sufficient uplift conditions along with a general construction of the twist matrix are left to future work.

\section*{Acknowledgements}
We would like to thank Martin Cederwall, Benedikt K\"onig, Emanuel Malek, Hermann Nicolai, Jakob Palmkvist and Henning Samtleben for discussions.
AK is grateful to \'Ecole Polytechnique for its warm hospitality during the early stages of this paper.
Part of this work was carried out at the workshop on Higher Structures, Gravity and Fields at the Mainz Institute for Theoretical Physics of the DFG Cluster of Excellence PRISMA+ (Project ID 39083149). We would like to thank the institute for its hospitality. 
This work has received funding from the European Research Council (ERC) under the European Union’s Horizon 2020 research and innovation programme (grant agreement No 740209).

\appendix
\section{Closure of the generalised Scherk--Schwarz ansatz}\label{app:closure}

This appendix collects some details on the generality and closure  of the gSS ansatz in E$_9$ ExFT.

\subsection{Useful expressions}

Combining the commutator of two $\hevirm$ elements $X,\,Y$ with shift operators, one finds a convenient expression
\begin{align}\label{eq:commutator shifts}
[X\,,\,Y]-[\cS_{-1}(X)\,,\,\cS_{+1}(Y)] =&\,\Big( \eta^{\alpha\beta}X_\alpha Y_\beta
+ X_\dK Y_0 + X_0 Y_\dK +\frac{c_\mathfrak{vir}}{4}\sum\limits_{k\in\mathbb Z} X_q Y_{-q} \,q(q-1)\Big) \,\dK\nonumber\\
&\,+\sum_{k\in\mathbb Z}\Big(X_{-k} \cS_{-k}(Y)+Y_{-k} \cS_{-k}(X)\Big)\,,
\end{align}
where $\eta^{\alpha\beta}$ is the inverse of $\eta_{\alpha\beta}$ within $\hat{\mathfrak e}_8\oleft\langle L_0\rangle$ and vanishes along the rest of $\vir$.
This relation is easily checked component by component. A similar formula is 
\bea &&  \eta_{n+1\, \alpha\beta} [ T^\alpha , T^\gamma ] \otimes T^\beta-\eta_{n\, \alpha\beta} [ T^\alpha , \mathcal{S}_1(T^\gamma)] \otimes T^\beta \label{eq: id A2} \\
&=& \dK \otimes \mathcal{S}_{n+1}(T^\gamma) -\mathcal{S}_{n+1}(T^\gamma)  \otimes \dK + \sum_m \delta^\gamma_{L_m} \eta_{n+m+1\, \alpha\beta} T^\alpha\otimes T^\beta - \frac{c_\mathfrak{vir}}{4} \delta^\gamma_{L_{-1-n}} n(n+1) \dK \otimes \dK \; . \nonumber \eea
It allows us to manipulate \eqref{eq:Theta1} using
\begin{align}\label{eq:commutator shifts for W}
\eta_{\alpha\beta}\bra{W_\gamma}\big[ T^\beta, T^\gamma \big] =\ &
\eta_{-1\,\alpha\beta}\bra{W_\gamma}\big[ T^\beta, \cS_{+1}(T^\gamma) \big] 
+ \bra{W_\alpha} - \delta_\alpha^\dK \bra{W_\gamma}T^\gamma
+\sum_{k=0}^\infty \eta_{-k\,\alpha\beta}\bra{W_{-k}}T^\beta\,.
\end{align}

\subsection{\texorpdfstring{On the role of $\bra{\tilde{h}_\alpha}$}{On the role of halpha}}
\label{app:ha}

In this appendix, we analyse an important technical point related to the generalised Scherk--Schwarz presented in Section~\ref{sec:gSScond} and in particular the role of the undetermined piece $\bra{\tilde{h}_\alpha}$ appearing in the expansion of the general embedding tensor~\eqref{eq:Theta2} that we reproduce here for convenience:
\begin{align}
\label{eq:Theta2app}
\bra{\Theta_\alpha} -\delta_\alpha^\dK\bra\vartheta = -\eta_{\alpha\beta}\bra{\vartheta}T^\beta
-\eta_{-1\,\alpha\beta}\bra{\theta}T^\beta 
+\bra{\tilde h_\alpha}\,,
\end{align}
Unlike the first two terms, $\bra{\tilde{h}_\alpha}$ is not in agreement with the general expansion \eqref{eq:Theta virm exp}, but we will argue that it can be set to zero without loss of generality. 

At face value, $\bra{\tilde h_\alpha}$ does not necessarily admit the same kind of expansion as~\eqref{eq:Theta virm exp} and could also contain other irreducible representations in the tensor product $\overline{R(\Lambda_0)_{-1}}\otimes(\hevirm)^*$.
The transformation of $\bra{w^+}$ under the gauged supergravity rigid $\hevirm$ is determined so that $\bra\theta$ transforms as a covector, compatibly with its expected transformation property from \eqref{eq:embtens rigid trf}. 
Covariance under $\hevirm$ also implies that $\bra{\tilde h_\alpha}$ transforms indecomposably with respect to $\bra\vartheta$ and $\bra\theta$.\footnote{There is also an ambiguity in the definition of $\bra{w^+}$, because it can be entirely reabsorbed into $\bra{\tilde h_\alpha}$. This ambiguity is lifted by identifying $\bra{w^+}$ as the object appearing in the reduction ansatz for the constrained scalar field $\bra{\tilde\chi_1}$.}

We may then wonder whether $\bra{\Theta_\alpha}$ or equivalently $\bra{\tilde h_\alpha}$ are further constrained by the ancillary components of \eqref{eq:gss cond e9}.
We now show that this component reduces to the quadratic constraint \eqref{eq:Theta QC} for constant $\bra{\Theta_\alpha}$.
We define the difference of the two sides of \eqref{eq:gss cond e9}:
\begin{equation}
\prepi\,\bbGamma_{\!\!12}\ =\ \prepi\bbLambda_1 \circ \prepi\bbLambda_2 + \Braket{\Theta_\alpha{-}\delta_\alpha^\dK\vartheta}{\lambda_1} \big(\,r^{-1}\cU^{-1} T^\alpha \ket{\lambda_2}\,,\, \bra{H_\beta}T^\alpha\ket{\lambda_2}  \,\big)\,.
\end{equation}
Taking $\bra{\Theta_\alpha}$ to be \emph{defined} by the generalised vector component of \eqref{eq:gss cond e9}, even if non-constant, we can compute the ancillary component of $\prepi\,\bbGamma_{\!\!12}$ and see how it differs from zero.
Equivalently, we can compute its projection $[\bbGamma_{\!\!12}]_\alpha$.
In order to compute $[ \bbLambda_1\circ\bbLambda_2 ]_\alpha$, we consider the generalised vector component of the Leibniz identity \eqref{eq:dorf leib proj} which we can rewrite as
\begin{align}
T^\alpha\ket{\Lambda_3}\big[ \bbLambda_1\circ\bbLambda_2 \big]_\alpha
=
\big[\cL_{\bbLambda_2}\,,\,\cL_{\bbLambda_1}\big]\ket{\Lambda_3}
+\ket{\Lambda_3}\Braket{\overset{\scriptscriptstyle\leftarrow}{\partial}{-}\overset{\scriptscriptstyle\rightarrow}{\partial}}{\cL_{\bbLambda_1}{\Lambda_2}}\,,
\end{align}
then, assuming all gauge parameters factorise according to \eqref{eq:gssgenvec} and using the gSS condition \eqref{eq:gss cond e9} we deduce the identity 
\begin{align}
\big[ \bbLambda_1\circ\bbLambda_2 \big]_\alpha\,\cU T^\alpha \cU^{-1} =\ & 
-\big[T^\alpha\,,\,T^\beta\big]\braket{\Theta_\alpha}{\lambda_1}\braket{\Theta_\beta}{\lambda_2}
\\\nonumber&
+2 \braket{\partial_\Theta}{\Lambda_{[1}}\Braket{\Theta_\alpha{-}\delta_\alpha^\dK\vartheta}{\lambda_{2]}}\,T^\alpha
+\bra{\partial_\Theta}r^{-1}\cU^{-1}T^\alpha\ket{\lambda_2}\Braket{\Theta_\alpha{-}\delta_\alpha^\dK\vartheta}{\lambda_1}
\\\nonumber&
+\Braket{\Theta_\beta{-}\delta_\beta^\dK\vartheta}{\lambda_1}\bra{W_\alpha}T^\beta\ket{\lambda_2} T^\alpha
-\Braket{\Theta_\beta{-}\delta_\beta^\dK\vartheta}{\lambda_1}\bra{\vartheta} T^\beta\ket{\lambda_2} \dK\,.
\end{align}
To compute the projection of the last term of $\prepi\,\bbGamma_{\!\!12}$, we can use the identity \eqref{eq:bbLambda red} with the substitution
\begin{equation}
\ket\lambda \to T^\beta\ket{\lambda_2}\Braket{\Theta_\beta{-}\delta^\dK_\beta\vartheta}{\lambda_1}\,,
\end{equation}
recalling however that \eqref{eq:bbLambda red} holds when $\ket\lambda$ is $y$-independent, hence an extra term proportional to the derivative of $\bra{\Theta_\beta{-}\delta^\dK_\beta\vartheta}$ must be computed separately.
One then arrives at the expression
\begin{align}\label{eq:bbGamma alpha}
[\bbGamma_{\!\!12}]_\alpha\,\cU T^\alpha \cU^{-1} =\ &
-\big[T^\alpha\,,\,T^\beta\big]\braket{\Theta_\alpha}{\lambda_1}\braket{\Theta_\beta}{\lambda_2}
\\\nonumber&
+\Braket{\Theta_\beta{-}\delta_\beta^\dK\vartheta}{\lambda_1}\bra{\Theta_\alpha}T^\beta\ket{\lambda_2} T^\alpha
-\Braket{\Theta_\beta{-}\delta_\beta^\dK\vartheta}{\lambda_1}\bra{\vartheta} T^\beta\ket{\lambda_2} \dK
\\\nonumber&
+2 \braket{\partial_\Theta}{\Lambda_{[1}}\Braket{\Theta_\alpha{-}\delta_\alpha^\dK\vartheta}{\lambda_{2]}}\,T^\alpha
+\bra{\partial_\Theta}r^{-1}\cU^{-1}T^\alpha\ket{\lambda_2}\Braket{\Theta_\alpha{-}\delta_\alpha^\dK\vartheta}{\lambda_1}
\\\nonumber&
+\eta_{\alpha\gamma}\bra{\partial_{\Theta,\vartheta}}T^\gamma r^{-1}\cU^{-1}T^\beta\ket{\lambda_2}
\Braket{\Theta_\beta{-}\delta^\dK_\beta\vartheta}{\lambda_1}\cU T^\alpha\cU^{-1}
\end{align}
where the extra term is in the last line.
Notice that there are no terms left depending on $\bra{W_\alpha}$ and that all derivatives act on the embedding tensor.
We see that if $\bra{\Theta_\alpha}$ is constant, the last two lines vanish and the gSS condition on ancillary parameters becomes the quadratic constraint \eqref{eq:Theta QC} for the embedding tensor, the last term in the second line being proportional to $\bra{\Theta_\alpha}\otimes\,\bbdelta^\alpha\!\bra{\vartheta}$.
We conclude that there is no obvious way in which the ancillary component of the condition \eqref{eq:gss cond e9} can restrict the representation content of $\bra{\tilde h_\alpha}$, as one might in principle be able to construct embedding tensors violating the representation constraint but satisfying the quadratic one.

An analysis of supersymmetry would likely clarify the role of $\bra{\tilde h_\alpha}$. Since we do not have access to it, we follow a different path to argue that there is no loss of generality in heavily constraining the representation content of $\bra{\tilde h_\alpha}$ and that one may in fact set it to vanish.
First, we notice that the requirement that $\bra{\tilde h_\alpha}$ must admit an expansion as in \eqref{eq:halpha expansion} implies that there is always a choice of parametrisation of the scalar manifold in which $\bra{\tilde h_\alpha}$ gauges only the shift symmetries of dual potentials but has no effect on the physical scalar fields.
In other words, there will always be a choice of physical Lagrangian in which $\bra{\tilde h_\alpha}$ does not affect the kinetic and topological terms.
We will also see in the next sections that the reduction ansatz for ancillary parameters does not enter directly in the reduction of the scalar potential.
Only the constrained object $\bra{w^+}$ introduced in \eqref{eq:shiftweitz} will enter, through the reduction ansatz for the constrained scalar $\bra{\tilde\chi_1}$.
Therefore, $\bra{\tilde h_\alpha}$ has no effect on the physical Lagrangian and can be safely set to vanish.
{One may wonder about the effect of $\bra{\tilde h_\alpha}$ on duality equations that do not descend from a Lagrangian, such as \eqref{FieldStrengthDuality}. Precisely because such equations do not affect the physical properties of the theory, any effect $\bra{\tilde h_\alpha}$ will be removable by a combination of field redefinitions and gauge transformations.}

Setting $\bra{\tilde h_\alpha} = 0$ breaks the formal $\virm$ covariance of the embedding tensor, because we truncate the expansion \eqref{eq:Theta virm exp} to the gaugings of $L_0$ and $L_{-1}$, but it is sufficient to capture all gauged maximal supergravities admitting a (geometric) uplift.
{Covariance under gauged supergravity $\virm$ transformations can be preserved by restricting $\bra{\tilde h_\alpha}$ to satisfy the irrep expansion \eqref{eq:Theta virm exp} rather than setting it to vanish entirely: $\bra{\tilde h_\alpha} = -\sum_{k\ge1}\eta_{-k\,\alpha\beta}\bra{\tilde h_{-k}}T^\beta$. The embedding tensor then reads $\bra{\Theta_{0}}=\bra{\vartheta}$, $\bra{\Theta_{-1}}=\bra{\theta}+\bra{\tilde h_{-1}}$ and $\bra{\Theta_{-k}}=\bra{\tilde h_{-k}}$ for $k\ge2$. Notice that contrary to \eqref{eq:extended theta to unextended}, we are not identifying $\bra{\Theta_{-1}}$ with $\bra{\theta}$. One then checks that the transformation properties of $\bra{\tilde h_{-k}}$ are compatible with \eqref{eq:embtens rigid trf} as well as with the expansion \eqref{eq:halpha expansion}. However, it must be stressed that the resulting gauged supergravity is nonetheless physically equivalent to the one obtained by setting $\bra{\tilde h_{-k}}=0$ for all $k\ge1$. In particular, the constraint \eqref{eq:halpha expansion} combined with constancy of the embedding tensor imply that $\bra{\tilde h_{-k}}$ must be constant for $k\ge2$ and any non-constant component of $\bra{\tilde h_{-1}}$ must be reabsorbed into $\bra{w^+}$. Having done so, $\bra\vartheta$ and $\bra\theta$ are constant and by our computation in Appendix~\ref{ClosureDorf} satisfy the quadratic constraint by themselves, so that the physical gauge couplings are the same as the ones that can be obtained setting $\bra{\tilde h_\alpha} = 0$.}

In the main body of the paper we therefore work with  $\bra{\tilde h_\alpha}=0$.

\subsection{Closure}
\label{ClosureDorf} 

In this section we prove \eqref{eq:bbGamma minimal}. The $\Lambda$ component of this equation is obvious, so we will only explain its $\Sigma$ component.  Using \eqref{eq:Dorf1}, we need to prove that any pair of gauge parameters satisfying the gSS ansatz \eqref{eq:gssgenvec minimal} verifies  
\bea
&&
  \cL_{\bbLambda_1}\Sigma_{2} 
+  \eta_{1\,\alpha\beta}\bra{\partial_{\Lambda_1}}T^\beta\ket{\Lambda_1}\,T^\alpha\ket{\Lambda_2}\bra{\partial_{\Lambda_1}}
+ \eta_{\alpha\beta}\Tr(T^\alpha\Sigma_1 ) \, T^\beta\ket{\Lambda_2}\bra{\partial_{\Sigma_1}} 
- \ket{\Lambda_2}\bra{\partial_{\Sigma_1}}\Sigma_1 \CR
&=& \bigl(  \eta_{-1\, \alpha\beta} \langle \theta | T^\alpha |\lambda_1\rangle +\eta_{\alpha\beta} \langle \vartheta | T^\alpha |\lambda_1\rangle  \bigr)   \cU^{-1}T^\gamma T^\beta \ket{\lambda_2}\bra{W^+_\gamma }r\cU  \CR
&&
-\big(\eta_{\alpha\beta}\bra\theta T^\alpha \ket{\lambda_1} + \eta_{1\,\alpha\beta }\bra\vartheta T^\alpha  \ket{\lambda_1}\big)\  \cU^{-1} T^\beta  \ket{\lambda_2} \bra{\partial_{\theta,\vartheta}} \,. \label{Sigma12toprove}  \eea
The first step is to compute $  \cL_{\bbLambda_1}\Sigma_{2} $ as 
\bea 
&&\hspace{-2mm}  \cL_{\bbLambda_1} \Big(\,\cU^{-1}T^\alpha\ket{\lambda_2}\bra{W^+_\alpha}r\cU\,\Big) 
\CR
\hspace{-2mm} &=&\hspace{-2mm}  \bigl(  \eta_{-1\, \alpha\beta} \langle \theta | T^\alpha |\lambda_1\rangle +\eta_{\alpha\beta} \langle \vartheta | T^\alpha |\lambda_1\rangle  \bigr)  
 \Big(\,\cU^{-1}T^\beta T^\gamma \ket{\lambda_2}\bra{W^+_\gamma }r\cU
-\cU^{-1}T^\gamma \ket{\lambda_2}\bra{W^+_\gamma }T^\beta r\cU \,\Big) 
\CR
&&\hspace{-2mm}
-\braket{\theta}{\lambda_1}\,\Big(\, \cU^{-1}T^\alpha\ket{\lambda_2}\bra{W_\alpha}r\cU  
  -\cU^{-1}\ket{\lambda_2}\bra{W_0}r\cU \,\Big)
\CR
&&\hspace{-2mm}
-\braket{\vartheta}{\lambda_1}\ \cU^{-1}T^\alpha\ket{\lambda_2}\bra{W^+_\alpha}r\cU
+\langle \partial_{W^+} | \cU^{-1} | \lambda_1\rangle \ \cU^{-1}T^\alpha\ket{\lambda_2}\bra{W^+_\alpha} \cU \; . 
\eea
This formula suggests to consider the difference 
\bea \Xi &\equiv&  \cL_{\bbLambda_1}\Sigma_{2} 
+  \eta_{1\,\alpha\beta}\bra{\partial_{\Lambda_1}}T^\beta\ket{\Lambda_1}\,T^\alpha\ket{\Lambda_2}\bra{\partial_{\Lambda_1}}
+ \eta_{\alpha\beta}\Tr(T^\alpha\Sigma_1 ) \, T^\beta\ket{\Lambda_2}\bra{\partial_{\Sigma_1}} 
- \ket{\Lambda_2}\bra{\partial_{\Sigma_1}}\Sigma_1 \CR
&& -\bigl(  \eta_{-1\, \alpha\beta} \langle \theta | T^\alpha |\lambda_1\rangle +\eta_{\alpha\beta} \langle \vartheta | T^\alpha |\lambda_1\rangle  \bigr)   \cU^{-1}T^\gamma T^\beta \ket{\lambda_2}\bra{W^+_\gamma }r\cU  \CR
 &=& \bigl(  \eta_{-1\, \alpha\beta} \langle \theta | T^\alpha |\lambda_1\rangle +\eta_{\alpha\beta} \langle \vartheta | T^\alpha |\lambda_1\rangle  \bigr)  
 \Big(\,\cU^{-1} [ T^\beta ,  T^\gamma ] \ket{\lambda_2}\bra{W^+_\gamma }r\cU-\cU^{-1}T^\gamma \ket{\lambda_2}\bra{W^+_\gamma }T^\beta r\cU \,\Big)  \CR
&& -\braket{\theta}{\lambda_1}\,\Big(\, \cU^{-1}T^\alpha\ket{\lambda_2}\bra{W_\alpha}r\cU  
  -\cU^{-1}\ket{\lambda_2}\bra{W_0}r\cU \,\Big)
\CR
&& -\braket{\vartheta}{\lambda_1}\ \cU^{-1}T^\alpha\ket{\lambda_2}\bra{W^+_\alpha}r\cU
+\langle \partial_{W^+} | \cU^{-1} | \lambda_1\rangle \ \cU^{-1}T^\alpha\ket{\lambda_2}\bra{W^+_\alpha} \cU \CR
&& - \eta_{1\, \alpha\beta} \langle W_\gamma | T^\alpha  T^\gamma  |\lambda_1 \rangle \mathcal{U}^{-1} T^\beta  |\lambda_2 \rangle \langle {\partial}_W | -\eta_{1\, \alpha\beta} \langle W_\gamma | [ T^\delta , T^\alpha ] T^\gamma  |\lambda_1 \rangle \mathcal{U}^{-1} T^\beta |\lambda_2 \rangle \langle W_\delta  |r \mathcal{U} \CR
&& + \eta_{1\, \alpha\beta} \langle W_0 | T^\alpha   |\lambda_1 \rangle  \mathcal{U}^{-1} T^\beta |\lambda_2 \rangle \langle {\partial}_W |   + \eta_{1\, \alpha\beta} \langle W_0 | [ T^\gamma , T^\alpha ]  |\lambda_1 \rangle \mathcal{U}^{-1} T^\beta |\lambda_2 \rangle \langle W_\gamma  | r \mathcal{U} \CR
&& +  \eta_{\alpha\beta}   \langle W^+_\gamma | T^\alpha  T^\gamma |\lambda_1 \rangle \mathcal{U}^{-1} T^\beta |\lambda_2 \rangle\bigl(  \langle \partial_{W^+} |- \langle W_0| r \mathcal{U} \bigr)   \CR
&& + \eta_{\alpha\beta}  \langle W^+_\gamma | [ T^\delta , T^\alpha ] T^\gamma |\lambda_1 \rangle \mathcal{U}^{-1} T^\beta |\lambda_2 \rangle \langle W_\delta | r \mathcal{U} \CR
&& -\bigl(  \langle \partial_{W^+} |\mathcal{U}^{-1} {-} r \langle W_0| \bigr)  T^\alpha |\lambda_1 \rangle \mathcal{U}^{-1} |\lambda_2 \rangle \langle W^+_\alpha | \mathcal{U} \CR
&& + \langle W_\alpha | T^\alpha T^\beta |\lambda_1 \rangle \mathcal{U}^{-1} |\lambda_2 \rangle \langle W^+_\beta | r \mathcal{U}  -  \langle W_\alpha |  T^\beta |\lambda_1 \rangle \mathcal{U}^{-1} |\lambda_2 \rangle \langle W^+_\beta | T^\alpha r \mathcal{U}  \; . \label{XiisZero}
\eea
We then collect all the terms involving the derivative of the Weitzenb\"{o}ck connection in
\bea
&& 
\langle \partial_{W^+} | \cU^{-1} | \lambda_1\rangle \ \cU^{-1}T^\alpha\ket{\lambda_2}\bra{W^+_\alpha} \cU - \langle \partial_{W^+} |\mathcal{U}^{-1}   T^\alpha |\lambda_1 \rangle \; \mathcal{U}^{-1} |\lambda_2 \rangle \langle W^+_\alpha | \mathcal{U}\CR
&& +  \eta_{\alpha\beta}   \langle W^+_\gamma | T^\alpha  T^\gamma |\lambda_1 \rangle \; \mathcal{U}^{-1} T^\beta |\lambda_2 \rangle  \langle \partial_{W^+} |    \CR
&& - \eta_{1\, \alpha\beta} \langle W_\gamma | T^\alpha  T^\gamma  |\lambda_1 \rangle \; \mathcal{U}^{-1} T^\beta  |\lambda_2 \rangle \langle {\partial}_W |  + \eta_{1\, \alpha\beta} \langle W_0 | T^\alpha   |\lambda_1 \rangle  \; \mathcal{U}^{-1} T^\beta |\lambda_2 \rangle \langle {\partial}_W |   \CR
&=& \hspace{-2mm} -\big(\eta_{\alpha\beta}\bra\theta T^\alpha \ket{\lambda_1} + \eta_{1\,\alpha\beta }\bra\vartheta T^\alpha  \ket{\lambda_1}\big)\  \cU^{-1} T^\beta  \ket{\lambda_2} \bra{\partial_{\theta,\vartheta}} \CR
&&- \langle W_{\alpha}^{+} | \lambda_1\rangle  \; T^\alpha | \lambda_2\rangle  \langle {\partial_{W^+}} |  +\langle {\partial_{W^+}}  | \mathcal{U}^{-1} |\lambda_1\rangle  \; T^\alpha | \lambda_2\rangle  \langle  W_{\alpha}^{+} | \mathcal{U} \CR
&&  +\langle W_{\alpha}^{+} |T^\alpha  |\lambda_1\rangle   \; |\lambda_2\rangle  \langle {\partial_{W^+}} |  -\langle {\partial_{W^+}}  | \mathcal{U}^{-1}  T^\alpha|\lambda_1\rangle \;  | \lambda_2\rangle \langle  W_{\alpha}^{+} |  \mathcal{U} \; . 
 \eea
The first line is precisely the term we want to show to be equal to $\Xi$ in \eqref{XiisZero} above for \eqref{Sigma12toprove} to be satisfied. The four remaining terms combine such that the  component $\langle W_{\dK}^+|= \langle w^+|$ drops out and they only involve the curl of the Weitzenb\"{o}ck connection which can be simplified using the Bianchi identity
\bea && \langle {\partial}_W | r^{-1} \mathcal{U}^{-1}  \otimes \langle W_\alpha | -  \langle W_\alpha | \otimes \langle {\partial}_W | r^{-1} \mathcal{U}^{-1} \\
 &=& \langle W_0  | \otimes \langle W_\alpha | -  \langle W_\alpha | \otimes \langle W_0 |- \langle W_\beta | \otimes \langle W_\alpha | T^\beta +  \langle W_\alpha | T^\beta  \otimes \langle W_\beta | + f^{\beta\gamma}{}_\alpha \langle W_\beta | \otimes \langle W_\gamma | \nonumber \label{BianchiU} \; . \eea
More precisely, we need to use the shift operator $\mathcal{S}_1$ on the Bianchi identity to obtain
\bea   \hspace{-2mm}  && \hspace{-2mm}  - \langle W_{\alpha}^{+} | \otimes T^\alpha \otimes \langle {\partial}_W | r^{-1} \mathcal{U}^{-1}   +\langle {\partial}_W | r^{-1} \mathcal{U}^{-1}   \otimes T^\alpha \otimes \langle  W_{\alpha}^{+} | \CR
 \hspace{-2mm}  && \hspace{-2mm}   +\langle W_{\alpha}^{+} |T^\alpha  \otimes  \dK  \otimes \langle {\partial}_W | r^{-1} \mathcal{U}^{-1}   -\langle {\partial}_W | r^{-1} \mathcal{U}^{-1}  T^\alpha \otimes \dK \otimes \langle  W_{\alpha}^{+} | \CR
 \hspace{-2mm}  &=& \hspace{-2mm} -\langle W_\alpha | \otimes T^\beta \otimes \langle W_\beta^{+} | T^\alpha+\langle W^{+}_\alpha | T^\beta  \otimes T^\alpha \otimes \langle W_\beta |  + \langle W_\alpha | \otimes [ T^\alpha , T^\beta ] \otimes \langle W^+_\beta | \CR
 \hspace{-2mm}  && \hspace{-2mm}  +2 \langle W_0  | \otimes T^\alpha \otimes \langle W_\alpha^{+} | -2 \langle W_0 | T^\alpha \otimes \dK \otimes \langle W_\alpha^{+} | +   \langle W_\alpha^{+} | T^\alpha \otimes \dK \otimes  \langle W_0| -  \langle W_\alpha^{+} | \otimes T^\alpha \otimes \langle W_0 | \CR
 \hspace{-2mm}  && \hspace{-2mm}  + \langle W_\alpha | T^\beta  \otimes \dK \otimes \langle W_\beta^{+} | T^\alpha -\langle W_\alpha^{+} |T^\beta T^\alpha \otimes \dK \otimes \langle W_\beta | -\langle W_\alpha |[ T^\alpha ,T^\beta]  \otimes \dK \otimes \langle W^+_\beta |    \; . \eea
 Note that the fact that the $\langle W_{\dK}^+|$ component drops out in this equation also ensures that the cocycle terms that  appear in passing the shift operator through the commutator also drop out. 
 
 Combing these results one obtains 
 \bea   \hspace{-2mm}& &\hspace{-2mm}  \cU \Bigl(  \Xi  +\big(\eta_{\alpha\beta}\bra\theta T^\alpha \ket{\lambda_1} + \eta_{1\,\alpha\beta }\bra\vartheta T^\alpha  \ket{\lambda_1}\big)\   \mathcal{U}^{-1}  T^\beta  \ket{\lambda_2} \bra{\partial_{\theta,\vartheta}}  \Bigr)  r^{-1} \cU^{-1}   \CR
\hspace{-2mm}& =&\hspace{-2mm}\bigl(  \eta_{-1\, \alpha\beta} \langle \theta | T^\alpha |\lambda_1\rangle +\eta_{\alpha\beta} \langle \vartheta | T^\alpha |\lambda_1\rangle  \bigr)  
 \big(\, [ T^\beta ,  T^\gamma ] \ket{\lambda_2}\bra{W^+_\gamma } -T^\gamma \ket{\lambda_2}\bra{W^+_\gamma }T^\beta  \,\big)  \CR
 &&  -\eta_{1\, \alpha\beta} \langle W_\gamma | [ T^\delta , T^\alpha ] T^\gamma  |\lambda_1 \rangle \; T^\beta |\lambda_2 \rangle \langle W_\delta  |  + \eta_{1\, \alpha\beta} \langle W_0 | [ T^\gamma , T^\alpha ]  |\lambda_1 \rangle \;  T^\beta |\lambda_2 \rangle \langle W_\gamma  | \CR
&& -  \eta_{\alpha\beta}   \langle W^+_\gamma | T^\alpha  T^\gamma |\lambda_1 \rangle \;  T^\beta |\lambda_2 \rangle  \langle W_0|   + \eta_{\alpha\beta}  \langle W^+_\gamma | [ T^\delta , T^\alpha ] T^\gamma |\lambda_1 \rangle \;  T^\beta |\lambda_2 \rangle \langle W_\delta | \CR
&& -  \langle \theta   |\lambda_1\rangle \;   T^\beta  |\lambda_2\rangle \langle W_\beta | - \langle \vartheta  |\lambda_1\rangle  \;   T^\beta |\lambda_2\rangle  \langle W_\beta^+ | \CR
&&  - \langle W_0 | T^\alpha  |\lambda_1\rangle \;   |\lambda_2\rangle \langle W_\alpha^+ |  + 2\langle W_0 |   \lambda_1\rangle  \;   T^\alpha  |\lambda_2\rangle  \langle W_\alpha^+ | -\langle W_\alpha^+ |   \lambda_1\rangle  \;  T^\alpha  |\lambda_2\rangle \langle W_0 | \CR
&&+\langle W_\alpha  | T^\beta  T^\alpha  |\lambda_1\rangle  \;    |\lambda_2\rangle  \langle W_\beta^+ |  + \langle W_\alpha  |  \lambda_1\rangle  \;   [ T^\alpha , T^\beta ]   |\lambda_2\rangle  \langle W_\beta^+ |   - \langle W_\alpha |  \lambda_1\rangle \;   T^\beta  |\lambda_2\rangle \langle W_\beta^+ | T^\alpha   \CR
&& + \langle W_\alpha^+ | T^\beta  |\lambda_1\rangle  \;  T^\alpha  |\lambda_2\rangle \langle W_\beta |   - \langle W_\alpha^+ | T^\beta T^\alpha   |\lambda_1\rangle  \;    |\lambda_2\rangle  \langle W_\beta | \CR
\hspace{-2mm}& =&\hspace{-2mm}  0  \; . 
\eea
To prove this last identity one expands 
\bea \hspace{-2mm}  && \hspace{-2mm}  \bigl( \eta_{-1\, \alpha\beta} \langle \theta | T^\alpha   + \eta_{\alpha\beta} \langle \vartheta | T^\alpha \bigr) \otimes \Bigl( [ T^\beta , T^\gamma ] \otimes \langle W_\gamma^+ | - T^\gamma \otimes \langle W^+_\gamma | T^\beta \Bigr) \CR
\hspace{-2mm}  &= & \hspace{-2mm} \eta_{\alpha\beta} \langle W_\gamma^+| T^\gamma [ T^\alpha , T^\delta ] \otimes T^\beta \otimes \langle W_\delta | -  \eta_{\alpha\beta} \langle W_\gamma | T^\gamma [ T^\alpha , T^\delta ] \otimes T^\beta \otimes \langle W_\delta^+ |\CR
&& +\eta_{\alpha\beta} \langle W_\gamma^+ | T^\gamma T^\alpha \otimes T^\beta\otimes \langle W_0| - \eta_{1\, \alpha\beta} \langle W_\gamma | T^\gamma T^\alpha \otimes T^\beta \otimes \langle W_0| \CR
&& - \langle W_\alpha^+| T^\alpha \otimes T^\beta \otimes \langle W_\beta | + \langle W_\alpha^+| T^\alpha T^\beta  \otimes \dK  \otimes \langle W_\beta |- \langle W_\alpha^+| T^\beta    \otimes T^\alpha   \otimes \langle W_\beta |\CR
&& + \langle W_\alpha | \otimes T^\beta \otimes \langle W_\beta^+| T^\alpha + \langle W_\alpha^+| \otimes T^\alpha \otimes \langle W_0| -  \langle W_0| \otimes T^\alpha \otimes \langle W_\alpha^+ | \; , 
\eea
as well as  
\bea \hspace{-2mm}  && \hspace{-2mm}   -\eta_{1\, \alpha\beta} \langle W_\gamma | [ T^\delta , T^\alpha ] T^\gamma  \otimes T^\beta \otimes   \langle W_\delta  |  + \eta_{1\, \alpha\beta} \langle W_0 | [ T^\gamma , T^\alpha ]  \otimes T^\beta \otimes  \langle W_\gamma  | \CR
&& -  \eta_{\alpha\beta}   \langle W^+_\gamma | T^\alpha  T^\gamma \otimes T^\beta \otimes    \langle W_0|   + \eta_{\alpha\beta}  \langle W^+_\gamma | [ T^\delta , T^\alpha ] T^\gamma \otimes T^\beta \otimes   \langle W_\delta | \CR
\hspace{-2mm}  &=& \hspace{-2mm}    \eta_{\alpha\beta} \langle W^+_\gamma | [ T^\delta , T^\alpha ] T^\gamma \otimes T^\beta \otimes \langle W_\delta|  - \eta_{\alpha\beta} \langle W_\gamma | [ T^\delta , T^\alpha ] T^\gamma \otimes T^\beta \otimes \langle W_\delta^+| \CR
&& + \eta_{1\, \alpha\beta} \langle W_\gamma | T^\gamma T^\alpha  \otimes T^\beta \otimes \langle W_0| - \eta_{\alpha\beta} \langle W_\gamma^+| T^\gamma T^\alpha \otimes T^\beta \otimes \langle W_0| \CR
&& + \eta_{\alpha\beta} \langle W_0 | [ T^\gamma , T^\alpha ] \otimes T^\beta \otimes \langle W_\gamma^+| - \langle W_\alpha^+ | T^\alpha \otimes \dK \otimes \langle W_0| + \langle W_\alpha^+ |  \otimes T^\alpha \otimes \langle W_0 | \CR
&& - \langle W_\alpha | T^\beta T^\alpha \otimes \dK \otimes \langle W_\beta^+| + \langle W_0| T^\alpha \otimes \dK \otimes \langle W_\alpha^+| - \langle W_0|  \otimes T^\alpha \otimes \langle W_\alpha^+| \; ,  \eea
and then use the identity 
\bea \hspace{-2mm}  && \hspace{-2mm}  \eta_{\alpha\beta} \langle W_\gamma^+| \bigl[ T^\gamma , [ T^\alpha , T^\delta ] \bigr] \otimes T^\beta \otimes \langle W_\delta | -  \eta_{\alpha\beta} \langle W_\gamma | \bigl[ T^\gamma ,  [ T^\alpha , T^\delta ] \bigr] \otimes T^\beta \otimes \langle W_\delta^+ |\CR
 \hspace{-2mm}  &=& \hspace{-2mm} - \langle W_\alpha | \otimes [ T^\alpha , T^\beta ] \otimes \langle W_\beta^+| - \langle W^+_\alpha |  [ T^\alpha , T^\beta ]  \otimes \dK \otimes \langle W_\beta| \CR
&&  - \langle W_\alpha^+| \otimes T^\alpha \otimes \langle W_0| +  \langle W_\alpha^+| T^\alpha \otimes \dK  \otimes \langle W_0|  + \eta_{\alpha\beta} \langle W_0| [ T^\alpha , T^\gamma] \otimes T^\beta \otimes \langle W_\gamma^+| \; .\quad   \eea 
 This concludes the proof of  \eqref{eq:bbGamma minimal}.

One can now extract an integrability condition from the Leibniz identity. For this purpose it is convenient to introduce the notation 
\be \bbLambda_{1(23)} \equiv  \bbLambda_1 \circ \bigl(  \bbLambda_2 \circ \bbLambda_3\bigr)\; , \quad \bbLambda_{(12)3} \equiv  \bigl(  \bbLambda_1 \circ \bbLambda_2\bigr) \circ  \bbLambda_3 \; ,  \ee
such that the Leibniz identity reads 
\be  \bbLambda_{1(23)}  =  \bbLambda_{(12)3} +   \bbLambda_{2(13)}\; . \ee 
We will only compute the $\Lambda$ component. Using \eqref{eq:Dorf1} and  \eqref{eq:bbGamma minimal} one gets 
\bea | \Lambda_{1(23)} \rangle  \hspace{-2mm} &=&\hspace{-2mm}  \mathcal{L}_{\bbLambda_1}  \Bigl( \bigl(  \eta_{-1\, \alpha\beta}  \langle \theta | T^\alpha | \lambda_2\rangle +\eta_{\alpha\beta}  \langle \vartheta | T^\alpha | \lambda_2\rangle\bigr)  r^{-1} \mathcal{U}^{-1}  T^\beta |\lambda_3\rangle \Bigr)  \\
\hspace{-2mm} &=&\hspace{-2mm}   \langle {\partial}_{\theta,\vartheta} | r^{-1}  \mathcal{U}^{-1}  |\lambda_1\rangle  \bigl( \eta_{-1\, \alpha\beta} \langle \theta | T^\alpha |\lambda_2\rangle {+} \eta_{\alpha\beta} \langle \vartheta | T^\alpha |\lambda_2\rangle  \bigr)  r^{-1} \mathcal{U}^{-1}  T^\beta |\lambda_3\rangle  \CR
&& \hspace{-2mm}  {+}  \bigl( \eta_{-1\, \alpha\beta} \langle \theta | T^\alpha|\lambda_1\rangle  + \eta_{\alpha\beta} \langle \vartheta | T^\alpha |\lambda_1\rangle  \bigr)  \bigl( \eta_{-1\, \gamma\delta} \langle \theta | T^\gamma|\lambda_2\rangle  {+} \eta_{\gamma\delta} \langle \vartheta | T^\gamma|\lambda_2\rangle   \bigr) r^{-1} \mathcal{U}^{-1}   T^\beta T^\delta   |\lambda_3\rangle \nonumber \eea
whereas
\bea 
| \Lambda_{(12)3} \rangle  \hspace{-3mm} &=&\hspace{-3mm}  \mathcal{L}_{\bbLambda_1\circ \bbLambda_2 }   \Bigl(  r^{-1} \mathcal{U}^{-1}   |\lambda_3\rangle \Bigr)  \\
  \hspace{-3mm} &=& \hspace{-3mm} - \bigl( \eta_{-1\, \alpha\beta} \langle \theta | T^\alpha|\lambda_1\rangle  {+} \eta_{\alpha\beta} \langle \vartheta | T^\alpha  |\lambda_1\rangle \bigr)   \eta_{\gamma\delta} \langle {\partial}_{\theta,\vartheta}| r^{-1} \mathcal{U}^{-1}   T^\gamma T^\beta | \lambda_2\rangle  r^{-1} \mathcal{U}^{-1}  T^\delta|\lambda_3\rangle   \CR
\hspace{-3mm} && \hspace{-3mm}  - \bigl( \eta_{-1\, \alpha\beta} \langle \theta | T^\alpha|\lambda_1\rangle  {+} \eta_{\alpha\beta} \langle \vartheta | T^\alpha  |\lambda_1\rangle \bigr)   \langle {\partial}_{\theta,\vartheta}| r^{-1}  \mathcal{U}^{-1}   T^\beta | \lambda_2\rangle r^{-1} \mathcal{U}^{-1} | \lambda_3\rangle     \CR
\hspace{-3mm} &&  \hspace{-3mm} +  \bigl( \eta_{-1\, \alpha\beta} \langle \theta | T^\alpha |\lambda_1\rangle {+} \eta_{\alpha\beta} \langle \vartheta | T^\alpha |\lambda_1\rangle  \bigr)  \bigl( \eta_{-1\, \gamma\delta} \langle \theta | T^\gamma T^\beta |\lambda_2\rangle {+} \eta_{\gamma\delta} \langle \vartheta | T^\gamma T^\beta  |\lambda_2\rangle \bigr)   (r \mathcal{U})^{-1} T^\delta |\lambda_3\rangle  \CR
\hspace{-3mm} && \hspace{-3mm} {+} \bigl( \eta_{ \alpha\beta} \langle \theta | T^\alpha|\lambda_1\rangle  {+} \eta_{1\, \alpha\beta} \langle \vartheta | T^\alpha|\lambda_1\rangle   \bigr)  \eta_{-1\, \gamma\delta} \langle {\partial}_{\theta,\vartheta}| r^{-1} \mathcal{U}^{-1}   T^\gamma T^\beta |\lambda_2\rangle  r^{-1} \mathcal{U}^{-1}   T^\delta |\lambda_3\rangle  \nonumber\; .  
\eea
Using these expressions one obtains 
\bea \hspace{-2mm}&&\hspace{-2mm}   | \Lambda_{1(23)} \rangle  -  | \Lambda_{(12)3}\rangle  -   | \Lambda_{2(13)} \rangle  \\
\hspace{-2mm}&=&\hspace{-2mm}  - \eta_{-1\, \alpha\beta}   \langle \hspace{-1mm}\langle   \mathcal{I}_1|\hspace{-0.5mm}|  \bigl( |\lambda_1 \rangle \otimes T^\alpha |\lambda_2\rangle\bigr)  \, r^{-1} \mathcal{U}^{-1} T^\beta |\lambda_3\rangle  - \eta_{\alpha\beta}   \langle \hspace{-1mm}\langle   \mathcal{I}_0|\hspace{-0.5mm}| \bigl(  |\lambda_1 \rangle \otimes T^\alpha |\lambda_2\rangle \bigr) \, r^{-1} \mathcal{U}^{-1} T^\beta |\lambda_3\rangle\; , \nonumber \eea 
where we use the notation that 
\be  \langle \hspace{-1mm}\langle   A |\hspace{-0.5mm}|   \bigl(  |\lambda_1 \rangle \otimes |\lambda_2\rangle\bigr)  =\bigl(  \langle A_1 | \otimes \langle A_2 | \bigr) \bigl( |\lambda_1 \rangle \otimes |\lambda_2\rangle\bigr) = \langle A_1 | \lambda_1\rangle \langle A_2 | \lambda_2\rangle \; , \ee
and  
\bea 
  \langle \hspace{-1mm}\langle   \mathcal{I}_0|\hspace{-0.5mm}|   \hspace{-2mm} &=& \hspace{-2mm}  \eta_{-1\, \alpha\beta} \langle \theta | T^\alpha \otimes \langle \vartheta | T^\beta  + \eta_{ \alpha\beta} \langle \vartheta | T^\alpha \otimes \langle \vartheta | T^\beta  \\
& & \hspace{-2mm} - \Bigl( \eta_{-1\, \alpha\beta} \langle \theta | T^\alpha +  \eta_{ \alpha\beta} \langle \vartheta | T^\alpha\Bigr) \otimes \langle {\partial}_{\theta,\vartheta} | r^{-1} \mathcal{U}^{-1}   T^\beta + \langle \vartheta | \otimes \langle {\partial}_{\vartheta} | r^{-1} \mathcal{U}^{-1}   - \langle {\partial}_{\vartheta} | r^{-1} \mathcal{U}^{-1}   \otimes \langle \vartheta | \CR
  \langle \hspace{-1mm}\langle   \mathcal{I}_1|\hspace{-0.5mm}|   \hspace{-2mm} &=& \hspace{-2mm}  \eta_{-1\, \alpha\beta} \langle \theta | T^\alpha \otimes \langle \theta | T^\beta  + \eta_{ \alpha\beta} \langle \vartheta | T^\alpha \otimes \langle \theta | T^\beta + \langle \vartheta | \otimes \langle \theta | -  \langle \theta | \otimes \langle \vartheta | \CR
& & \hspace{-2mm} + \Bigl( \eta_{\alpha\beta} \langle \theta | T^\alpha +  \eta_{1\, \alpha\beta} \langle \vartheta | T^\alpha\Bigr) \otimes \langle {\partial}_{\theta,\vartheta} | r^{-1} \mathcal{U}^{-1}   T^\beta + \langle \theta | \otimes \langle {\partial}_{\theta} | r^{-1} \mathcal{U}^{-1}   -  \langle {\partial}_{\theta} | r^{-1} \mathcal{U}^{-1}   \otimes \langle \theta |\; . \nonumber \eea

\section{Shifted Maurer--Cartan equation for gauged supergravity}

In order to write the shifted Maurer--Cartan equation for gauged supergravity  compactly, we need to introduce some shifted versions of the embedding tensor.
In analogy with the definition of $[\bbF]_\alpha^{(m)}$ in \eqref{eq:shift F} and \eqref{eq:hat F}, we can define for any gauged supergravity 
\begin{equation}
\Bra{\Theta_\alpha^{(m)}}\otimes T^\alpha =
\bra{\Theta_\alpha}\otimes \cS^{\sugraupgamma}_m(T^\alpha) + \Bra{\widehat\Theta_m}\otimes\dK\,, 
\end{equation}
where the last term is the $\dK$ completion of the shifted embedding tensor.
It reads
\begin{equation}
\Bra{\widehat{\Theta}_m} = \bra\theta\cS^\sugraupgamma_m(L_{-1}) \,.
\end{equation}
We can also immediately generalise this expression for any embedding tensor, regardless of whether it admits an uplift in terms of a gSS ansatz.
Using then the expansion \eqref{eq:Theta virm exp}, one has $\Bra{\widehat\Theta_m} = 
\sum_{k=0}^\infty \bra{\Theta_{-k}}\cS^{\sugraupgamma}_{m}(L_{-k})$.

The gauged Maurer--Cartan equation for $P_\mu$ reads
\begin{equation}
DP + [Q\,,\,P] + \braket{\Theta_\alpha}{F}\frac12\big(  VT^\alpha V^{-1}+\mathrm{h.c.}\big) = 0\,,
\end{equation}
where in the second term we omitted the wedge product and the commutator is graded.
Applying a shift operator to this expression and following steps entirely analogous to Section~4.2 of \cite{Bossard:2021jix}, we then find 
\begin{align}\label{eq:shiftedMC gsugra}
\frac{1}{2\varrho}\cL^{\text{top}}_{\text{gsugra}}\,\dK =\ &
DP^{(1)} + [Q\,,\,P^{(1)}] 
+ \sum_{k=1}^\infty P_{k}\big( P^{(k+1)}-P^{(k-1)} \big)
+\frac12\Braket{\Theta^{(1)}_\alpha{+}\,\Theta^{(-1)}_\alpha}{F} \, V\,T^\alpha\,V^{-1}
\nonumber\\&
+\frac{c_\vir}{12}\,\sum_{k=2}^\infty (k^3-k) P_k\big( P_{k+1} +P_{k-1} \big)\ \dK\,.
\end{align}
The term in the second line is both $K(\mathrm{E}_9)$ and gauge invariant by itself and does not descend from the shifted Maurer--Cartan equation.
It is added for consistency of the equations of motion, following the same reasoning as in \cite{Bossard:2021jix}.
As in the rest of this paper, the constant $c_\vir$ denotes the Virasoro central charge in the representation in which $P_\mu$ are defined, which we always take to be the basic representation (so that $c_\vir=8$).
Equation \eqref{eq:shiftedMC gsugra} defines the topological term in a form entirely independent of the choice of basis in which we expand the currents.
This will become important in the companion paper \cite{SO9}, as different parametrisations of the currents are related to different duality frames.

\section{Gauge invariance of the Virasoro-extended scalar potential}\label{app:gaugeextpot}

We want to check the invariance of the potential \eqref{eq:extpot} under extended generalised diffeomorphisms. We will as usual decompose a gauge variation $\bbdelta_\mathbbm\Lambda$ as
\begin{equation}
\bbdelta_\mathbbm\Lambda=\cL_\mathbbm\Lambda+\Delta_\mathbbm\Lambda\,,
\end{equation} 
where $\Delta_\mathbbm\Lambda$ denotes the non-covariant part of the variation. The covariant part of the variation, which is captured by the generalised Lie derivative, reduces to a total internal derivative. This is ensured by the fact that the extended potential is an $\widehat{\mathrm{E}}_8\rtimes \text{Vir}^-$ scalar density, which scales under $L_0$ according to \eqref{eq:hevirmextpot}. This implies
\begin{equation}
\cL_\mathbbm\Lambda V_{\text{ExFT}}=\bra{\partial}\Big(\ket{\Lambda} V_{\text{ExFT}}\Big)\,.\label{eq:covvar pot}
\end{equation}
In the following, we will therefore only focus on the non-covariant part of the variations. For the scalar fields, these variations were presented in \cite{Bossard:2021jix} and we recall them here explicitly for convenience.  The coset scalars $\cV$ are gauge covariant, and we therefore simply have
\begin{equation} 
\Delta_\mathbbm\Lambda\mathcal{V}=0\,,\label{eq:gauge var coset}
\end{equation}
while the non-covariant variation of $\bra{\tilde \chi_1}$ can be written as
\begin{align}
2\,\Delta_{\mathbbm\Lambda} \bra{\tilde \chi_1}=&\, \Big(\hbm{\Lambda}^\upgamma_{ 1} + \hbm{\Lambda}^\upgamma_{- 1} + \omega^\alpha(\mathcal{V})\,[\mathbbm{\Lambda}]_\beta \,\bigl[ \cS_{ 1}^{\upgamma}(T^\beta)\bigr]_\alpha+ \omega^\alpha(\mathcal{V})\,[\mathbbm{\Lambda}]_\beta \,\bigl[ \cS_{- 1}^{\upgamma}(T^\beta)\bigr]_\alpha\Big)\bra{\partial_\mathbbm\Lambda}\nonumber\\[1mm]
&-\rho^{-1}\bra{\partial_\Sigma}\Sigma^{(1)}+\rho^{-1}\text{Tr}(\Sigma^{(1)})\bra{\partial_\Sigma}\,,\label{eq: gauge var internal chi}
\end{align}
in terms of the (Virasoro scalar-dependent) series of shift operators $\cS_{\pm 1}^\upgamma$ defined in \eqref{eq:Sgamma def}, as well as of the group cocycle $\omega^\alpha(\cV)$ defined in \eqref{eq:group cocycle def}. The combinations of scalars and gauge parameters $\hbm{\Lambda}^\upgamma_{\pm1}$ read
\begin{equation}
\hbm{\Lambda}^\upgamma_{\pm1}=-\bra{\partial_{\Lambda}} \cS^\upgamma_{\pm1}(L_0) \ket\Lambda 
- \sum_{q=1}^\infty \Tr\Big(\,\Sigma^\ord{n} \cS^\upgamma_{\pm1}(L_{-q})  \,\Big)\,.\label{eq:defhatL}
\end{equation}
It is important to note that in \eqref{eq: gauge var internal chi}, the derivatives only act on the parameters $\mathbbm\Lambda=(\ket{\Lambda},\Sigma^{(k)})$ and not on the scalars. The combinations \eqref{eq:defhatL} also correspond to the $\dK$ completions of the shifted projections
\begin{equation}
[\mathbbm{\Lambda}]^{(\pm 1)}_\alpha\, T^ \alpha=[\mathbbm{\Lambda}]_\alpha\,\cS^{\upgamma}_{\pm 1}(T^\alpha)+\hbm{\Lambda}^\upgamma_{\pm 1}\mathsf K\,,\label{eq:Kcomp Lambda}\,,
\end{equation}
where $[\mathbbm\Lambda]_\alpha$ was defined in \eqref{eq:squareProj}. As usual, they ensure covariance of the above expressions under rigid $\hevirm$ transformations. Note that this is in direct analogy with the shifted projections of field strengths \eqref{eq:shift F}.

In order to write down the non-covariant variations of the scalar currents, which follow from the variations \eqref{eq: gauge var internal chi} and \eqref{eq:gauge var coset}, let us first introduce an  ``underlined notation'' to denote the conjugation by $\mathcal{V}$. In analogy with Section~\ref{sec:GSS pot}, this notation will simplify the upcoming computations by preventing the explicit appearance of cocycles. More precisely, we define
\begin{align}
[\mathbbm{\underline\Lambda}]_\alpha \, T^\alpha=&\,\frac12[\mathbbm{\Lambda}]_\alpha \, \mathcal{V}T^\alpha\mathcal{V}^{-1}\,.\label{eq:ulambda}\\[1mm]
[\mathbbm{\underline\Lambda}]_\alpha^{(\pm 1)} \,T^\alpha=&\,\frac12 [\mathbbm{\Lambda}]^{(\pm 1)}_\alpha\,\cV T^\alpha\cV^{-1}\,,\label{eq:tildeLhatL}
\end{align}
These two definitions are related by  
\begin{equation}
[\mathbbm{\underline\Lambda}]^{(\pm 1)}_\alpha \, T^\alpha=[\mathbbm{\underline\Lambda}]_\alpha\,\cS_{\pm 1}\big(T^\alpha\big)+\tilde{\mathbbm{\Lambda}}_{\pm 1} \dK\,.\label{eq:ushiftL}
\end{equation}
where the $\dK$ completions $\tilde{\mathbbm{\Lambda}}_{\pm 1}$ ensure that the above expressions transform with the commutator under $K(\mathfrak e_9)$. They read
\begin{equation}
\tilde{\mathbbm{\Lambda}}_{\pm 1}=\frac12 \hbm{\Lambda}^\upgamma_{\pm 1} +\frac12 \omega^\alpha(\mathcal{V})\,[\mathbbm{\Lambda}]_\beta \,\cS_{\pm 1}^{\upgamma}(T^\beta)_\alpha\,.\label{eq:tildelambda}
\end{equation} 
in terms of the combinations \eqref{eq:defhatL}, and they can be used to rewrite the non-covariant variation \eqref{eq: gauge var internal chi} of $\bra{\tilde \chi_1}$ as 
\begin{equation}
\Delta_{\mathbbm\Lambda}\bra{\tilde \chi_1}=\big(\tilde{\mathbbm{\Lambda}}_{1}+\tilde{\mathbbm{\Lambda}}_{- 1}\big)\bra{\partial_\mathbbm\Lambda}-\tfrac12\rho^{-1}\bra{\partial_\Sigma}\Sigma^{(1)}+\tfrac12\rho^{-1}\text{Tr}(\Sigma^{(1)})\bra{\partial_\Sigma}
\end{equation}
Using the underlined notation, the non-covariant variations of the scalar currents \eqref{eq:P internal} and \eqref{eq: P internal} can then be simply written as
\begin{align}
\Delta_\mathbbm{\Lambda}\bra{\cP_\alpha}\otimes T^ \alpha=&\,[\mathbbm{\underline\Lambda}]_\alpha\,\bra{\partial_\mathbbm{\Lambda}}\otimes\big(T^\alpha+T^{\alpha\dagger}\big) \,,\label{eq:noncov P}\\[.5mm]
\Delta_\mathbbm{\Lambda}\bra{\cP^{( 1)}_\alpha}\otimes T^\alpha=&\,[\mathbbm{\underline\Lambda}]^{( 1)}_\alpha\,\bra{\partial_\mathbbm{\Lambda}}\otimes T^\alpha+[\mathbbm{\underline\Lambda}]^{(-1)}_\alpha\,\bra{\partial_\mathbbm{\Lambda}}\otimes T^{\alpha\dagger}\nonumber\\[.5mm]
&-\tfrac12\rho^{-1}\big(\bra{\partial_\Sigma}\Sigma^{(1)}-\text{Tr}(\Sigma^{(1)})\bra{\partial_\Sigma}\big)\otimes \dK\,.\label{eq:varshiftP}
\end{align}
We emphasise that the derivatives $\bra{\partial_\mathbbm\Lambda}$ only act on the gauge parameters and not on the scalars. At this point let us also remark that, unlike $\bra{\cP_\alpha}\otimes T^\alpha$, $[\mathbbm{\underline\Lambda}]_\alpha\, T^\alpha$ is only valued in $\mathfrak{\hat e}_8\oleft \mathfrak{vir}^-$ and is therefore not Hermitian. This also implies that $[\mathbbm{\underline\Lambda}]_\alpha^{(\pm 1)}\,T^{\alpha\dagger}\neq[\mathbbm{\underline\Lambda}]^{(\mp 1)}_\alpha\, T^\alpha$ and $\tilde{\mathbbm{\Lambda}}_{ 1}\neq\tilde{\mathbbm{\Lambda}}_{- 1}$.

Let us finally present two relations which will be useful for the upcoming computations. By using \eqref{eq:tildeLhatL}, \eqref{eq:Kcomp Lambda}, \eqref{eq:Sgamma expansion}, followed by the identity \eqref{eq:etaSid} and the section constraint, one can derive 
\begin{align}
[\mathbbm{{\underline\Lambda}}]^{(-1)}_\alpha\,\bra{\partial_*}\,\cV^{-1}T^\alpha\cV=&\,0\label{eq:noncovid1}\\
[\mathbbm{{\underline\Lambda}}]^{(+1)}_\alpha\,\bra{\partial_\mathbbm{\Lambda}}\,\cV^{-1}T^\alpha\cV=&-\tfrac12\rho^{-1}\text{Tr}(\Sigma^{(1)})\bra{\partial_\Sigma}+\tfrac12\rho^{-1}\bra{\partial_\Sigma}\Sigma^{(1)}\label{eq:noncovid2}\,,
\end{align}
where here $\bra{\partial_*}$ denotes any constrained `bra', and where $\bra{\partial_\mathbbm\Lambda}$ only acts on the gauge parameters.

We now consider the non-covariant variations of the terms in the extended potential \eqref{eq:extpot} one by one. We start from the simplest variations, which stem from $V_5$. Using \eqref{eq:noncov P}, we directly compute
\begin{equation}
\Delta_\mathbbm\Lambda V_5=2\, c_{\mathfrak{vir}}\sum\limits_{q=2}^{\infty}q(q-1)\,[\mathbbm{\underline\Lambda}]_{-q}\bra{\partial_\mathbbm{\Lambda}}\mathcal{M}^{-1}\ket{\cP_q}\,.\label{eq:gvarV5}
\end{equation}
We continue with the variation of $V_4$, which gives
\begin{align}
\Delta_{\mathbbm{\Lambda}} V_4&=\,[\mathbbm{\underline\Lambda}]_\alpha\bra{\cP_0}\mathcal{V}^{-1}(T^\alpha+T^\alpha{}^\dagger)\mathcal{V}^{-\dagger}\ket{\partial_{\mathbbm{\Lambda}}}-2\langle\partial_\Lambda|\Lambda\rangle\bra{\partial_\Lambda}\mathcal{V}^{-1}\,T^\alpha\,\mathcal{V}^{-\dagger}\ket{\cP_\alpha}\nonumber\\[.5mm]
&=\tfrac12\,[\mathbbm{\Lambda}]_\alpha\bra{\cP_0}T^\alpha\mathcal{M}^{-1}\ket{\partial_{\mathbbm{\Lambda}}}+\tfrac12[\mathbbm{\Lambda}]_\alpha\bra{\cP_0}\mathcal{M}^{-1}T^\alpha{}^\dagger\ket{\partial_{\mathbbm{\Lambda}}}-\tfrac12\langle\partial_\Lambda|\Lambda\rangle\bra{\partial_\Lambda}T^\alpha \mathcal{M}^{-1}\ket{\cJ_\alpha}\nonumber\\[.5mm]
&=\,-\tfrac14\langle\partial_\Lambda|\Lambda\rangle\bra{\cJ_0}\mathcal{M}^{-1}\ket{\partial_{\Lambda}}+\tfrac14\langle \cJ_0|\Lambda\rangle\bra{\partial_\Lambda}\mathcal{M}^{-1}\ket{\partial_{\Lambda}}-\tfrac12\langle\partial_\Lambda|\Lambda\rangle\bra{\partial_\Lambda}T^\alpha \mathcal{M}^{-1}\ket{\cJ_\alpha}\,,\label{eq:gvarV4}
\end{align}
where in the first line we used $\Delta_{\mathbbm \Lambda}\bra{\cP_0}=-\langle\partial_\Lambda|\Lambda\rangle\bra{\partial_\Lambda}$. To go from the second to the third line we applied the section constraint. In particular the second term vanishes. We also introduced the non-Hermitian current
\begin{equation}
\bra{\cJ_\alpha}\otimes T^\alpha=2\bra{\cP_\alpha}\otimes\cV^{-1}T^\alpha\cV=\bra{\partial_\cM}\otimes \cM^{-1}\cM\,.
\end{equation}
which we use to write terms that will ultimately combine into internal total derivatives. After a gauge-fixing of the Virasoro scalars, the current $\bra{\cJ_\alpha}\otimes T^\alpha$ also allows for more convenient comparison of the non-covariant variations with those of the unextended case presented in \cite{Bossard:2018utw}. Note that $\bra{\cJ_0}=2\bra{\cP_0}=-2\rho^{-1}\bra{\partial\rho}$. Let us now look at $V_2$. The two currents give the same contribution so we can write directly
\begin{align}
\Delta_{\mathbbm{\Lambda}} V_2&=\,2\,[\mathbbm{\underline\Lambda}]_\alpha\bra{\partial_\mathbbm{\Lambda}}\mathcal{V}^{-1}\,T^\beta(T^\alpha+T^\alpha{}^\dagger)\mathcal{V}^{-\dagger}\ket{\cP_\beta}\label{eq:gvarV2}\\[.5mm]
&=\,2\,[\mathbbm{\underline\Lambda}]_\alpha\bra{\partial_\mathbbm{\Lambda}}\mathcal{V}^{-1}\,[T^\beta,T^\alpha]\mathcal{V}^{-\dagger}\ket{\cP_\beta}+\tfrac12[\mathbbm{\Lambda}]_\alpha\bra{\partial_\mathbbm{\Lambda}}T^\alpha T^\beta \mathcal{M}^{-1}\ket{\cJ_\beta}+\tfrac12[\mathbbm{\Lambda}]_\alpha\bra{\partial_\mathbbm{\Lambda}}T^\beta \mathcal{M}^{-1}T^\alpha{}^\dagger\ket{\cJ_\beta}\nonumber\\[.5mm]
&=\,2\,[\mathbbm{\underline\Lambda}]_\alpha\bra{\partial_\mathbbm{\Lambda}}\mathcal{V}^{-1}\,[T^\beta,T^\alpha]\mathcal{V}^{-\dagger}\ket{\cP_\beta}-\tfrac12\langle\partial_\Lambda|\Lambda\rangle\bra{\partial_\Lambda}T^\beta \mathcal{M}^{-1}\ket{\cJ_\beta}+\tfrac12\langle \cJ_\beta|\Lambda\rangle\bra{\partial_\Lambda}T^\beta \mathcal{M}^{-1}\ket{\partial_\Lambda}\,.\nonumber
\end{align}
We again used the section constraint to move from the second to the third line. One also recovers directly the unextended results by gauge-fixing the Virasoro. Now on to $V_3$. Once again, the two shifted currents give the same contribution so we get
\begin{align}
\Delta_{\mathbbm{\Lambda}} V_3=&\,2\,[\mathbbm{\underline{\Lambda}}]^{(+1)}_\alpha\bra{\partial_{\mathbbm{\Lambda}}}\mathcal{V}^{-1}\,T^{\beta\dagger} T^\alpha\,\mathcal{V}^{-\dagger}\ket{\cP^{(1)}_\beta}+2\,[\mathbbm{\underline{\Lambda}}]^{(-1)}_\alpha\bra{\partial_{\mathbbm{\Lambda}}}\mathcal{V}^{-1}\,T^{\beta\dagger} T^\alpha{}^\dagger\,\mathcal{V}^{-\dagger}\ket{\cP^{(1)}_\beta}\nonumber\\[.5mm]
&\,-\rho^{-1}\bra{\partial_{\Sigma}}\Sigma^{(1)}\mathcal{V}^{-1}\,T^{\beta\dagger}\,\mathcal{V}^{-\dagger}\ket{\cP^{(1)}_\beta}+\rho^{-1}\,\text{Tr}(\Sigma^{(1)})\bra{\partial_{\Sigma}}\mathcal{V}^{-1}\,T^{\beta\dagger}\,\mathcal{V}^{-\dagger}\ket{\cP^{(1)}_\beta}\nonumber\\[.5mm]
=&\,2\,[\mathbbm{\underline{\Lambda}}]_\alpha\bra{\partial_{\mathbbm{\Lambda}}}\mathcal{V}^{-1}\,[\cS_{-1}(T^\beta), \cS_{1}(T^\alpha)]\,\mathcal{V}^{-\dagger}\ket{\cP_\beta}\nonumber\\[.5mm]
&\,+2\,[\mathbbm{\underline\Lambda}]^{(+1)}_\alpha\bra{\partial_{\mathbbm{\Lambda}}}\cV^{-1}T^\alpha T^{\beta\dagger}\cV^{-\dagger}\ket{\cP^{(1)}_\beta}+2\,[\mathbbm{\Lambda}]^{(-1)}_\alpha\bra{\partial_{\mathbbm{\Lambda}}}\cV^{-1}T^{\beta\dagger}T^\alpha{}^\dagger\cV^{-\dagger}\ket{\cP^{(1)}_\beta}\nonumber\\[.5mm]
&\,-\rho^{-1}\bra{\partial_{\Sigma}}\Sigma^{(1)}\mathcal{V}^{-1}\,T^{\beta\dagger}\,\mathcal{V}^{-\dagger}\ket{\cP^{(1)}_\beta}+\rho^{-1}\,\text{Tr}(\Sigma^{(1)})\bra{\partial_{\Sigma}}\mathcal{V}^{-1}\,T^{\beta\dagger}\,\mathcal{V}^{-\dagger}\ket{\cP^{(1)}_\beta}\nonumber\\[.5mm]
=&\,2\,[\mathbbm{\underline{\Lambda}}]_\alpha\bra{\partial_{\mathbbm{\Lambda}}}\mathcal{V}^{-1}\,[\cS_{-1}(T^\beta), \cS_{1}(T^\alpha)]\,\mathcal{V}^{-\dagger}\ket{\cP_\beta}\,.
\end{align}
To go from the second expression to the third we used \eqref{eq:noncovid1} and \eqref{eq:noncovid2}. To handle the remaining commutator of shift operators, we use the identity \eqref{eq:commutator shifts}. 
Note that this identity involves the Virasoro central charge in the basic representation $c_\mathfrak{vir}$, because the commutator is acting on a `bra' (or a `ket'). This will cancel against the $V_5$ contribution \eqref{eq:gvarV5}. The non-covariant variation of $V_3$ can therefore be written as 
\begin{align}
\Delta_{\mathbbm{\Lambda}} V_3=&\,2\,[\mathbbm{\underline{\Lambda}}]_\alpha\bra{\partial_{\mathbbm{\Lambda}}}\mathcal{V}^{-1}\,[T^\beta, T^\alpha]\,\mathcal{V}^{-\dagger}\ket{\cP_\beta}-2\,\eta^{\alpha\beta}[\mathbbm{\underline\Lambda}]_\alpha\bra{\partial_\mathbbm{\Lambda}}\mathcal{M}^{-1}\ket{\cP_\beta}\nonumber\\
&\,-2\,[\mathbbm{\underline\Lambda}]_0\bra{\partial_\mathbbm{\Lambda}}\mathcal{M}^{-1}\ket{\cP_\mathsf{K}}-2\,[\mathbbm{\underline\Lambda}]_\mathsf{K}\bra{\partial_\mathbbm{\Lambda}}\mathcal{M}^{-1}\ket{\cP_0}-\frac{c_{\mathfrak{vir}}}{2}\sum\limits_{q=2}^{\infty}q(q-1)\,[\mathbbm{\underline\Lambda}]_{-q}\bra{\partial_\mathbbm{\Lambda}}\mathcal{M}^{-1}\ket{\cP_q}\nonumber\\
&-2\sum\limits_{q\in\mathbb{Z}}[\mathbbm{\underline\Lambda}]_\alpha\bra{\partial_\mathbbm{\Lambda}}\mathcal{V}^{-1}\cS_q(T^\alpha)\mathcal{V}^{-\dagger}\ket{\cP_q}-2\sum\limits_{q=0}^{\infty}[\mathbbm{\underline\Lambda}]_{-q}\bra{\partial_\mathbbm{\Lambda}}\mathcal{V}^{-1}\cS_{-q}(T^\alpha)\mathcal{V}^{-\dagger}\ket{\cP_\alpha}\,.\label{eq:gvarV3}
\end{align}
Finally, let us look at $V_1$. We find 
\begin{align}
\Delta_{\mathbbm{\Lambda}} V_1=&\,4\,\eta^{\alpha\beta}[\mathbbm{\underline\Lambda}]_\alpha\bra{\partial_\mathbbm{\Lambda}}\mathcal{M}^{-1}\ket{\cP_\beta}\nonumber\\
&\,+4\sum\limits_{q=1}^\infty[\mathbbm{\underline\Lambda}]_{\alpha}\bra{\partial_\mathbbm{\Lambda}}\mathcal{V}^{-1}\big(\cS_q(T^\alpha)+\cS_{-q}(T^\alpha)^\dagger\big)\mathcal{V}^{-\dagger}\ket{\cP_q}\nonumber\\
&\,+4\sum\limits_{q=1}^\infty[\mathbbm{\underline\Lambda}]_{-q}\bra{\partial_\mathbbm{\Lambda}}\mathcal{V}^{-1}\cS_{-q}(T^\alpha)\mathcal{V}^{-\dagger}\ket{\cP_\alpha}\,.\label{eq:gvarV1}
\end{align}
Putting \eqref{eq:gvarV3} and \eqref{eq:gvarV1} together, we obtain 
\begin{align}
\Delta_{\mathbbm{\Lambda}} (2\,V_3+V_1)=&\,4\,[\mathbbm{\underline{\Lambda}}]_\alpha\bra{\partial_{\mathbbm{\Lambda}}}\mathcal{V}^{-1}\,[T^\beta, T^\alpha]\,\mathcal{V}^{-\dagger}\ket{\cP_\beta}-c_\mathfrak{vir}\sum\limits_{q=2}^{\infty}q(q-1)\,[\mathbbm{\underline\Lambda}]_{-q}\bra{\partial_\mathbbm{\Lambda}}\mathcal{M}^{-1}\ket{\cP_q}\label{eq:gvarV13}\\
&\,+4\sum\limits_{q=1}^\infty[\mathbbm{\underline{\Lambda}}]_\alpha\bra{\partial_{\mathbbm{\Lambda}}}\mathcal{V}^{-1}\big(\cS_{-q}(T^\alpha)^\dagger-\cS_{-q}(T^\alpha)\big)\mathcal{V}^{-\dagger}\ket{\cP_q}+\langle\partial_\Lambda|\Lambda\rangle\bra{\partial_{\Lambda}}T^\alpha\mathcal{M}^{-1}\ket{\cJ_\alpha}\,.\nonumber
\end{align}
To write the above last line, we have split the sums over $q$ in the last line of \eqref{eq:gvarV3} into their $q=0$ contribution and the rest. In particular, the $q=0$ contribution of the last term in \eqref{eq:gvarV3} gives the last term of \eqref{eq:gvarV13} after using the section constraint. One can subsequently observe that the two cocycles that get generated by moving $\cV$ inside of the shift operators in the first two terms of the second line cancel each other. Using the identity \eqref{eq:etaSid} and the section constraint, one can furthermore show that the remaining terms vanish, such that the first two term of the second line fully vanish.

By combining \eqref{eq:gvarV5}, \eqref{eq:gvarV4}, \eqref{eq:gvarV2}, and \eqref{eq:gvarV13}, we see that all non-covariant variations cancel except those written in terms of the current $\bra{\cJ_\alpha}\otimes T^\alpha$. Taking into account that $\Delta_\mathbbm\Lambda\,\rho=0$, we obtain the final result 
\begin{align}
\Delta_\mathbbm\Lambda V_{\text{ExFT}}=& \,\rho^{-1}\,\langle\partial_\Lambda|\Lambda\rangle\bra{\partial_\Lambda}T^\alpha\cM^{-1}\ket{\cJ_\alpha}-\tfrac12\rho^{-1}\langle\partial_\Lambda|\Lambda\rangle\langle\cJ_0|\cM^{-1}\ket{\partial_\Lambda}\nonumber\\[.5mm]
&- \rho^{-1}\,\langle\cJ_\alpha|\Lambda\rangle\bra{\partial_\Lambda}T^\alpha\cM^{-1}\ket{\partial_\Lambda}+\tfrac12 \rho^{-1}\langle \cJ_0|\Lambda\rangle\bra{\partial_\Lambda}\cM^{-1}\ket{\partial_\Lambda}\nonumber\\[.5mm]
=&\,\bra{\partial}\big(\rho^{-1}\ket{\Lambda}\bra{\partial_\Lambda}\cM^{-1}\ket{\partial_\Lambda}-\rho^{-1}\cM^{-1}\ket{\partial_\Lambda}\langle\partial_\Lambda|\Lambda\rangle\big)\,.
\end{align}
Together with \eqref{eq:covvar pot}, this proves that the extended scalar potential is gauge invariant up to internal total derivatives. After gauge fixing $\mathcal{V}\in\widehat{\mathrm{E}}_8\rtimes (\mathbb R^+_{L_0}\ltimes \mathbb{R}_{L_{-1}})$, the total derivatives match those identified in the unextended case \cite{Bossard:2018utw}.


\providecommand{\href}[2]{#2}\begingroup\raggedright\endgroup

\end{document}